\documentclass[journal]{IEEEtran}
\usepackage{enumitem}

\usepackage{amssymb,amsmath,color}
\usepackage{subfigure}
\usepackage{cite}
\usepackage{verbatim}
\usepackage{algorithm}
\usepackage{algorithmic}
\usepackage{graphicx,epstopdf}
\usepackage{url}
\usepackage{bm}
\usepackage{bbm}
\usepackage{amssymb}
\usepackage{amsmath}
\usepackage{dsfont}
\usepackage{setspace}

\newtheorem{theorem}{Theorem}

\newtheorem{lemma}{Lemma}

\newcommand{\minitab}[2][l]{\begin{tabular}{#1}#2\end{tabular}}

\usepackage{multirow}
\usepackage{multicol}

\begin{document}

\title{Power-Delay Tradeoff with Predictive Scheduling in Integrated Cellular and Wi-Fi Networks}

\author{Haoran Yu, Man Hon Cheung, Longbo Huang, and Jianwei Huang
\thanks{Manuscript received April 15, 2015; revised September 14, 2015; accepted December 11, 2015. 
The work of H. Yu, M. H. Cheung, and J. Huang was supported by the General Research Funds (Project Number CUHK 412713 and 14202814) established under the University Grant Committee of the Hong Kong Special Administrative Region, China.
The work of L. Huang was supported in part by the National Basic Research Program of China Grant 2011CBA00300, 2011CBA00301, the National Natural Science Foundation of China Grant 61033001, 61361136003, 61303195, Tsinghua Initiative Research Grant, Microsoft Research Asia Collaborative Research Award, and the China Youth 1000-talent Grant.
Part of this paper was presented in \cite{yu2014predictive}.}
\thanks{Haoran Yu, Man Hon Cheung, and Jianwei Huang (\{yh012, mhcheung, jwhuang\}@ie.cuhk.edu.hk) are with the Department of Information Engineering, the Chinese University of Hong Kong, Hong Kong, China. Longbo Huang (longbohuang@tsinghua.edu.cn) is with the Institute for Interdisciplinary Information Sciences, Tsinghua University, Beijing, China.}
}

\maketitle

\thispagestyle{empty}

\normalsize

\begin{abstract}
The explosive growth of global mobile traffic has lead to a rapid growth in the energy consumption in communication networks. In this paper, we focus on the energy-aware design of the network selection, subchannel, and power allocation in cellular and Wi-Fi networks, while taking into account the traffic delay of mobile users. The problem is particularly challenging due to the two-timescale operations for the network selection (large timescale) and subchannel and power allocation (small timescale). Based on the two-timescale Lyapunov optimization technique, we first design an online \emph{Energy-Aware Network Selection and Resource Allocation} (\textsf{ENSRA}) algorithm. The \textsf{ENSRA} algorithm yields a power consumption within $O\left( {\frac{1}{V}} \right)$ bound of the optimal value, and guarantees an $O\left( V \right)$ traffic delay for any positive control parameter $V$. Motivated by the recent advancement in the accurate estimation and prediction of user mobility, channel conditions, and traffic demands, we further develop a novel predictive Lyapunov optimization technique to utilize the predictive information, and propose a \emph{Predictive Energy-Aware Network Selection and Resource Allocation} (\textsf{P-ENSRA}) algorithm. We characterize the performance bounds of \textsf{P-ENSRA} in terms of the power-delay tradeoff theoretically. To reduce the computational complexity, we finally propose a \emph{Greedy Predictive Energy-Aware Network Selection and Resource Allocation} (\textsf{GP-ENSRA}) algorithm, where the operator solves the problem in \textsf{P-ENSRA} approximately and iteratively. {{Numerical results show that \textsf{GP-ENSRA} significantly improves the power-delay performance over \textsf{ENSRA} in the large delay regime. For a wide range of system parameters, \textsf{GP-ENSRA} reduces the traffic delay over \textsf{ENSRA} by $20\sim30$\% under the same power consumption.}}
\end{abstract}



\normalsize
\section{Introduction} \label{sec:introduction}
\IEEEPARstart{W}{ith} the explosive growth of global mobile data traffic, the energy consumption in communication networks has increased significantly. According to \cite{fehske2011global}, the information and communications technology industry constituted 2\% of global ${\rm CO}_2$ emissions. In addition, the high energy consumption in communication networks accounts for a significant proportion of the operational expenditure (OPEX) to the mobile operators \cite{oh2011toward}. Therefore, mobile operators have the incentives to reduce the energy consumption, through innovations in several areas such as novel hardware design, efficient resource management, and dynamic base station activations \cite{tutorialA,tutorialB}.

In this paper, we focus on the problem of energy-aware \emph{network selection} and \emph{resource allocation} (i.e., \emph{subchannel and power allocation}). First, since Wi-Fi networks often consume less energy than the macrocell network due to their smaller coverages and shorter communication distances \cite{ismail2011network}, the operator of an integrated cellular and Wi-Fi network can significantly reduce the system energy consumption by {offloading} part of the cellular traffic to the Wi-Fi networks.
Second, within the cellular network, the operator can reduce the transmission power while maintaining the system throughput by allocating the subchannels and power to the cellular users with good channel conditions. Since the reduction of the transmission power leads to the power reduction at the amplifiers and cooling systems, an efficient resource allocation can substantially reduce the macrocell network's total power consumption \cite{auer2011much}.

There are three major challenges in our problem. First, we consider a stochastic system where users' locations, channel conditions, and traffic demands change over time. This requires the operator to design an online algorithm that dynamically selects networks and allocates resources for users based on limited information of the future. Second, the resource allocation (regarding subchannel allocation and power allocation) is often performed much more frequently than the network selection. This requires the operator to determine the network selection and resource allocation in two different timescales, which makes the problem different from the often studied single-timescale control (\emph{e.g.}, \cite{neely2006energy}). Third, the operator needs to reduce the total power consumption while providing delay guarantees to all users. This requires the operator to keep a good balance between the power consumption and fairness among users.

\begin{table*}[t]\scriptsize
\centering
\vspace{-0.25cm}
\caption{Summary of Algorithm Design}
\vspace{-0.25cm}
\begin{tabular}{|c|c|c|c|c|}
\hline
{\minitab[c]{Algorithm}} & {\minitab[c]{Randomness Information}} & {\minitab[c]{Methodology}} & {\minitab[c]{Complexity}}& {\minitab[c]{Performance}}\\
\hline
{\minitab[c]{\textsf{ENSRA} (Sec. {\ref{sec:withoutprediction}})}} & {\minitab[c]{Current frame}} & {\minitab[c]{Two-timescale Lyapunov opt.}} & {\minitab[c]{Low}} & {\minitab[c]{Theoretical}}\\
\hline
{\minitab[c]{\textsf{P-ENSRA} (Sec. \ref{sec:withprediction})}} & {\minitab[c]{Current \& future frames}} & {\minitab[c]{Two-timescale Lyapunov opt.\\\!\!\!\!\!$+$ predictive Lyapunov opt.}} & {\minitab[c]{High}}& {\minitab[c]{Theoretical}}\\
\hline
{\minitab[c]{\textsf{GP-ENSRA} (Sec. \ref{sec:withprediction})}} & {\minitab[c]{Current \& future frames}}  & {\minitab[c]{Two-timescale Lyapunov opt.\\\!\!\!\!\!$+$ predictive Lyapunov opt.}} & {\minitab[c]{Low (heavy traffic)\\Medium (non-heavy traffic)}}& {\minitab[c]{Numerical}}\\
\hline
\end{tabular}\label{table:summary}
\vspace{-0.6cm}
\end{table*}
In the first part of this paper, we apply the two-timescale Lyapunov optimization technique \cite{yao2012data} to design an online \emph{Energy-Aware Network Selection and Resource Allocation} (\textsf{ENSRA}) algorithm.{\footnote{{Lyapunov optimization is widely used for solving scheduling and resource allocation problems in stochastic networks, mainly due to its low computational complexity even for a stochastic system with a large number of system states. Moreover, Lyapunov optimization does not require the prior knowledge on the statistical information of the system randomness. 
}}} 
We show that \textsf{ENSRA} yields a power consumption that can be pushed arbitrarily close to the optimal value, at the expense of an increase in the average traffic delay.

In the second part of this paper, motivated by the recent advancement of accurate estimation of users' mobilities \cite{nicholson2008breadcrumbs}, traffic demands \cite{paul2011understanding}, and channel conditions \cite{arslan2007channel}, we improve the performance of \textsf{ENSRA} by incorporating the prediction of the system randomness into the algorithm design.
{{The main idea is that if the operator knows that the users will experience good channel conditions or be covered by high-capacity Wi-Fi networks in the next few frames, the operator will not serve the users by the macrocell network in the current frame. This can reduce the time average power consumption and achieve a better power-delay tradeoff.}}
However, designing such a predictive algorithm is challenging, as the state space grows exponentially with the size of the information window.{\footnote{{{We define the system state of a frame as the realization of the random events (\emph{i.e.}, users' channel conditions, locations, and traffic arrivals) during the frame. Then the state space of a frame is simply the set of all possible realizations of the random events during the frame. Suppose the state space of a frame has a size of $J$. Then if we consider the predictive scheduling over a window with $W$ frames, the state space of the window will have a size of $J^{W}$, which is exponentially increasing in $W$. Different from dynamic programming, Lyapunov optimization does not need to consider all possible states when making decisions for the current frame. Hence, the exponential increase in the state space does not significantly increase the complexity of the predictive algorithm design under Lyapunov optimization.}}}} This makes it infeasible to apply the common dynamic programming technique.
Instead, we design a \emph{Predictive Energy-Aware Network Selection and Resource Allocation} (\textsf{P-ENSRA}) algorithm through a novel predictive Lyapunov optimization technique
. {{Different from the previous Lyapunov optimization techniques in \cite{neely2010stochastic, yao2012data}, we introduce a novel control parameter $\theta$ to optimize the operations within the entire information window.}} By properly adjusting $\theta$, we can balance the variance of queue length within each information window, and significantly improve the delay performance. We also characterize the performance bounds of \textsf{P-ENSRA} as functions of $\theta$.

To reduce the computational complexity of \textsf{P-ENSRA}, we further propose a  \emph{Greedy Predictive Energy-Aware Network Selection and Resource Allocation} (\textsf{GP-ENSRA}) algorithm, where the operator solves the optimization problem in \textsf{P-ENSRA} approximately and iteratively. {{Our numerical results show that \textsf{GP-ENSRA} achieves a much better power-delay tradeoff than \textsf{ENSRA} in the large delay regime, and the improvement increases with the prediction window size.}}

To the best of our knowledge, this is the first work that proposes energy-aware network selection and resource allocation algorithms in the stochastic cellular and Wi-Fi networks. We summarize our algorithms in Table \ref{table:summary}. The main contributions of this paper are as follows:

\begin{itemize}
\item \emph{Online two-timescale scheduling}: 
    {{We study the two-timescale online network selection and resource allocation problem for a stochastic multi-user and multi-network system.}}

\item \emph{Novel predictive Lyapunov optimization technique}: 
    {{We develop a novel predictive Lyapunov optimization technique, 
    and characterize the power and delay tradeoff theoretically.}}

\item \emph{Performance improvement with prediction}: {{Simulation results show that the predictive future information significantly improves the power-delay performance in the large delay regime.}} For a wide range of system parameters, the predictive algorithm reduces the traffic delay by $20\sim30$\% over the non-predictive algorithm under the same power consumption.
\end{itemize}

There are many literatures studying either energy-aware network selection or energy-aware resource allocation problems. For example,
Venturino \emph{et al.} in \cite{venturino2014energy} studied energy-efficient resource allocation and base station coordination in a static downlink cellular system.
Xiong \emph{et al.} in \cite{xiong2012energy} investigated energy-efficient resource allocation under quality-of-service constraints, in a static cellular system with both downlink and uplink communications.
Meshkati \emph{et al.} in \cite{meshkati2009energy} used a game-theoretic approach to analyze the energy-efficient power and rate control problem.
However, none of these literatures studied joint energy-aware network selection and resource allocation in the stochastic cellular and Wi-Fi networks, which is the focus of our work.

{{There are also some literatures using Lyapunov optimization to design scheduling algorithms for wireless networks considering the power-delay tradeoff. Neely in \cite{neely2007optimal} analyzed the power allocation with the optimal power-delay tradeoff for a multi-user wireless downlink system.
Li \emph{et al.} in \cite{li2014energy} and \cite{li2015throughput} studied the energy-efficient power allocation in interference-free wireless networks based on Lyapunov optimization.
Lakshminarayana \emph{et al.} in \cite{lakshminarayanatransmit} investigated the transmit power minimization problem in small cell networks. 
However, these references focused on the power allocation, without the detailed consideration of the channel allocation and network selection. Moreover, none of them studied the predictive algorithm design with the future information.}}

The rest of the paper is organized as follows. In Sections \ref{sec:systemmodel} and \ref{sec:problemformulation}, we introduce the system model and formulate the problem. In Sections \ref{sec:withoutprediction} and \ref{sec:withprediction}, we study the non-predictive and predictive network selection and resource allocation, respectively. We present the numerical results in Section \ref{sec:simulation}, and conclude the paper in Section \ref{sec:conclusion}.
\vspace{-0.3cm}
\section{System Model} \label{sec:systemmodel}
\begin{figure}[t]
  \centering
  \includegraphics[scale=0.33]{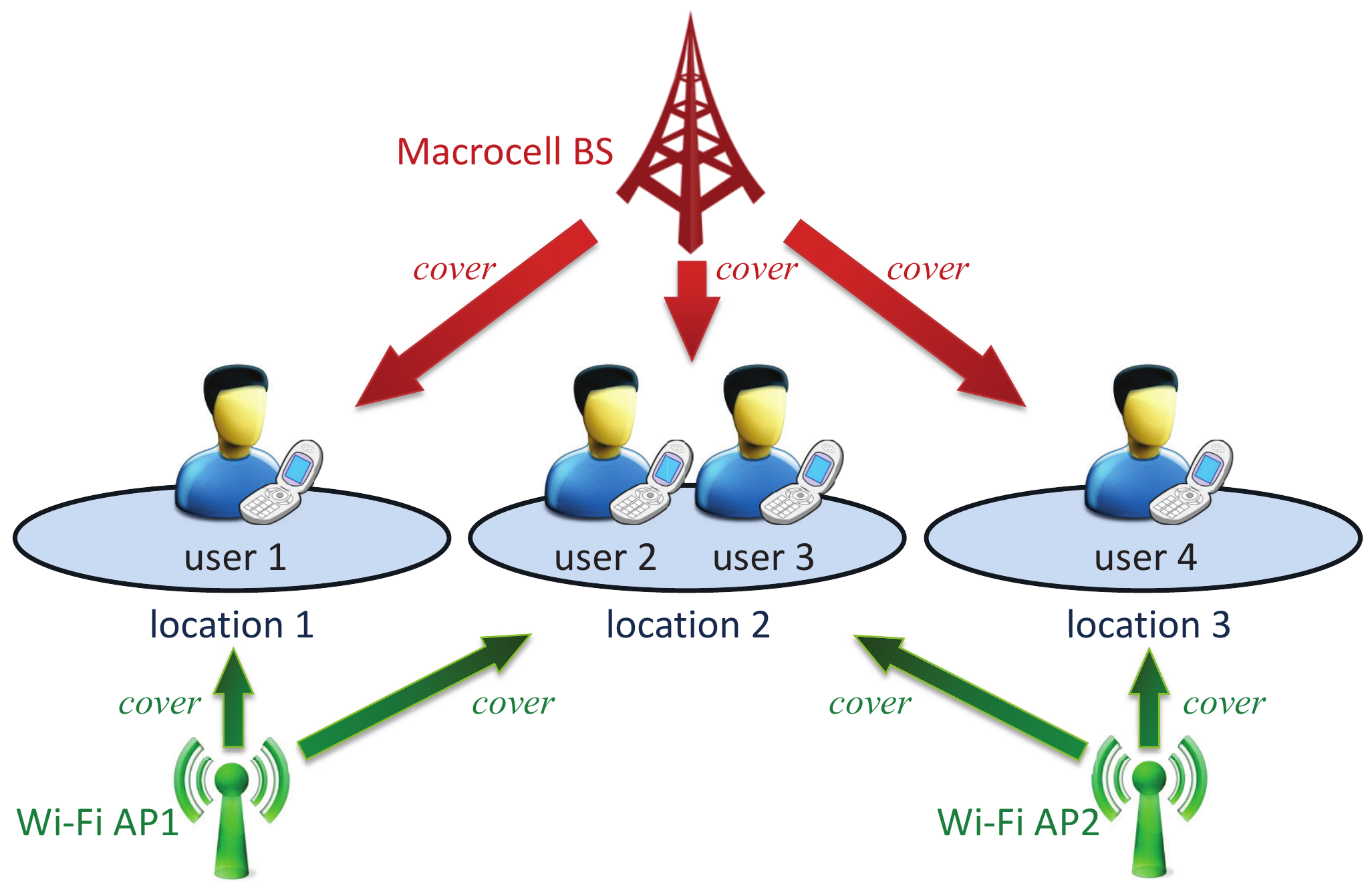}
  \vspace{-0.2cm}
  \centering
  \caption{{An example of the system model, where user 1, 2, 3, and 4 are moving within the set of locations ${\cal S} = \left\{ {1,2,3} \right\}$. The macrocell covers all locations. Each location is covered by a set of Wi-Fi networks, \emph{e.g.}, ${{\cal N}_1} = \left\{ 1\right\}$, ${{\cal N}_2} = \left\{ 1,2\right\}$, ${{\cal N}_3} = \left\{ 2\right\}$.}}
  \label{fig:Systemfigure}
  \vspace{-0.3cm}
\end{figure}

We consider the downlink transmission in a slotted system, indexed by $t \in \left\{0,1,\ldots\right\}$. We focus on the monopoly case, where the single operator serves users by its own macrocell and Wi-Fi networks.{\footnote{For example, AT\&T serves its users with both the cellular network and more than 30,000 Wi-Fi hotspots in the US \cite{ATTWiFi}.}} We introduce the following notations:
\begin{itemize}
\item ${\cal L} \triangleq \left\{ {1,2, \ldots ,L} \right\}$: set of the users;
\item ${\cal N} \triangleq \left\{ {1,2, \ldots ,N} \right\}$: set of the Wi-Fi networks;
\item ${\cal S} \triangleq \left\{ {1,2, \ldots ,S} \right\}$: set of the locations.
\end{itemize}
We assume that the macrocell base station covers all $S$ locations, and we use ${{\cal N}_s} \subseteq {\cal N}$ to denote the set of available Wi-Fi networks at location $s \in {\cal S}$. We illustrate the system model through an example in Figure \ref{fig:Systemfigure}.
\vspace{-0.1cm}
\subsection{Two-timescale operations}\label{subsec:two-timescale}
The operator aims at reducing the total power consumption through the network selection, subchannel allocation, and power allocation.
We assume that the network selection is operated in a larger timescale than the subchannel and power allocation. This is because a frequent switch among different networks interrupts the data delivery and incurs a nonnegligible cost (\emph{e.g.}, in the form of energy consumption, quality-of-service degradation, and delays).

\begin{figure}[t]
  \centering
  \includegraphics[scale=0.38]{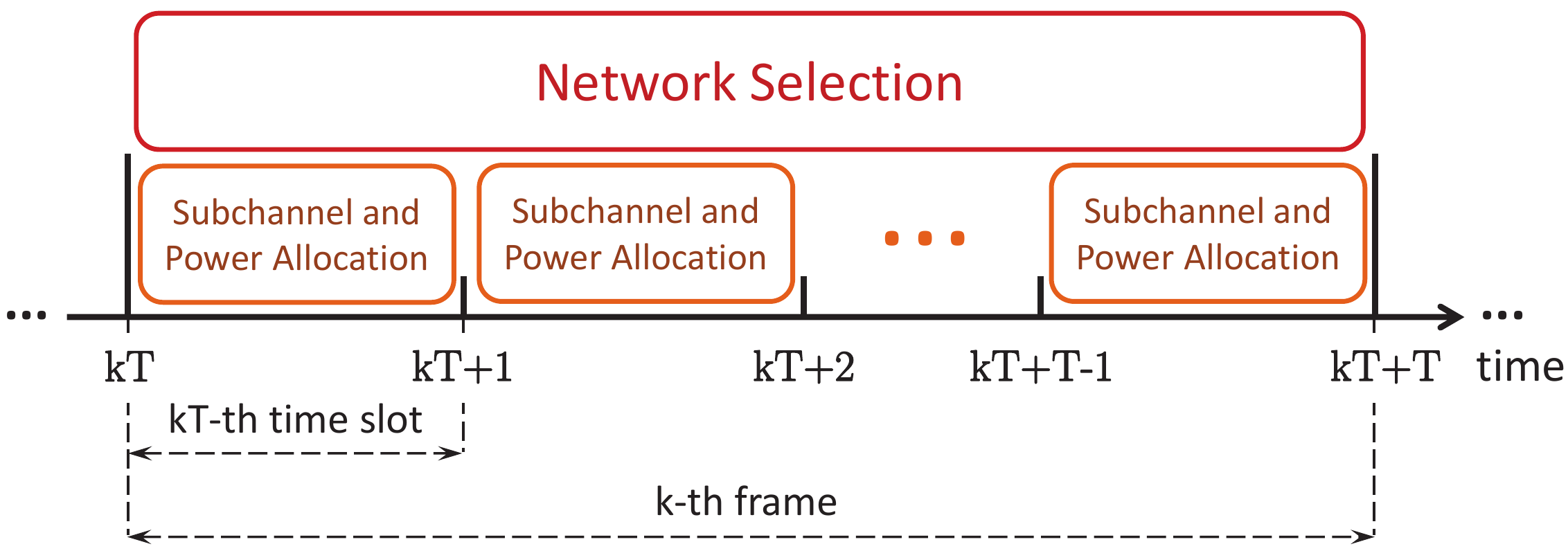}
  \vspace{-0.2cm}
  \caption{Two-timescale operations: (a) at time $t=kT$, \emph{e.g.}, the beginning of the $k$-th frame, the operator determines the network selection for the $k$-th frame; (b) at time $t\in{\cal T}_k$, the operator determines the subchannel and power allocation for time slot $t$.}
 \label{fig:TwoScale:a}
 \vspace{-0.3cm}
\end{figure}

We refer every $T$ time slots as a \emph{frame},\footnote{{In our simulation in Section \ref{sec:simulation}, we choose $1$ time slot to be $10$ milliseconds and $1$ frame to be $1$ second.}}
and define the $k$-th frame ($k \in {\mathbb N}$) as the time interval that contains a set ${{\cal T}_k}\triangleq\left\{ {kT,kT + 1, \ldots ,kT + T - 1} \right\}$ of time slots. We assume that:
\begin{itemize}
\item the operator determines network selection at the beginning of every frame (\emph{large-timescale});
\item the operator determines subchannel and power allocation at the beginning of every time slot (\emph{small-timescale}).
\end{itemize}
We illustrate such a two-timescale structure in Figure \ref{fig:TwoScale:a}.
\vspace{-0.1cm}
\subsection{Frame-based network selection}\label{subsec:select}
At time slot $t=kT$, \emph{i.e.}, the beginning of the $k$-th frame, the operator determines the network selection for the $k$-th frame. We denote the network selection by ${\bm \alpha }\left( kT \right) = \left(\alpha_l\left(kT\right),\forall l\in\mathcal{L}\right)$, where ${\alpha _l}\left( kT \right)$ indicates the network that user $l$ is connected to during the $k$-th frame. Let the random variable ${S_l}\left( kT \right) \in {\cal S}$ be user $l$'s location during the $k$-th frame, and define ${\bm S }\left( kT \right) = \left(S_l\left(kT\right),\forall l\in\mathcal{L}\right)$.{\footnote{User locations ${\bm S}\left( kT \right)$ do not change during the frame. The reason is that the user location usually changes much less frequently than the other types of randomness, \emph{e.g.}, the channel condition in the macrocell network.}} Since the availabilities of Wi-Fi networks are location-dependent, we have the following constraint for ${\bm \alpha }\left( kT \right)$:
\begin{equation}
{\alpha _l}\left( kT \right) \in {{\cal N}_{{S_l}\left( kT \right)}} \cup \left\{ 0 \right\},{\rm{~}}  \forall l \in \mathcal{L},k=0,1,\ldots,\label{equ:feasibility}
\end{equation}
where selection ${\alpha _l}\left( kT \right) = 0$ indicates that user $l$ is connected to the macrocell network.
\vspace{-0.3cm}
\subsection{Macrocell network model}\label{subsec:cellular}
We consider an Orthogonal Frequency Division Multiplexing (OFDM) system for the macrocell network,{\footnote{OFDM is one of the core technologies of the 4G cellular network \cite{docomo5G}.}} following the standard model as used in \cite{huang2009downlink,shen2005adaptive}.
\subsubsection{Subchannel allocation}
Let ${\cal M}\triangleq \left\{1,2,\ldots,M\right\}$ be the set of subchannels, and denote the subchannel allocation by ${\bm x }\left( t \right) = \left(x_{lm}\left(t\right),\forall l\in{\mathcal{L}}, m\in{\mathcal {M}}\right)$. Variable $x_{lm}\left(t\right)\in\left\{0,1\right\}$ for all $l$ and $m$: if user $l$ is allocated with subchannel $m$, $x_{lm}\left(t\right)=1$; otherwise, $x_{lm}\left(t\right)=0$. We assume that each subchannel can at most be allocated to one user:
\begin{equation}
\sum\limits_{l=1}^{L} {x_{lm}}\left(t\right)  \le 1, \forall m\in\cal M.\label{equ:subchannelconstraint}
\vspace{-0.15cm}
\end{equation}
Different from the frame-based network selection ${\bm \alpha }\left( kT \right)$, the operator determines the subchannel allocation ${\bm x }\left( t \right)$ every time slot. Since the operator can only allocate subchannels to those users who are connected to the cellular network, we have the following constraint for ${\bm x }\left( t \right)$:
\begin{equation}
{\alpha_l\left(t_T\right)}{x_{lm}\left(t\right)}=0, \forall l\in{\cal L}, m\in{\cal M},t\ge0.\label{equ:constraintforresource:a}
\end{equation}
Here, ${t_T} \triangleq {\left\lfloor {\frac{t}{T}} \right\rfloor}T$ is the beginning of the frame that time slot $t$ belongs to, and network selection ${\alpha_l\left(t_T\right)}$ indicates user $l$'s associated network during the frame.
\subsubsection{Power allocation}
We denote the power allocation by ${\bm p }\left( t \right) = \left(p_{lm}\left(t\right),\forall l\in{\mathcal{L}}, m\in{\mathcal {M}}\right)$. Variable $p_{lm}\left(t\right)\ge0$ denotes the power allocated to user $l$ on subchannel $m$. We have the following power budget constraint:
\begin{equation}
\sum\limits_{m=1}^{M}{\sum\limits_{l=1}^{L}{p_{lm}\left(t\right)}}\le P_{\max}^C, \forall t\ge0.\label{equ:powerbudeget}
\end{equation}
Similar as (\ref{equ:constraintforresource:a}), the operator can only allocate the power to those users who are connected to the cellular network. We have the following constraint for ${\bm p }\left( t \right)$:{\footnote{{It is possible to explicitly write out the constraint that if $x_{lm}\left(t\right)=0$, then $p_{lm}\left(t\right)=0$, \emph{i.e.}, the power cannot be allocated to a user-channel pair unless the channel is assigned to that user. However, such a constraint is automatically satisfied by all decisions made under our algorithms, as choosing $p_{lm}\left(t\right) > 0$ with $x_{lm}\left(t\right)=0$ only increases the power consumption but does not serve users' traffic.}}}
\begin{equation}
{\alpha_l\left(t_T\right)}{p_{lm}\left(t\right)}=0, \forall l\in{\cal L}, m\in{\cal M},t\ge0.\label{equ:constraintforresource:b}
\end{equation}
\subsubsection{Macrocell transmission rate}\label{subsubsec:macrocellrate}
We use ${\bm H}\left(t\right)=\left(H_{lm}\left(t\right),\forall l\in{\cal L},m\in{\cal M}\right)$ to denote the channel conditions, where $H_{lm}\left(t\right)$ is a random variable that represents the channel condition for user $l$ on subchannel $m$ at time slot $t$. Given the subchannel allocation ${\bm x}^l\left(t\right)=\left(x_{lm}\left(t\right),\forall m\in{\mathcal {M}}\right)$ and power allocation ${\bm p}^l\left(t\right)=\left(p_{lm}\left(t\right),\forall m\in{\mathcal {M}}\right)$, the transmission rate of a cellular user $l$ (\emph{i.e.}, ${\alpha_l\left(t_T\right)}=0$) at time slot $t$ is{\footnote{{Since we study the problem within the coverage of one macrocell base station, we do not consider the interference from neighboring cells. Similar interference-free assumption has been commonly used in prior literatures on the study of the single cell transmission problem \cite{huang2009downlink,li2014energy,li2015throughput}.}}}
\begin{equation}
{r_l^{C}}\bigl( {\bm x}^l\left(t\right), {\bm p}^l\left(t\right)\bigr) = \!\!\frac{B}{M}\!\sum\limits_{m = 1}^M {{x_{lm}\left(t\right)}{\log_2 \left( {1 \!+ \!\frac{{{p_{lm}\left(t\right)}H_{lm}^2\left( t \right)}}{{{N_0}\frac{B}{M}}{}}} \right)}},\label{equ:macrocellrate}
\end{equation}
where $B$ is the total bandwidth and $N_0$ is the noise power spectral density.
\subsubsection{Macrocell power consumption}
According to \cite{auer2011much}, the power consumption of the macrocell base station contains two components: the first component is a fixed term that measures the radio frequency (RF) and baseband unit power consumptions; the second component corresponds to the transmission power. Since the first component is fixed,{\footnote{{Some references, \emph{e.g.}, \cite{oh2013dynamic}, considered turning off the macrocell base stations to save the RF and baseband unit power consumptions when no user is connected to the macrocell networks. Nevertheless, in our work, we consider one macrocell network. According to the simulation, the operator serves at least one user in the macrocell network for most of the time. Even if the macrocell network is idle for a short time period (\emph{e.g.}, several frames), turning the macrocell network off during such a short period does not significantly save the power consumption, and the turning on/off process incurs some switching costs in practice. Therefore, we do not consider the potential power saving by dynamically turning on and off the base station.}}} in our model, {{we focus on minimizing the time average of the second component,}} which is given by
\begin{equation}
P^{C}\left({{\bm p}\left(t\right)}\right)=\kappa  \sum\limits_{m=1}^M{\sum\limits_{l=1}^L{p_{lm}\left(t\right)}}.\label{equ:macropower}
\end{equation}
Here, parameter $\kappa $ is the scale factor that depends on the power amplifier efficiency and the losses incurred by the antenna feeder, power supply, and cooling \cite{auer2011much}.
\vspace{-0.3cm}
\subsection{Wi-Fi network model}\label{subsec:Wi-Fi}
Let ${\rho_n}$ be the number of users associated with Wi-Fi network $n$. We assume that Wi-Fi network $n$'s total transmission rate and power consumption are functions of ${\rho_n}$, and we denote them by $R_n\left({\rho_n}\right)$ and $P_n^W\left({\rho_n}\right)$, respectively. We further assume that $R_n\left({\rho_n}\right)$ and $P_n^W\left({\rho_n}\right)$ are non-negative bounded functions, \emph{i.e.}, there exist positive constants ${R_{n,\max }}$ and $P_{n,\max }^W$ such that
\vspace{-0.1cm}
\begin{equation}
0 \le {R_n}\left( {\rho_n}  \right) \le {R_{n,\max }} {\rm ~and~}0 \le P_n^W\left( {\rho_n}  \right) \le P_{n,\max }^W\label{equ:boundedWiFi}
\end{equation}
for all ${\rho_n}=0,1,2,\ldots$.

We allow general functions of $R_n\left({\rho_n}\right)$ and $P_n^W\left({\rho_n}\right)$ that satisfy (\ref{equ:boundedWiFi}) in our algorithm design in Sections \ref{sec:withoutprediction} and \ref{sec:withprediction}. In Section \ref{sec:simulation}, we apply the transmission rate function $R_n\left({\rho_n}\right)$ defined in \cite{bianchi2000performance}, and the power consumption function $P_n^W\left({\rho_n}\right)$ defined in \cite{jung2012adaptive} for simulation.
\subsubsection{Wi-Fi transmission rate}\label{subsubsec:WiFirate}
Given function $R_n\left({\rho_n}\right)$ and network selection ${\bm \alpha}\left(t_T\right)$, we can compute the transmission rate of a Wi-Fi user $l$ (\emph{i.e.}, $\alpha_l\left(t_T\right)>0$) at time slot $t$ by \cite{Michael2014}:{\footnote{We assume that all users in the same Wi-Fi network compete on the same channel, and different close-by Wi-Fi networks choose different channels. Hence the transmission rate of a user in Wi-Fi only depends on the total number of users competing for the same Wi-Fi. We will consider the interferences among Wi-Fi networks in the future.}}
\begin{equation}
r_l^W\left( {{\bm \alpha} \left( {{t_T}} \right)} \right) = \frac{{{R_{\alpha_l\left(t_T\right)}}\left( {\sum\limits_{k = 1}^L {{{\mathds 1}_{\left\{{{\alpha _k}\left( {{t_T}} \right) = {\alpha _l}\left( {{t_T}} \right)}\right\}}}} }  \right)}}{{\sum\limits_{k = 1}^L {{{\mathds 1}_{\left\{{\alpha _k}\left( {{t_T}} \right) = {\alpha _l}\left( {{t_T}} \right)\right\}}}} } }.\label{equ:userWiFirate}
\end{equation}
Here, summation ${{\sum\limits_{k = 1}^L {{{\mathds 1}_{{\alpha _k}\left( {{t_T}} \right) = {\alpha _l}\left( {{t_T}} \right)}}} } }$ returns the number of users in the Wi-Fi network that user $l$ is associated with.{\footnote{${\mathds 1}_{\left\{\cdot\right\}}$ is the indicator function, which equals $1$ if the event in the brace is true, and equals $0$ if the event is false.}}
\subsubsection{Wi-Fi power consumption}
Given function $P_n^W\left({\rho_n}\right)$ and network selection ${\bm \alpha}\left(t_T\right)$, we can compute the power consumption of all Wi-Fi networks as:
\begin{equation}
{P^W}\left( {{\bm \alpha} \left( {{t_T}} \right)} \right) = \sum\limits_{n = 1}^N {P_n^W\left( {\sum\limits_{l = 1}^L {{{\mathds 1}_{\left\{ {{\alpha _l}\left( {{t_T}} \right) = n} \right\}}}} } \right)}.\label{equ:allWiFipower}
\end{equation}
\subsection{Users' traffic model}\label{subsec:traffic}
We assume that the users randomly generate traffic, and the traffic generation is not affected by the operator's operations. We use a random variable $A_l\left(t\right)$ to denote the traffic arrival rate of user $l\in{\cal L}$ at time slot $t$, and let ${\bm A}\left(t\right)=\left(A_l\left(t\right),l\in{\cal L}\right)$. We assume that there exists a positive constant $A_{\max}$ such that
\begin{equation}
0 \le {A_l}\left( t \right) \le {A_{\max }}, \forall l\in{\cal L}, t \geq 0.\label{equ:arrivalbound}
\end{equation}
\subsection{Summary}\label{subsec:summary}
\subsubsection{Macrocell $+$ Wi-Fi transmission rate}
If a user is associated with the macrocell network, its transmission rate is given by ${{r_l^{C}}\left( {\bm x}^l\left(t\right), {\bm p}^l\left(t\right)\right)}$ in (\ref{equ:macrocellrate}); if it is associated with Wi-Fi networks, its transmission rate is given by ${r_l^W\left( {{\bm \alpha} \left( {{t_T}} \right)} \right)}$ in (\ref{equ:userWiFirate}). To summarize, user $l$'s transmission rate at time slot $t$ is given by
\begin{align}
{r_l}\bigl( {{\bm \alpha} \left( {{t_T}} \right)\!,\!{{\bm x}^l}\left( t \right)\!,\!{{\bm p}^l}\left( t \right)} \bigr) \!=\! \left\{ {\begin{array}{*{20}{l}}
{{r_l^{C}}\!\bigl( {\bm x}^l\!\left(t\right)\!, {\bm p}^l\!\left(t\right)\bigr),}&{{\rm if~\!}{\alpha_l\!\left(t_T\!\right)\!=0,}}\\
{r_l^W\bigl( {{\bm \alpha} \left( {{t_T}} \right)} \bigr),}&{{\rm otherwise}.}
\end{array}} \right.\label{equ:totalrate}
\end{align}
Because of the power budget constraint (\ref{equ:powerbudeget}) in the macrocell network, function ${{r_l^{C}}\left( {\bm x}^l\left(t\right), {\bm p}^l\left(t\right)\right)}$ is upper bounded. Furthermore, since Wi-Fi networks' total transmission rates are upper bounded as in (\ref{equ:boundedWiFi}), function $r_l^W\left( {{\bm \alpha} \left( {{t_T}} \right)} \right)$ is also upper bounded. As a result, there exists a positive constant $r_{\max}$ such that
\begin{equation}
0 \le {{r_l}\bigl( {{\bm \alpha} \left( {{t_T}} \right),{{\bm x}^l}\left( t \right),{{\bm p}^l}\left( t \right)} \bigr)} \le {r_{\max }}\label{equ:ratebound}
\end{equation}
for all $l\in{\cal L}$ and ${{\bm \alpha} \left( {{t_T}} \right),{{\bm x}^l}\left( t \right),{{\bm p}^l}\left( t \right)}$ satisfying (\ref{equ:feasibility}), (\ref{equ:subchannelconstraint}), (\ref{equ:constraintforresource:a}), (\ref{equ:powerbudeget}), and (\ref{equ:constraintforresource:b}).
\subsubsection{Macrocell $+$ Wi-Fi power consumption}
The operator considers the power consumption in both the macrocell and Wi-Fi networks. The macrocell network's power consumption is given by $P^C\left({{\bm p}\left(t\right)}\right)$ in (\ref{equ:macropower}), and Wi-Fi networks' total power consumption is given by $P^W\left({\bm \alpha}\left(t_T\right)\right)$ in (\ref{equ:allWiFipower}). Therefore, the operator's total power consumption at time slot $t$ is given by
\begin{equation}
P\bigl( {{\bm \alpha}\left(t_T\right),{\bm p}\left(t\right)} \bigr)=P^C\left({{\bm p}\left(t\right)}\right)+P^W\left({\bm \alpha}\left(t_T\right)\right).
\end{equation}
According to the cellular power budget constraint (\ref{equ:powerbudeget}) and the bounded Wi-Fi power consumption condition (\ref{equ:boundedWiFi}), it is easy to find that $P\left( {{\bm \alpha}\left(t_T\right),{\bm p}\left(t\right)} \right)$ is bounded:
\begin{equation}
0 \le P\bigl( {{\bm \alpha}\left(t_T\right),{\bm p}\left(t\right)} \bigr) \le P_{\max}, \forall t\ge0,\label{equ:boundedtotalpower}
\end{equation}
where $P_{\max}\triangleq \kappa P_{\max}^C + \sum\limits_{n=1}^{N} {P_{n,\max}^W}$.
\subsubsection{Randomness}
\begin{figure}[t]
  \centering
  \includegraphics[scale=0.38]{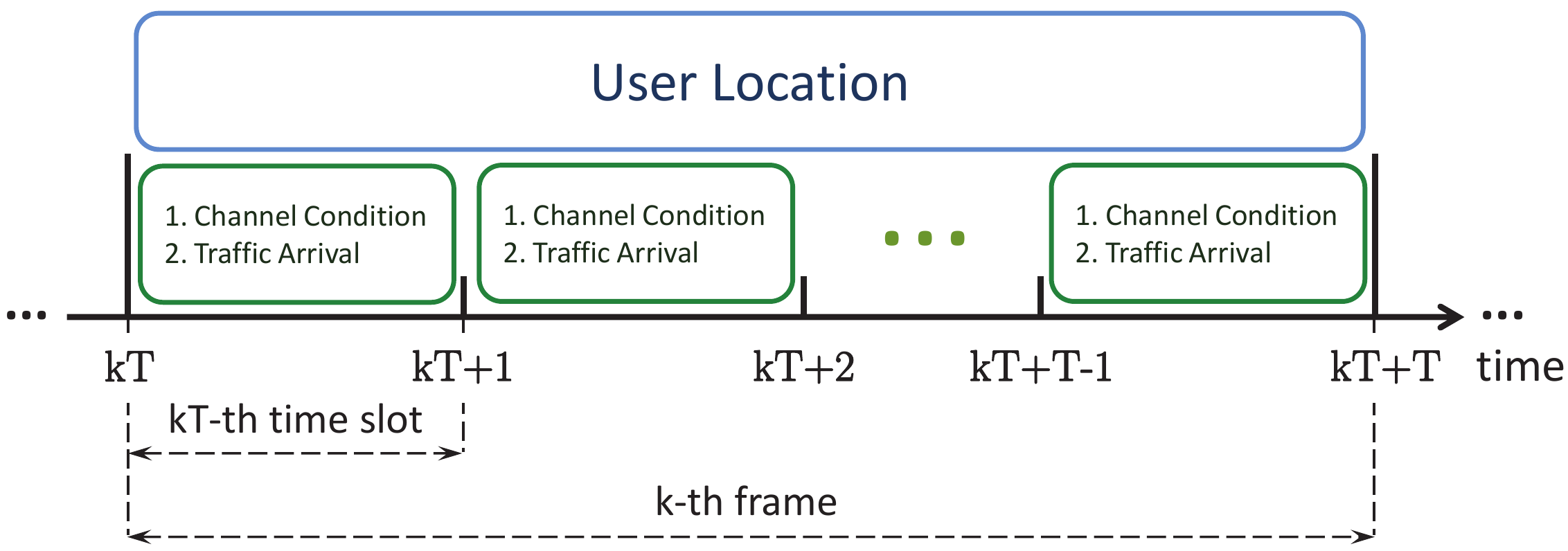}
  \vspace{-0.2cm}
  \caption{Two-timescale randomness: (a) users' locations change every frame; (b) channel conditions and users' traffic arrivals change every time slot.}
 \label{fig:TwoScale:b}
 \vspace{-0.4cm}
\end{figure}
There are three kinds of randomness in the system:
\begin{itemize}
\item Users' locations ${\bm S}\left(kT\right)$, introduced in Section \ref{subsec:select};
\item The macrocell network's channel conditions ${\bm H}\left(t\right)$, introduced in Section \ref{subsubsec:macrocellrate};
\item Users' traffic arrivals ${\bm A}\left(t\right)$, introduced in Section \ref{subsec:traffic}.
\end{itemize}
As we assumed in Section \ref{subsec:select}, ${\bm S}\left(kT\right)$ changes at the beginning of each frame, while ${\bm H}\left(t\right)$ and ${\bm A}\left(t\right)$ change every time slot. The two-timescale randomnesses is in Figure \ref{fig:TwoScale:b}.
\section{Problem formulation}\label{sec:problemformulation}
We assume that each user has a data queue, the length of which denotes the amount of unserved traffic. Let ${\bm Q}\left( t \right) = \left( Q_l\left(t\right),\forall l \in \cal{L} \right)$ be the queue length vector, where ${Q_l}\left(t\right)$ is user $l$'s queue length at time slot $t$. We assume that all queues are initially empty, \emph{i.e.},
\vspace{-0.25cm}
\begin{equation}
{Q_l}\left( 0 \right) = 0, {\rm~} \forall l \in \cal{L}.
\vspace{-0.25cm}
\end{equation}
The queue length evolves according to the traffic arrival rate and transmission rate as
\begin{equation}
{Q_l}\!\left( {t\! +\! 1} \right)\!\!=\!\! \left[{Q_l}\!\left( t \right)\!-\!{{r_l}\bigl( {{\bm \alpha}\! \left( {{t_T}} \!\right)\!,\!{{\bm x}^l}\left( t \right)\!,\!{{\bm p}^l}\!\left( t \right)}\! \bigr)} \right]^{+} \!\!+\! {A_l}\!\left( t \right),\!\forall l \!\in\!\mathcal{L},t\!\geq\! 0.\label{equ:queueing}
\end{equation}
Here ${\left[ x \right]^ + } = \max \left\{ {x,0} \right\}$ is due to the fact that the actual amount of served packets cannot exceed the current queue size.

The objective of the operator is to design an online network selection and resource allocation algorithm that minimizes the expected time average power consumption,{\footnote{``Online'' emphasizes that the algorithm relies on limited or no future
information, as opposed to an ``offline'' algorithm which requires complete
future information. We focus on the study of the online algorithm, as it is not practical for the operator to know all future information on the system randomness.}} while keeping the network stable. This can be formulated as the following optimization problem:
\begin{align}
\begin{split}
&\min {~~} {\overline P} \triangleq \mathop {\limsup}\limits_{K \to \infty } \frac{1}{{KT}}\sum\limits_{t = 0}^{KT - 1} {{{\mathbb E}\left\{ {P\left({\bm \alpha}\left(t_T\right)\right),{\bm p}\left(t\right)} \right\}} }\\
&{\rm s.t.~~} {\overline {Q_l}}\triangleq \mathop {\lim \sup }\limits_{K \to \infty } \frac{1}{KT}\sum\limits_{t= 0}^{KT - 1} {\mathbb{E}{\left\{ {Q_l}\left(t\right) \right\}}} <\infty{\rm{,~}}\forall l \in {\cal L},\\
&{~~~~~~} {\rm constraints~~ (\ref{equ:feasibility}),(\ref{equ:subchannelconstraint}),(\ref{equ:constraintforresource:a}),(\ref{equ:powerbudeget}),(\ref{equ:constraintforresource:b})},\\
&{\rm{var.}}~~~{{\bm \alpha}\left(t_T\right)}, {{\bm x}\left(t\right)}, {{\bm p}\left(t\right)},\forall t \ge 0.
\end{split}\label{equ:stochasticopt}
\end{align}
Here, ${\overline {Q_l}}$ is user $l$'s time average queue length, and constraint ${\overline {Q_l}}<\infty$ for all $l\in{\cal L}$ ensures the stability of the network. According to Little's law, ${\overline {Q_l}}$ is proportional to user $l$'s time average traffic delay. We will show that our algorithms guarantee upper bounds for ${\overline {Q_l}}$ and thus achieve bounded traffic delay.
\vspace{-0.3cm}
\section{Network Selection and Resource Allocation Without Prediction} \label{sec:withoutprediction}
\begin{algorithm}[t]\small
\caption{Energy-Aware Network Selection and Resource Allocation (ENSRA)}
\begin{algorithmic}[1]\label{algo:ENSRA}
\STATE Set $t=0$ and ${\bm Q}\left(0\right)={\bm 0}$;
\STATE {\bf while} {$t<t_{end}$} {\bf do} \label{line:ENSRAwhile}
\STATE $ \ \ \ $ {\bf if} $\mod\left(t,T\right)=0$ $//$ \emph{Compute the operations for the frame if $t$ is the beginning time slot of the frame.}
\STATE $ \ \ \ \ \ \ $ Set $k=\frac{t}{T}$ and solve problem (\ref{equ:ENSRA}) to determine ${\bm \alpha}\left( kT\right)$, ${\bm x}\left( \tau\right)$, ${\bm p}\left( \tau\right), \forall \tau\in{\cal T}_k$;\label{line:finish}
\STATE $ \ \ \ $ {\bf end if}
\STATE $ \ \ \ $ Update ${\bm Q}\left(t+1\right)$, according to (\ref{equ:queueing});
\STATE $ \ \ \ $ $t\leftarrow t+1$.
\STATE {\bf end while}
\end{algorithmic}
\end{algorithm}

We study the situation where the operator cannot predict the system randomness for the future frames. In Sections \ref{subsec:ENSRA} and \ref{subsec:SolveENSRA}, we assume that the operator has the complete information for the channel conditions within the current frame (but not the future frames), and propose \textsf{ENSRA} algorithm to generate a power consumption that can be pushed arbitrarily close to the optimal value of problem (\ref{equ:stochasticopt}). In Section \ref{subsec:ENSRAperform}, we analyze the performance of \textsf{ENSRA}. In Section \ref{subsec:Implementation}, we relax the assumption on the complete channel condition information, and discuss the implementation of \textsf{ENSRA}.
\vspace{-0.2cm}
\subsection{Energy-aware network selection and resource allocation (ENSRA) algorithm}\label{subsec:ENSRA}
\vspace{-0.2cm}
We assume that the operator has the complete information for the channel conditions within the current frame, \emph{i.e.}, at time slot $t=kT$ (the beginning of the $k$-th frame), the operator has the information of ${\bm H}\left(\tau\right)$ for all $\tau \in {{{\cal T}_{k}}}$.

\begin{figure*}[t]
\begin{align}
\begin{split}
&\min~V\sum\limits_{\tau = kT}^{kT + T - 1} {P\bigl({\bm \alpha}\left(kT\right),{\bm p}\left(\tau\right)\bigr)}  - \sum\limits_{l = 1}^L {{Q_l}\left( {kT} \right)\sum\limits_{\tau = kT}^{kT + T - 1} {r_l\bigl( {{\bm \alpha} \left( {kT} \right),{{\bm x}^l}\left( \tau \right),{\bm p}^l\left( \tau \right)} \bigr)} } \\
&{\rm{s.t.}}~~~~~{\rm constraints~~ (\ref{equ:feasibility}),(\ref{equ:subchannelconstraint}),(\ref{equ:constraintforresource:a}),(\ref{equ:powerbudeget}),(\ref{equ:constraintforresource:b})},\\
&{\rm{var.}}~~~~{\bm \alpha}\left( kT\right), {\bm x}\left( \tau\right), {\bm p}\left( \tau\right), \forall \tau\in{\cal T}_k.
\end{split}\label{equ:ENSRA}
\end{align}
\hrulefill
\end{figure*}

\begin{figure}[t]
  \centering
  \includegraphics[scale=0.35]{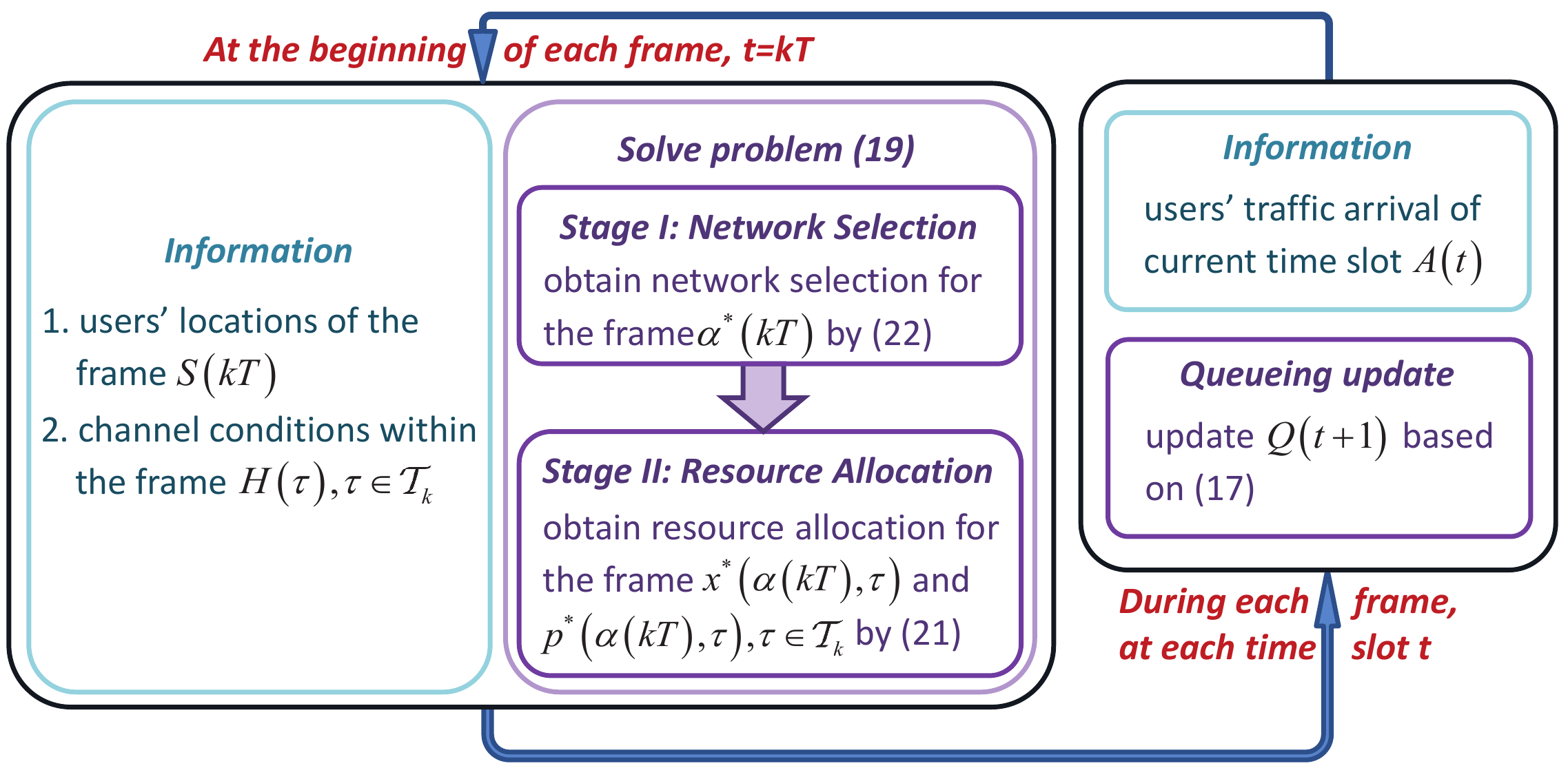}
  \vspace{-0.2cm}
  \caption{Flowchart of \textsf{ENSRA} algorithm: (a) at the beginning of each frame, the operator determines the network selection and resource allocation for the frame; (b) during each frame, the operator updates users' traffic queues.}
 \label{fig:ENSRA}
 \vspace{-0.3cm}
\end{figure}

We present \textsf{ENSRA} in Algorithm \ref{algo:ENSRA} and illustrate its flowchart in Figure \ref{fig:ENSRA}.{\footnote{In line \ref{line:ENSRAwhile}, we use $t_{end}$ to denote the number of running time slots for \textsf{ENSRA}. 
{{{As we will explain in Section \ref{subsec:ENSRAperform}, problem (\ref{equ:ENSRA}) is designed to minimize a ``drift-plus-penalty'' term for each frame, which characterizes the tradeoff between the power and the traffic delay.}}}}}
The intuition behind problem (\ref{equ:ENSRA}) in \textsf{ENSRA} can be understood as follows:{\footnote{{Notice that the unit of the control parameter $V$ is ${{{\rm Mb}^2}}/{\rm W \cdot s}$, and both terms in the objective function of problem (\ref{equ:ENSRA}) have the same units, \emph{i.e.}, ${{\rm Mb}^2}/{s}$.}}}

\begin{itemize}
\item If user $l$'s queue length $Q_l\left(kT\right)$ is small, the operator will focus less on term $-{Q_l}\left( {kT} \right)\cdot\sum\limits_{\tau = kT}^{kT + T - 1} {r_l\bigl( {{\bm \alpha} \left( {kT} \right),{{\bm x}^l}\left( \tau \right),{\bm p}^l\left( \tau \right)} \bigr)} $ and more on term $V\sum\limits_{\tau = kT}^{kT + T - 1} {P\bigl({\bm \alpha}\left(kT\right),{\bm p}\left(\tau\right)\bigr)}$ to minimize the objective function in problem (\ref{equ:ENSRA}). This implies that the operator will wait for those good channels or low power cost Wi-Fi networks to serve user $l$. Since $Q_l\left(kT\right)$ is small, suspending user $l$'s traffic in several time slots does not heavily increase the average queue length. According to Little's law, this also does not incur much delay;
\item If user $l$'s queue length $Q_l\left(kT\right)$ is large, the operator will focus more on term $-{Q_l}\left( {kT} \right)\cdot\sum\limits_{\tau = kT}^{kT + T - 1} {r_l\bigl( {{\bm \alpha} \left( {kT} \right),{{\bm x}^l}\left( \tau \right),{\bm p}^l\left( \tau \right)} \bigr)} $. This implies that there exists a big ``pressure'' to push the operator to serve user $l$ immediately, even when the user has a poor channel condition or the power needs to serve this user is high. As a result, user $l$'s queue length is reduced and the operator avoids a severe traffic delay.
\end{itemize}

In summary, by adjusting the control parameter $V>0{}$, the operator can achieve a good tradeoff between the power consumption and the traffic delay under \textsf{ENSRA}.
\subsection{Solving problem (\ref{equ:ENSRA}) in ENSRA algorithm}\label{subsec:SolveENSRA}
We solve problem (\ref{equ:ENSRA}) in \textsf{ENSRA} in two stages, for the $k$-th frame.
\begin{itemize}
\item {Stage I} (Network Selection): The operator determines the network selection ${\bm \alpha}^*\left( kT\right)$.
\item {Stage II} (Resource Allocation): Given any network selection ${\bm \alpha}\left( kT\right)$, the operator determines the subchannel allocation ${\bm x}^*\left({\bm \alpha}\left( kT\right), \tau\right)$ and power allocation ${\bm p}^*\left({\bm \alpha}\left( kT\right), \tau\right)$ for the macrocell network in each time slot $\tau\in{\cal T}_k$.
\end{itemize}

We solve the two-stage problem by backward induction, and start the analysis from {Stage II}.
\subsubsection{Stage II (Resource Allocation)}
Based on the given network selection ${\bm \alpha}\left( kT\right)$, we define ${\cal L}_0\triangleq \left\{l\in{\cal L}:\alpha_l\left( kT\right)=0\right\}$ as the set of users associated with the macrocell network. Due to constraints (\ref{equ:constraintforresource:a}) and (\ref{equ:constraintforresource:b}), we only need to study the subchannel and power allocation for these cellular users.{\footnote{For $l \notin {\cal L}_0$, we simply set ${x_{lm}\left(\tau\right)}={p_{lm}\left(\tau\right)}=0$ for all $m\in{\cal M},\tau\in {\cal T}_k$.}} According to (\ref{equ:ENSRA}), we have the following problem in Stage II:
\begin{align}
\begin{split}
&\min~V\!\!\!\!\sum\limits_{\tau = kT}^{kT + T - 1} {\!\!\!\!P^C\!\!\left({\bm p}\!\left(\tau\right)\right)}  \!-\! \sum\limits_{l\in{\cal L}_0} \!{{Q_l}\!\left( {kT} \right)\!\!\!\!\!\sum\limits_{\tau = kT}^{kT + T - 1} \!\!{r_l^C\!\left( {{{\bm x}^l}\!\left( \tau \right)\!,{\bm p}^l\!\left( \tau \right)} \right)} } \\
&{\rm{s.t.}}~~~~~{\rm constraints~~ (\ref{equ:subchannelconstraint}),(\ref{equ:powerbudeget})},\\
&{\rm{var.}}~~~~{\bm x}^l\left( \tau\right), {\bm p}^l\left( \tau\right), \forall l\in{\cal L}_0,\tau\in{\cal T}_k.
\end{split}\label{equ:stageII}
\end{align}

It is easy to observe from (\ref{equ:stageII}) that the resource allocations for different time slots $\tau\in{\cal T}_k$ are fully decoupled. For a particular time slot $\tau\in{\cal T}_k$, we expand function ${P^C\left({\bm p}\left(\tau\right)\right)}$ by (\ref{equ:macrocellrate}), function ${r_l^C\left( {{{\bm x}^l}\left( \tau \right),{\bm p}^l\left( \tau \right)} \right)}$ by (\ref{equ:macropower}), and obtain the following problem:{\footnote{In order to compare problem (\ref{equ:resourceallocation}) with the problem in \cite{huang2009downlink}, we arrange (\ref{equ:resourceallocation}) into a maximization problem.}}
\begin{align}
\begin{split}
&\max~\frac{B}{M}\!\sum\limits_{m=1}^{M} \!{\sum\limits_{l\in{\cal L}_0} {{Q_l}\!\left( kT \right){x_{lm}}\!\left( \tau \right)\log_2 \left( {1 \!+ \!\frac{{{p_{lm}}\left( \tau \right)H_{lm}^2\left( \tau \right)}}{{{N_0}\frac{B}{M}}}} \!\right)} }\\
& {~~~~~~~}- V \kappa \sum\limits_{m=1}^M {\sum\limits_{l\in{\cal L}_0} {{p_{lm}}\left( \tau \right)} }\\
&{\rm{s.t.}}~~~~\sum\limits_{m=1}^M {\sum\limits_{l\in{\cal L}_0} {{p_{lm}}\left( \tau \right)} }  \le {P_{\max }^C},\sum\limits_{l\in{\cal L}_0} {{x_{lm}}\left( \tau \right)}  \le 1,\forall m,\\
&{\rm{var.}}~~~~{x_{lm}}\left( \tau \right) \in \left\{ {0,1} \right\},{p_{lm}}\left( \tau \right) \ge 0,\forall l\in {\cal L}_0,m\in{\cal M}.
\end{split}\label{equ:resourceallocation}
\end{align}

Problem (\ref{equ:resourceallocation}) is similar to the weighted sum throughput maximization problem studied in \cite{huang2009downlink}, but with an extra linear power term $-V \kappa \sum\limits_{m=1}^M {\sum\limits_{l\in{\cal L}_0} {{p_{lm}}\left( \tau \right)} }$ in the objective function. {{According to \cite{huang2009downlink}, the complexity of solving problem (\ref{equ:resourceallocation}) is $O\left(LM\right)$.}}
We leave the detailed analysis and solutions to problem (\ref{equ:resourceallocation}) in \cite{haoran2015JSAC}.
\subsubsection{Stage I (Network Selection)}
We use ${\bm p}^*\bigl({{\bm \alpha}\left(kT\right)},\tau\bigr)$ and ${\bm x}^*\bigl({{\bm \alpha}\left(kT\right)},\tau\bigr)$ to denote the optimal resource allocation at time slot $\tau\in{{\cal T}_k}$ under network selection ${{\bm \alpha}\left(kT\right)}$. We have obtained ${\bm p}^*\bigl({{\bm \alpha}\left(kT\right)},\tau\bigr)$ and ${\bm x}^*\bigl({{\bm \alpha}\left(kT\right)},\tau\bigr)$ for all $\tau\in{{\cal T}_k}$ in Stage II.
Based on (\ref{equ:ENSRA}), the problem in Stage I is formulated as:
\begin{align}
\begin{split}
&\min~V\sum\limits_{\tau = kT}^{kT + T - 1} {P\Bigl({\bm \alpha}\left(kT\right),{\bm p}^*\big({{\bm \alpha}\left(kT\right)},\tau\bigl)\Bigr)}  -\\
&{~ }\sum\limits_{l = 1}^L \!{{Q_l}\!\left( {kT} \right)\!\!\!\!\!\sum\limits_{\tau = kT}^{kT + T - 1}\!\!\!\! {r_l\Bigl( {\!{\bm \alpha} \!\left( {kT} \right)\!,{{\bm x}^{l,*}}\bigr( {\bm \alpha}\left(kT\right)\!,\!\tau \bigl),{\bm p}^{l,*}\!\bigr(\! {\bm \alpha}\!\left(kT\right),\!\tau \!\bigl)\!} \Bigr)} } \\
&{\rm{s.t.}}~~~{\alpha _l}\left( kT \right) \in {{\cal N}_{{S_l}\left( kT \right)}} \cup \left\{ 0 \right\},{\rm{~}}  \forall l \in \mathcal{L}.
\end{split}\label{equ:combinatorial}
\end{align}

Problem (\ref{equ:combinatorial}) is a combinatorial optimization problem, and we apply the exhaustive search to pick the optimal network selection ${{\bm \alpha}^*\left(kT\right)}$.\footnote{There can be other low-complexity heuristic algorithms that solve the Stage I problem approximately. However, since the main contribution of this paper is to understand the impact of prediction (and hence the performance of the two algorithms to be proposed later), we will just use the exhaustive search method for \textsf{ENSRA} here.}
\subsection{Performance analysis of ENSRA}\label{subsec:ENSRAperform}
{{In this section, we prove the performance bounds of \textsf{ENSRA} in terms of the power-delay tradeoff.}}
For ease of exposition, we analyze the performance of \textsf{ENSRA} by assuming that the system randomness is independent and identically distributed (i.i.d.). 
Notice that with the technique developed in \cite{huang2010max}, we can obtain similar results under Markovian randomness.

We define the capacity region $\Lambda$ as the closure of the set of arrival vectors that can be stably supported, considering all network selection and resource allocation algorithms. We assume that the mean traffic arrival is strictly interior to $\Lambda$, \emph{i.e.}, there exists an $\eta>0$ such that
\begin{equation}
{\mathbb{E}}\left\{{{\bm A}\left(t\right)}\right\}+\eta \cdot {\bm 1} \in\Lambda.\label{equ:eta}
\end{equation}
This assumption is commonly used in the network stability literatures \cite{huang2010max,neely2010stochastic}. It guarantees that, we can find a network selection and resource allocation algorithm such that each user's expected transmission rate is greater than its mean traffic arrival rate.

We define the $T$-slot Lyapunov drift ${\Delta}_T\left( t \right)$ as
\begin{align}
{\Delta_T}\left( t \right)\triangleq {\mathbb {E}}\left\{ {\frac{1}{2}\!\sum\limits_{l = 1}^L {{Q_l}{{\left( t+T \right)}^2}} \!- \!\frac{1}{2}\sum\limits_{l = 1}^L {{Q_l}{{\left( t \right)}^2}} \!\left| {{\bm Q}\left( t \right)} \right.} \!\!\right\}.
\end{align}
Intuitively, the $T$-slot Lyapunov drift characterizes the expected change in the quadratic function of the queue length over every $T$ time slots. It will be used to show that \textsf{ENSRA} stabilizes the system and guarantees an upper bound on the time average queue length.

We define the ``drift-plus-penalty'' term for the $k$-th frame as
\begin{align}
\nonumber
D_T\left(kT\right)&\triangleq \Delta_T \left( kT \right) + \\
& V  {\mathbb E}\left\{ \sum\limits_{\tau=kT}^{kT+T-1} { {P\left({\bm \alpha}\left(kT\right),{\bm p}\left(\tau\right)\right) \left| {{\bm Q}\left( kT \right)} \right.}  }\right\}.\label{equ:DPPdefineframe}
\end{align}
The ``drift-plus-penalty'' term captures both the queue variance and the power consumption for the frame. In Lyapunov optimization, we minimize the upper bound of the ``drift-plus-penalty'' term, which is established in the following lemma (the proofs of all lemmas and theorems can be found in \cite{haoran2015JSAC}):
\begin{lemma}\label{lemma:nonpreDrift}
For any values of ${{\bm Q}\left( kT \right)}$, ${\bm \alpha} \left( kT \right)$, ${\bm x} \left( \tau \right)$, and ${\bm p} \left( \tau \right)$, $\tau\in{\cal T}_k$,
\begin{align}
\nonumber
& D_T\left(kT\right) \le {B_1}T + V {\mathbb E}\left\{\sum\limits_{\tau=kT}^{kT+T-1} {  {P\left({\bm \alpha}\left(kT\right),{\bm p}\left(\tau\right)\right) \left| {{\bm Q}\left( kT \right)} \right.}  }\right\}\\
&+\! {\mathbb E}\!\left\{ {{\!\sum\limits_{l = 1}^L{\!\!\sum\limits_{\tau=kT}^{kT+T-1}\!\!\! \!{ {{Q_l}\!\left(\! \tau \!\right)\!\left(\! {{A_l}\!\left( \tau \right)\! -\! {{r_l\!\left( {{\bm \alpha}\! \left( {kT} \right)\!,{{\bm x}^l}\left(\! \tau \right)\!,{\bm p}^l\left( \!\tau \right)} \!\right)}}} \!\right)}} \!\left| {{\bm Q}\!\left( kT \right)} \!\right.}}}\!\right\}\!,\label{equ:DPPupperbound1}
\end{align}
where $B_1 \triangleq \frac{1}{2}{{L\left( {A_{\max }^2 + r_{\max }^2} \right)}}$.{\footnote{Constants $A_{\max}$ and $r_{\max}$ are defined in (\ref{equ:arrivalbound}) and (\ref{equ:ratebound}), respectively.}}
\end{lemma}

In single-timescale control problems \cite{neely2006energy}, there is no frame structure (\emph{i.e.}, frame size $T=1$), so the upper bound given in Lemma \ref{lemma:nonpreDrift} is easy to minimize. However, our work studies two different timescales. Since the queue length ${\bm Q}\left(\tau\right)$ correlates users' transmission rates at time slot $\tau$ with the transmission rates during the time interval $\left[kT,kT+1,\ldots,\tau-1\right]$, it is difficult to directly minimize the upper bound in Lemma \ref{lemma:nonpreDrift}. Thus, we further relax the upper bound in Lemma \ref{lemma:nonpreDrift} in the following lemma.{\footnote{We obtain (\ref{equ:DPPupperbound2}) from (\ref{equ:DPPupperbound1}) by using the fact ${Q_l}\left( {kT} \right)-\left( {\tau\!-\!kT} \right){r_{\max }} \le {Q_l}\!\left( \tau  \right) \! \le {Q_l}\left( {kT} \right)+\!\left( {\tau\!-\!kT} \right){A_{\max }},\forall \in\!{\cal L}, \tau\in{\cal T}_k$.}}
\begin{lemma}\label{lemma:nonpreDrift:b}
For any values of ${{\bm Q}\left( kT \right)}$, ${\bm \alpha} \left( kT \right)$, ${\bm x} \left( \tau \right)$, and ${\bm p} \left( \tau \right)$, $\tau\in{\cal T}_k$,
\begin{align}
\nonumber
& D_T\left(kT\right) \le {B_2}T + V {\mathbb E}\left\{\sum\limits_{\tau=kT}^{kT+T-1} {  {P\left({\bm \alpha}\left(kT\right),{\bm p}\left(\tau\right)\right) \left| {{\bm Q}\left( kT \right)} \right.}  }\right\}\\
&\!+ \!{\mathbb E}\! \left\{ {{\sum\limits_{l = 1}^L\!{{Q_l}\!\left( kT \right)\! \!\!\!\!\!\sum\limits_{\tau=kT}^{kT+T-1}\!\!\!\!{\! {\left(\! {{A_l}\!\left( \tau \right)\!\! -\!\! {{r_l\left( {\!{\bm \alpha} \left( {kT} \right)\!,{{\bm x}^l}\!\left(\! \tau \!\right)\!,{\bm p}^l\!\left( \tau \right)\!} \right)}}} \right)}} \!\left| {{\bm Q}\!\left( kT \right)\!} \right.}}}\!\right\}\!,\label{equ:DPPupperbound2}
\end{align}
where $B_2 \triangleq \frac{1}{2}{{TL\left( {A_{\max }^2 + r_{\max }^2} \right)}}$.
\end{lemma}

The upper bound in Lemma \ref{lemma:nonpreDrift:b} is independent of ${\bm Q}\left(\tau\right)$. {{As formulated in (\ref{equ:ENSRA}), \textsf{ENSRA} essentially minimizes the right hand side of (\ref{equ:DPPupperbound2}) during every frame.}} We use $P_{av}^*$ to denote the optimal expected time average power consumption of problem (\ref{equ:stochasticopt}). The performance of \textsf{ENSRA} is described in the following theorem.
\begin{theorem}\label{theorem:ENSRA}
\textsf{ENSRA} achieves:
\vspace{-0.2cm}
\begin{align}
&P_{av}^{\textsf{ENSRA}} \!\triangleq \!\mathop {\limsup}\limits_{K \to \infty }\! \frac{1}{{KT}}\!\!\sum\limits_{t = 0}^{KT - 1}\!\! {{{\mathbb E}\!\left\{ \!{P\!\left({\bm \alpha}\!\left(t_T\right),{\bm p}\!\left(t\right)\right)} \!\right\}} } \!\le\! P_{av}^* \!+\! \frac{B_2}{V},\label{equ:pENSRA}\\
&Q_{av,T}^{\textsf {ENSRA}} \triangleq \mathop {\lim \sup }\limits_{K \to \infty } \frac{1}{K}\sum\limits_{l = 1}^{L} {\sum\limits_{k= 0}^{K-1} {\mathbb{E}{\left\{ {Q_l}\left(kT\right) \right\}}}}  \le \frac{{B_2 + V{P_{\max }}}}{\eta }.\label{equ:QENSRA}
\end{align}
where $B_2$ is defined in Lemma \ref{lemma:nonpreDrift:b}, $P_{\max}$ is defined in (\ref{equ:boundedtotalpower}), and $\eta$ is defined in (\ref{equ:eta}).
\end{theorem}

Here, notation $P_{av}^{\textsf{ENSRA}}$ is the expected time average power consumption of \textsf{ENSRA}, and notation $Q_{av,T}^{\textsf {ENSRA}}$ is the expected time average value of user queue length at the beginning of each frame. Based on (\ref{equ:QENSRA}), it is easy to show that the expected time average value of user queue length at each time slot is also upper bounded:
\begin{align}
\nonumber
Q_{av}^{\textsf {ENSRA}} & \triangleq  \mathop {\lim \sup }\limits_{K \to \infty } \frac{1}{KT}\sum\limits_{l = 1}^{L} {\sum\limits_{t= 0}^{KT-1} {\mathbb{E}{\left\{ {Q_l}\left(t\right) \right\}}}} \\
& \le \frac{{B_2 + V{P_{\max }}}}{\eta }+\!\frac{T-1}{2}LA_{\max}.\label{equ:QENSRA:every}
\end{align}

Theorem \ref{theorem:ENSRA} establishes the upper bounds of time average power consumption and queue length (or equivalently, average traffic delay). Theorem \ref{theorem:ENSRA} implies that, by increasing parameter $V$, the operator can push the power consumption arbitrarily close to the optimal value, \emph{i.e.}, $P_{av}^*$, but at the expense of the increase in the average traffic delay.{\footnote{{The performance bounds (\ref{equ:pENSRA}), (\ref{equ:QENSRA}), and (\ref{equ:QENSRA:every}) increase with the frame size $T$. This is because the resource allocation does not respond to instantaneous queue length values ${\bm Q}\left(t\right)$, and only considers the queue length values at the beginning of the frame ${\bm Q}\left(kT\right)$. When the frame size $T$ is larger, there are more time slots contained in each frame and the disadvantage of responding to ${\bm Q}\left(kT\right)$ instead of ${\bm Q}\left(t\right)$ becomes larger.}}}
\vspace{-0.3cm}
\begin{figure}[t]
  \centering
  \includegraphics[scale=0.45]{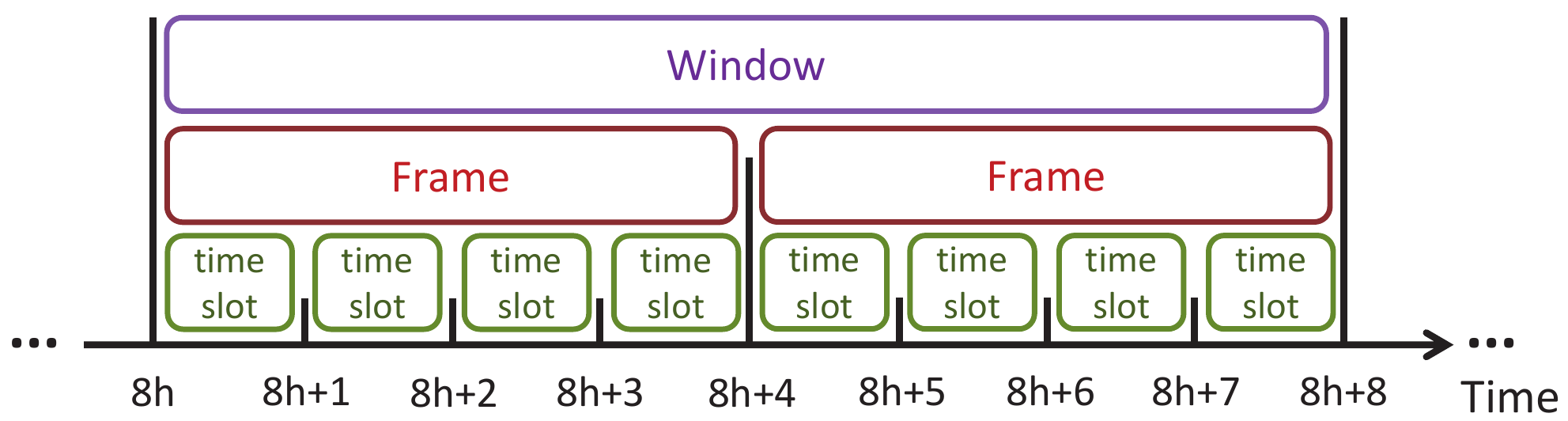}
  \vspace{-0.5cm}
  \centering
  \caption{{An example of the window structure, where the window size $W=2$, and the frame size $T=4$. The ${h}$-th window contains frames ${\cal T}_{2{h}}=\left\{8{h},8{h}+1,8{h}+2,8{h}+3\right\}$ and ${\cal T}_{2{h}+1}=\left\{8{h}+4,8{h}+5,8{h}+6,8{h}+7\right\}$. }}
  \label{fig:PredictStructure}
  \vspace{-0.2cm}
\end{figure}
\subsection{Implementation of ENSRA: Incomplete Channel Condition Information}\label{subsec:Implementation}
In Sections \ref{subsec:ENSRA} and \ref{subsec:SolveENSRA}, we assume that the operator has the complete channel condition information for the current frame. In practice, this complete information may not be available, which prevents us from directly solving problem (\ref{equ:ENSRA}) in \textsf{ENSRA}. To implement \textsf{ENSRA} in practice, we revise problem (\ref{equ:ENSRA}) by taking an expectation on the objective function with respect to the channel condition over the $k$-th frame. By minimizing the expected objective function, the operator can determine the network selection for the frame without knowing the channel conditions for the frame.{\footnote{The operator will observe the actual channel condition and determine the resource allocation at every time slot.}} To save space, we provide the detailed algorithm design in \cite{haoran2015JSAC}.{\footnote{{{Such a modified algorithm is optimal to the revised problem where the objective function is taken an expectation with respect to the channel condition over the frame, while it is not optimal to problem (\ref{equ:ENSRA}).
However, if the frame size $T$ is large (\emph{e.g.}, $T=100$ in our simulation), the channel conditions within each frame will average out and taking an expectation can well approximate the actual channel condition. In this case, the modified algorithm approximately solves problem (\ref{equ:ENSRA}).}}}}
\section{Network Selection and Resource Allocation With Prediction} \label{sec:withprediction}
We study the situation where the operator can predict the system randomness for the future frames. With the predictive future information, the operator is able to achieve better performance than \textsf{ENSRA}. In Section \ref{subsec:premodel}, we introduce the information prediction model. In Section \ref{subsec:P-ENSRA} and \ref{subsec:P-ENSRAperform}, we design and analyze \textsf{P-ENSRA} algorithm. In Section \ref{subsec:GP-ENSRA}, we design \textsf{GP-ENSRA} algorithm to reduce the computational complexity.
\subsection{Information prediction model}\label{subsec:premodel}
We consider the structure of the \emph{prediction window}, where the window size $W$ is the number of frames in a window. Thus, we define the ${h}$-th (${h} \in \left\{ {0,1, \ldots } \right\}$) window as the time interval that contains frames ${\cal T}_{{h}W},{\cal T}_{{h}W+1},\ldots,{\cal T}_{{h}W+W-1}$.
We use ${\cal W}_{h}\triangleq{{\cal T}_{{h}W}}\cup{{\cal T}_{{h}W+1}}\cup\ldots\cup{{\cal T}_{{h}W+W-1}}$ to define the set of time slots within the ${h}$-th window. Equivalently, we have ${\cal W}_{h}=\left\{{h}WT,{h}WT + 1, \ldots ,{h}WT + WT - 1\right\}$. We illustrate the window structure in Figure \ref{fig:PredictStructure}.

We assume that at time slot $t={h}WT$, \emph{i.e.}, the beginning of the ${h}$-th window, the operator accurately predicts the system randomness for the whole window:{\footnote{{{The perfect prediction assumption allows us to evaluate the fundamental benefits of having the future information in the scheduling algorithm design. This is an important first step towards understanding more general and practical scenarios of imperfect prediction. Similar perfect prediction assumption has been made in \cite{huang2013backpressure}.}}}} (a) ${\bm S}\left({h}WT+wT\right),w=0,1,\ldots,W-1$, where ${\bm S}\left({h}WT+wT\right)$ denotes users' locations during frame ${\cal T}_{{h}W+w}$; (b) ${\bm H}\left(\tau\right)$, ${\bm A}\left(\tau\right),\tau\in{\cal W}_{h}$, where ${\bm H}\left(\tau\right)$ and ${\bm A}\left(\tau\right)$ denote users' channel conditions and traffic arrivals at time slot $\tau$, respectively.

At time slot $t={h}WT$, with the predictive information, the operator runs \textsf{P-ENSRA} or \textsf{GP-ENSRA}, and determines the operations for the whole window: (a) ${\boldsymbol \alpha}\left( {h}WT+wT\right), w=0,1,\ldots,W-1$, where ${\boldsymbol \alpha}\left( {h}WT+wT\right)$ denotes the network selection during frame ${\cal T}_{{h}W+w}$; (b) ${\boldsymbol x}\left( \tau\right),{\boldsymbol p}\left( \tau\right),\tau\in{\cal W}_{h}$, where ${\boldsymbol x}\left( \tau\right)$ and ${\boldsymbol p}\left( \tau\right)$ are the subchannel allocation and power allocation at time slot $\tau$, respectively.

\subsection{Predictive energy-aware network selection and resource allocation (P-ENSRA) algorithm}\label{subsec:P-ENSRA}
\begin{figure*}[t]
\begin{align}
\begin{split}
&\min~V\sum\limits_{w = 0}^{W-1}{\sum\limits_{\tau = \left({h}W+w\right)T}^{\left({h}W+w+1\right)T - 1} {P\bigl({\boldsymbol \alpha}\left({h}WT+wT\right),{\boldsymbol p}\left(\tau\right)\bigr)}} + \sum\limits_{l = 1}^L {\sum\limits_{w = 0}^{W-1} {{{Q_l}\left( {{h}WT+wT} \right)}\sum\limits_{\tau = \left({h}W+w\right)T}^{\left({h}W+w+1\right)T- 1} {\left(A_l\left(\tau \right)+\theta\right)} }}\\
&{~~~~~~} - \sum\limits_{l = 1}^L {\sum\limits_{w = 0}^{W-1} {{{Q_l}\left( {{h}WT+wT} \right)}\sum\limits_{\tau = \left({h}W+w\right)T}^{\left({h}W+w+1\right)T- 1}{r_l\bigl({\boldsymbol \alpha} \left( {{h}WT+wT} \right), {{{\boldsymbol x}^l}\left( \tau \right),{\boldsymbol p}^l\left( \tau \right)} \bigr)} }} \\
&{\rm{s.t.}}~~~~~{\rm constraints~~ (\ref{equ:feasibility}),(\ref{equ:subchannelconstraint}),(\ref{equ:constraintforresource:a}),(\ref{equ:powerbudeget}),(\ref{equ:constraintforresource:b})},\\
&{\rm{var.}}~~~~{\boldsymbol \alpha}\left( {h}WT+wT\right), w=0,1,\ldots,W-1,{\boldsymbol x}\left( \tau\right), {\boldsymbol p}\left( \tau\right), \tau\in{\cal W}_{h}.
\end{split}\label{equ:P-ENSRA}
\end{align}
\hrulefill
\end{figure*}

\begin{algorithm}[t]\small
\caption{Predictive Energy-Aware Network Selection and Resource Allocation (P-ENSRA)}
\begin{algorithmic}[1]\label{algo:P-ENSRA}
\STATE Set $t=0$ and ${\bm Q}\left(0\right)={\bm 0}$;
\STATE {\bf while} {$t<t_{end}$} {\bf do}
\STATE $ \ \ \ $ {\bf if} $\mod\left(t,WT\right)=0$ $//$ \emph{Compute the operations for the window if $t$ is the beginning slot of the window.}
\STATE $ \ \ \ \ \ $Set \!${h}\!=\!\frac{t}{WT}$\! and solve problem \!(\ref{equ:P-ENSRA})\! to determine \!${\boldsymbol \alpha}\!\left( {h}WT\!+\!wT\right)\!, w\!=\!0,1,\ldots,W\!-\!1,{\boldsymbol x}\left( \tau\right), {\boldsymbol p}\left( \tau\right),\! \tau\!\in\!{\cal W}_{h}$;
\STATE $ \ \ \ $ {\bf end if}
\STATE $ \ \ \ $ Update ${\bm Q}\left(t+1\right)$, according to (\ref{equ:queueing});
\STATE $ \ \ \ $ $t\leftarrow t+1$.
\STATE {\bf end while}
\end{algorithmic}
\end{algorithm}

We propose \textsf{P-ENSRA} in Algorithm \ref{algo:P-ENSRA}. Recall that, the basic idea of \textsf{ENSRA} in Section \ref{sec:withoutprediction} is to minimize the upper bound of the ``drift-plus-penalty'' term for a frame.
As we will see in Section \ref{subsec:P-ENSRAperform}, different from \textsf{ENSRA}, \textsf{P-ENSRA} guarantees a $\theta$-controlled upper bound on the ``drift-plus-penalty'' term instead of minimizing the ``drift-plus-penalty'' term for a window. This is because \textsf{P-ENSRA} determines the network selection and resource allocation for several frames (\emph{i.e.}, a window), and it needs to use a novel control parameter $\theta>0$ to balance the queue lengths among different frames.{\footnote{{The unit of parameter $\theta$ is the same as traffic arrival $A_l\left(t\right)$, \emph{i.e.}, $\rm Mbps$.}}}
Through introducing $\theta$, we can assign larger weights to the transmission rates of the earlier frames than those of the latter frames within a prediction window. By doing this, we can reduce the time average queue length.
\subsection{Performance analysis of P-ENSRA}\label{subsec:P-ENSRAperform}
Similar as \textsf{ENSRA}, we characterize the performance of \textsf{P-ENSRA} under the i.i.d. system randomness and assume that the condition (\ref{equ:eta}) is satisfied. 
\subsubsection{Power consumption-delay tradeoff}
We define the ``drift-plus-penalty'' term for the ${h}$-th window as
\begin{align}
\nonumber
& D_{WT}\left( {h}WT \right) \triangleq \Delta_{WT} \left( {h}WT \right) \\
& \!\!+\! V \! {\mathbb E}\!\left\{ \!{ {{\sum\limits_{w = 0}^{W\!-\!1}{\sum\limits_{\tau = \left({h}W+w\right)T}^{\left({h}W+w+1\right)T\! -\! 1} {\!\!\!\!\!P\bigl({\boldsymbol \alpha}\left({h}WT\!+\!wT\right)\!,{\boldsymbol p}\left(\tau\right)\bigr)}}} \left| {{\bm Q}\left( {h}WT \right)} \right.}  }\!\!\!\right\} .
\end{align}

Next we introduce Lemma \ref{lemma:pre2} and Lemma \ref{lemma:pre3} to show that \textsf{P-ENSRA} guarantees a $\theta$-controlled upper bound on $D_{WT}\left( {h}WT \right)$. The introduction of parameter $\theta$ and the $\theta$-controlled upper bound is different from all previous Lyapunov optimization techniques.
\begin{lemma}\label{lemma:pre2}
For any values of ${{\bm Q}\left( {h}WT \right)}$, ${\boldsymbol \alpha}\left( {h}WT+wT\right)$, ${\boldsymbol x}\left( \tau\right)$, and ${\boldsymbol p}\left( \tau\right)$, $w=0,1,\ldots,W-1,\tau\in{\cal W}_{h}$, we have (\ref{equ:DPPupperboundforP}),
\begin{figure*}
\begin{align}
\nonumber
& D_{WT}\left( {h}WT \right) \le {B_2}WT + V  {\mathbb E}\left\{ { {{\sum\limits_{w = 0}^{W-1}{\sum\limits_{\tau = \left({h}W+w\right)T}^{\left({h}W+w+1\right)T - 1} {P\bigl({\boldsymbol \alpha}\left({h}WT+wT\right),{\boldsymbol p}\left(\tau\right)\bigr)}}} \left| {{\bm Q}\left( {h}WT \right)} \right.}  }\right\}\\
& +{\mathbb E} \left\{ {{\sum\limits_{l = 1}^L \sum\limits_{w=0}^{W-1}{{Q_l}\left( {h}WT+wT \right)\sum\limits_{\tau=\left({h}W+w\right)T}^{\left({h}W+w+1\right)T-1} { {\Bigl( {{A_l}\left( \tau \right) - {{r_l\bigl( {{\bm \alpha} \left( {{h}WT+wT} \right),{{\bm x}^l}\left( \tau \right),{\bm p}^l\left( \tau \right)} \bigr)}}} \Bigr)}} \left| {{\bm Q}\left( {h}WT \right)} \right.}}}\right\},\label{equ:DPPupperboundforP}
\end{align}
\hrulefill
\end{figure*}
where $B_2$ is the constant defined in Lemma \ref{lemma:nonpreDrift:b}.
\end{lemma}

For any $\theta \in \left( {0,\eta } \right]$,{\footnote{Parameter $\eta$ is defined in (\ref{equ:eta}).}} we define $P\left(\theta\right)$ as the minimum power consumption required to stabilize the traffic arrival vector ${\mathbb{E}}\left\{{{\bm A}\left(t\right)}\right\}+\theta \cdot {\bm 1}$, considering all network selection and resource allocation algorithms. Naturally, we have the following relation:{\footnote{$P_{av}^*$ is the minimum expected time average power consumption of problem (\ref{equ:stochasticopt}). {{The proof of the continuity of function $P\left(\theta\right)$ can be found in \cite{neely2010stochastic}.}}}}
\begin{equation}
\mathop {\lim }\limits_{\theta  \to 0} {P\left( \theta  \right)} = P_{av}^*.
\end{equation}

\begin{lemma}\label{lemma:pre3}
Let ${\boldsymbol \alpha}^*\left( {h}WT+wT\right)$, ${\boldsymbol x}^*\left( \tau\right)$, and ${\boldsymbol p}^*\left( \tau\right)$, $w=0,1,\ldots,W-1,\tau\in{\cal W}_{h}$, be the optimal solutions to problem (\ref{equ:P-ENSRA}). Inequality (\ref{equ:longequation:A}) holds for any $\theta  \in \left( {0,\eta } \right]$.
\begin{figure*}
\begin{align}
\nonumber
& {\mathbb E}\!\left\{\!{{\sum\limits_{l = 1}^L \!\sum\limits_{w=0}^{W-1}{\!{Q_l}\!\left( {h}WT\!+\!wT \right) \!\!\!\!\!\!\sum\limits_{\tau=\left({h}W+w\right)T}^{\left({h}W+w+1\right)T-1}\!\!\!\! { {\Bigl( {\!{A_l}\left( \tau \right)\!-\!{{r_l\bigl( {{{\bm \alpha}^*}\!\left( {{h}WT+wT} \right),{{\bm x}^{l,*}}\!\left( \tau \right),{\bm p}^{l,*}\!\left( \tau \right)} \bigr)}}} \Bigr)\!}} \left| {{\bm Q}\left( {h}WT \right)} \right.}}}\!\!\right\}\\
\nonumber
& + V  {\mathbb E}\left\{ { {{\sum\limits_{w = 0}^{W-1}{\sum\limits_{\tau = \left({h}W+w\right)T}^{\left({h}W+w+1\right)T - 1} {P\bigl({\boldsymbol \alpha}^*\left({h}WT+wT\right),{\boldsymbol p}^*\left(\tau\right)\bigr)}}} \left| {{\bm Q}\left( {h}WT \right)} \right.}  }\right\}\\
& \le -\theta T {\mathbb{E}}\left\{ {\sum\limits_{l = 1}^L {\sum\limits_{w = 0}^{W-1} { {{Q_l}\left( {h}WT+wT  \right)} }} \left| {{\bm Q}\left( {{h}WT} \right)} \right.} \right\} + VWTP\left( \theta  \right).\label{equ:longequation:A}
\end{align}
\hrulefill
\end{figure*}
\end{lemma}

According to Lemma \ref{lemma:pre2} and Lemma \ref{lemma:pre3}, \textsf{P-ENSRA} guarantees that
\begin{align}
\nonumber
& D_{WT}\left( {h}WT \right) \le {B_2}WT + VWTP\left( \theta  \right)\\
& -\theta T {\mathbb{E}}\left\{ {\sum\limits_{l = 1}^L {\sum\limits_{w = 0}^{W-1} { {{Q_l}\left( {h}WT+wT  \right)} }} \left| {{\bm Q}\left( {{h}WT} \right)} \right.} \right\} .\label{equ:speakclear}
\end{align}
Here, ${{Q_l}\left( {h}WT+wT  \right)}, w=0,1,\ldots,W-1,$ is user $l$'s queue length at the beginning of frame ${\cal T}_{{h}W+w}$ under \textsf{P-ENSRA}. Inequality (\ref{equ:speakclear}) shows the most important feature of \textsf{P-ENSRA}: it establishes the relation between the ``drift-plus-penalty'' term and the queue length generated by the algorithm. Based on this, we can prove the upper bound of the average queue length under \textsf{P-ENSRA} (Theorem \ref{theorem:PENSRA}). If we consider other algorithms (\emph{e.g.}, the algorithm that directly minimizes the right hand side of (\ref{equ:DPPupperboundforP}) for each window), it is difficult to find a relation similar as (\ref{equ:speakclear}) and prove the queue length bound. This shows the special design of \textsf{P-ENSRA}.

Based on (\ref{equ:speakclear}), we show the performance bounds of \textsf{P-ENSRA}. We define $P_{av}^{\textsf{P-ENSRA}}$ as the expected time average power consumption of \textsf{P-ENSRA}, and define $Q_{av,T}^{\textsf {P-ENSRA}}$ as the expected time average value of user queue length at the beginning of each frame under \textsf{P-ENSRA}. The performance of \textsf{P-ENSRA} is described in the following theorem.
\begin{theorem}\label{theorem:PENSRA}
\textsf{P-ENSRA} achieves
\begin{align}
\nonumber
P_{av}^{\textsf{P-ENSRA}} &\triangleq \mathop {\limsup}\limits_{H \to \infty } \frac{1}{{HWT}}\sum\limits_{t = 0}^{HWT - 1} {{{\mathbb E}\left\{ {P\left({\bm \alpha}\left(t_T\right),{\bm p}\left(t\right)\right)} \right\}} } \\
& \le P\left(\theta\right) + \frac{B_2}{V},\label{equ:pENSRA:pre}\\
\nonumber
Q_{av,T}^{\textsf {P-ENSRA}} &\triangleq \mathop {\lim \sup }\limits_{H \to \infty } \frac{1}{HW}\sum\limits_{l = 1}^{L} {\sum\limits_{{h}= 0}^{HW-1} {\mathbb{E}{\left\{ {Q_l}\left({h}T\right) \right\}}}}  \\
& \le \frac{{B_2 + V{P\left(\theta\right)}}}{\theta },\label{equ:QENSRA:pre}
\end{align}
for any $V>0{}$ and $\theta \in \left( {0,\eta } \right]$, where $B_2$ is defined in Lemma \ref{lemma:nonpreDrift:b}.{\footnote{{The performance bounds (\ref{equ:pENSRA:pre}) and (\ref{equ:QENSRA:pre}) do not depend on the window size $W$. This is because the impact of the window size $W$ heavily depends on the concrete settings of the system randomness. Intuitively, when the system randomness changes frequently, the impact of the window size $W$ is expected to be large. In our work, we only consider general i.i.d. system randomness, hence it is hard to evaluate the impact of $W$ without specifying the concrete distribution of the system randomness. We leave the study of the impact of the window size $W$ as our future work.}}}
\end{theorem}

Based on (\ref{equ:QENSRA:pre}), it is easy to show that the expected time average value of users queue length at each time slot under \textsf{P-ENSRA} is also bounded:
\begin{align}
\nonumber
Q_{av}^{\textsf {P-ENSRA}} & \triangleq \mathop {\lim \sup }\limits_{H \to \infty } \frac{1}{HWT}\sum\limits_{l = 1}^{L} {\sum\limits_{t= 0}^{HWT-1} {\mathbb{E}{\left\{ {Q_l}\left(t\right) \right\}}}}  \\
& \le \frac{{B_2 + V{P\left(\theta\right)}}}{\theta }+\frac{T-1}{2}LA_{\max},\label{equ:QENSRA:preevery}
\end{align}
\subsubsection{Comparison between \textsf{{ENSRA}} and \textsf{{P-ENSRA}}}
Comparing Theorem \ref{theorem:ENSRA} and Theorem \ref{theorem:PENSRA}, we find \textsf{{P-ENSRA}} achieves similar performance bounds as \textsf{{ENSRA}}. In particular:
\begin{itemize}
\item When $\theta$ approaches $0$, the bound for the power consumption achieved by \textsf{{P-ENSRA}} equals that of \textsf{ENSRA} in (\ref{equ:pENSRA}). That is,
\begin{equation}
\mathop {\lim }\limits_{\theta  \to 0} \left({P\left( \theta  \right)+\frac{B_2}{V}}\right) = P_{av}^*+\frac{B_2}{V};
\end{equation}
\item When $\theta=\eta$, the average queue length of \textsf{{P-ENSRA}} satisfies
\begin{equation}
Q_{av,T}^{\textsf {P-ENSRA}} \le \frac{{B_2 + V{P{\left(\eta\right) }}}}{\eta } \le \frac{{B_2 + V{P_{\max }}}}{\eta },
\end{equation}
where the right bound is the same as the one specified in (\ref{equ:QENSRA}) for \textsf{ENSRA}.{\footnote{From (\ref{equ:QENSRA:every}) and (\ref{equ:QENSRA:preevery}), it is easy to obtain the similar comparison between $Q_{av}^{\textsf {ENSRA}}$ and $Q_{av}^{\textsf {P-ENSRA}}$.}}
\end{itemize}

{~}{{\!\!\!The reason that the performance bounds of \textsf{{P-ENSRA}} in (\ref{equ:pENSRA:pre}) and (\ref{equ:QENSRA:pre}) are not better than those of \textsf{ENSRA} is because the performance bounds in (\ref{equ:pENSRA:pre}) and (\ref{equ:QENSRA:pre}) are valid for all delay regimes. As we will observe in Section VI, \textsf{{P-ENSRA}} cannot outperform \textsf{{ENSRA}} when the generated delay is restricted to a small value. In this case, the operator has to serve the traffic immediately even if the users' channel conditions and Wi-Fi availabilities in the future frames are better.
}}
\vspace{-0.3cm}
\subsection{Greedy predictive energy-aware network selection and resource allocation (GP-ENSRA)}\label{subsec:GP-ENSRA}
In problem (\ref{equ:P-ENSRA}) the network selections and resource allocations in different frames are tightly coupled by the queue lengths. Such coupling significantly increases the difficulty of directly solving problem (\ref{equ:P-ENSRA}). 
Here, we propose a greedy algorithm, \textsf{GP-ENSRA}, which approximately solves problem (\ref{equ:P-ENSRA}) for each window and significantly reduces the complexity.

\begin{algorithm}[t]\small
\caption{\small{Greedy Predictive Energy-Aware Network Selection and Resource Allocation (GP-ENSRA)}}
\begin{algorithmic}[1]\label{algo:GP-ENSRA}
\STATE Set $t=0$ and ${\bm Q}\left(0\right)={\bm 0}$;
\STATE {\bf while} {$t<t_{end}$} {\bf do}
\STATE $ \ \ \ $ {\bf if} $\frac{t}{{WT}}\in {\mathbb N}$ $//$ \emph{Compute the operations for the window if $t$ is the beginning slot of the window.}
\STATE $ \ \ \ \ \ $ Set ${h}=\frac{t}{WT}$, $i=0$, and $\bm \beta \left( {h}WT+wT  \right) =  {\bm 0}$, $\forall w=0,1,\ldots,W-1$;
\STATE $ \ \ \ \ \ \ $ {\bf while} {$i<2$ or ${F^{i-1}} - {F^{i}} > \epsilon$} {\bf do} $//$ \emph{Approximately solve problem (\ref{equ:P-ENSRA}).} \label{line:while}
\STATE $ \ \ \ \ \ \ \ \ \ $ $i\leftarrow i+1$;
\STATE $ \ \ \ \ \ \ \ \ \ $ {\bf for} {$w=0$ to $W-1$} {\bf do}
\STATE $ \ \ \ \ \ \ \ \ \ \ \ \ \ $ Minimize the objective function in problem (\ref{equ:P-ENSRA}) over $\bm \beta \left({h}WT\!+\!wT\right)$ (fix $\bm \beta \left({h}WT\!+\!w'T\right)$ for all $w'\!\ne\!w$);\label{line:A}
\STATE $ \ \ \ \ \ \ \ \ \ $ Update $\bm \beta \left({h}WT+wT\right)$ with the optimal solution obtained in line \ref{line:A};\label{line:C}
\STATE $ \ \ \ \ \ \ \ \ \ $ {\bf end for}
\STATE $ \ \ \ \ \ \ \ \ \ $ Denote the value of the objective function in (\ref{equ:P-ENSRA}) under $\bigl(\bm \beta \left( {h}WT+wT  \right), w=0,1,\ldots,W-1\bigr)$ by $F^{i}$;\label{line:B}
\STATE $ \ \ \ \ \ \ $ {\bf end while}\label{line:endwhile}
\STATE $ \ \ \ \ \ \ $ Output vector $\bm \beta \left( {h}WT+\!wT  \right), w\!=\!0,1,\ldots,W\! -\! 1,$ as the operations for the window;
\STATE $ \ \ \ $ {\bf end if}
\STATE $ \ \ \ $ Update ${\bm Q}\left(t+1\right)$, according to (\ref{equ:queueing});
\STATE $ \ \ \ $ $t\leftarrow t+1$.
\STATE {\bf end while}
\end{algorithmic}
\end{algorithm}

The basic idea of the greedy algorithm is that, instead of globally searching for the optimal solution to problem (\ref{equ:P-ENSRA}), the operator iteratively updates the operations for different frames within the window. For example, when updating the operations for frame ${\cal T}_{{h}W+w}$, the operator treats the operations for all other frames, \emph{i.e.}, ${\cal T}_{{h}W+w'},w'\ne w,w'=0,1,\ldots,W-1$, as given constants, and minimizes the objective function over the operations for frame ${\cal T}_{{h}W+w}$.

We present \textsf{GP-ENSRA} in Algorithm \ref{algo:GP-ENSRA}. In order to simplify the description, we use ${{\bm \beta}\left({h}WT\!+\!wT\right)}\!=\!\bigl({{\bm \alpha}\left({h}WT\!+\!wT\right)},{{\bm x}\left(\tau\right)},{{\bm p }\left(\tau\right)},\tau\!\in{\cal T}_{{h}W\!+\!w}\bigr)$ to represent the operator's operations (network selection and resource allocation) over frame ${\cal T}_{{h}W+w}$, $w=0,1,\ldots,W-1$.
From line \ref{line:while} to line \ref{line:endwhile}, the operator iteratively updates the operations for all frames within the window.{\footnote{{During each iteration, the operator updates the operations for frames ${\cal T}_{{h}W}$, ${\cal T}_{{h}W+1}$, $\ldots$, ${\cal T}_{{h}W+W-1}$ sequentially: when updating the operations for frame ${\cal T}_{{h}W+w}$, the operator treats the operations for all other frames, \emph{i.e.}, ${\cal T}_{{h}W+w'},w'\ne w,w'=0,1,\ldots,W-1$, as fixed, and minimizes the objective function in problem (\ref{equ:P-ENSRA}) over the operations for frame ${\cal T}_{{h}W+w}$.}}} As shown in line \ref{line:B}, we use $F^i$ to denote the value of the objective function in (\ref{equ:P-ENSRA}) under the $i$-th iteration. The condition for ending the iteration (line \ref{line:while}) implies that the decrease from $F^{i-1}$ to $F^{i}$ is no larger than a positive parameter $\epsilon$. Such a condition is guaranteed to be achievable, and we leave the detailed proof in \cite{haoran2015JSAC}. Briefly speaking, the updating rule (line \ref{line:A} and line \ref{line:C}) guarantees that $F^{i}$ is always non-increasing in $i$. Furthermore, we can prove that the objective function in (\ref{equ:P-ENSRA}) is both lower and upper bounded. As a result, it is easy to show that there exists a finite $i$ such that ${F^{i-1}} - {F^{i}} \le \epsilon$.

The complexity of \textsf{GP-ENSRA} mainly depends on how we solve the problem specified in line \ref{line:A}. In fact, if the initial queue vector of the window ${\bm Q}\left( {{h}WT} \right)$ satisfies the condition
\begin{align}
{{Q_l}\left( {{h}WT} \right)}\ge WT r_{\max},\forall l\in{\cal L},\label{equ:heavy}
\end{align}
then the problem in line \ref{line:A} can be solved as problem (\ref{equ:ENSRA}).{\footnote{We leave the detailed analysis in \cite{haoran2015JSAC}. In our simulation, \textsf{GP-ENSRA}'s computational time just polynomially increases with the window size $W$. For example, the actual time lengths required for \textsf{ENSRA} and \textsf{GP-ENSRA} with $W=5{}$ to compute the operations for $1,000$ time slots in MATLAB are $9{}$ seconds and $55{}$ seconds, respectively.}} The condition guarantees that, during the whole window, there are always enough packets in users' queues to be served. Such a condition is mild in a heavy traffic situation. {{We leave the complexity analysis of the case without condition (\ref{equ:heavy}) as our future work.}}
\section{Simulation} \label{sec:simulation}
In Section \ref{subsec:simusetting}, we explain the simulation settings. In \ref{subsec:simulationresults}, we introduce a heuristic network selection and resource allocation algorithm for comparison, and simulate the power and delay performance of the heuristic algorithm, \textsf{ENSRA}, and \textsf{GP-ENSRA}, respectively.
\subsection{Simulation settings}\label{subsec:simusetting}
{{We simulate the problem with $L=10$ users, $1$ macrocell network, $N=10$ Wi-Fi networks, and $S=100$ locations. Each location has a size of $15\times15~{\rm m^2}$.}} We set the time slot length to be $10$ milliseconds, and the frame length to be $1$ second, \emph{i.e.}, the frame size $T=100$. We run each experiment in MATLAB for $5,000$ frames.
\subsubsection{Macrocell network}
We assume that the macrocell network covers all locations. We consider $M=8$ subchannels, and assume that the channel gain ${H_{lm}}\left( t \right) = \frac{{{\xi _{lm}}\left( t \right)}}{{d_{l}^{1.5}\left(t\right)}}$ follows the Rayleigh fading,{\footnote{{Based on (\ref{equ:macrocellrate}), the SNR is proportional to ${H_{lm}^2}\left( t \right)$. Hence, if ${H_{lm}}\left( t \right) = \frac{{{\xi _{lm}}\left( t \right)}}{{d_{l}^{1.5}\left(t\right)}}$, the SNR is in inverse proportion to $d_{l}^3\left(t\right)$. Notice that a path loss exponent of $3$ is in line with the empirical channel measurements \cite{rappaport1996wireless}.}}} where ${\xi _{lm}}\left( t \right)$ follows a Rayleigh distribution \cite{gajic2014competition} and $d_l\left(t\right)$ is the distance between user $l$ and the macrocell base station. {{Distance $d_l\left(t\right)$ is computed as follows. We use function $g\left(s\right)$ to denote the distance between the macrocell base station and the users at location $s\in{\cal S}$. In Section II-B, we use $S_l\left(kT\right)$ to denote user $l$'s location during the $k$-th frame. Hence, if $t\in{\cal T}_k$, distance $d_{l}\left(t\right)$ is determined by $g\left(S_l\left(kT\right)\right)$.{\footnote{{Since we set the size of each location as $15\times15~{\rm m^2}$, it is reasonable to assume that all users at the same location have a similar distance to the base station.}}}}} Table \ref{table:simulation} summarizes other system parameters.
\subsubsection{Wi-Fi networks}
\begin{table}[t]\small
\centering
\caption{Simulation Parameters \cite{auer2011much,bianchi2000performance,jung2012adaptive}}
\vspace{-0.3cm}
\begin{tabular}{|c|c|c|c|c|}
\hline
{\minitab[c]{$P_{\max}^C$}} & {\minitab[c]{$20$ W}}& {\minitab[c]{$B$}} & {\minitab[c]{$2.5$ MHz}}\\
{\minitab[c]{$N_0$}} & {\minitab[c]{$10^{-7}$ W$\slash$MHz}}& {\minitab[c]{$\kappa$}} & {\minitab[c]{$4.7$}}\\
{\minitab[c]{$G$}} & {\minitab[c]{$800$ b}} & {\minitab[c]{$T_b$}} & {\minitab[c]{$28$ $\mu$s}}\\
{\minitab[c]{$T_s$}} & {\minitab[c]{$100$ $\mu$s}}&{\minitab[c]{$T_c$}} & {\minitab[c]{$100$ $\mu$s}}\\
{\minitab[c]{$E_b$}} & {\minitab[c]{$22.4$ $\mu$J}}&{\minitab[c]{$E_s$}} & {\minitab[c]{$180$ $\mu$J}}\\
{\minitab[c]{$E_{c,j}\left({\rho_n}\right)$}} & {\minitab[c]{$80{\rho_n}+100j+80$ $\mu$J}}&{\minitab[c]{}} & {\minitab[c]{}}\\
\hline
\end{tabular}\label{table:simulation}
\vspace{-0.5cm}
\end{table}

We assume that each Wi-Fi network is randomly distributed spatially, and each Wi-Fi network covers $1\sim4$ connected locations. We choose the transmission rate function from \cite{bianchi2000performance}, and define $R_n\left({\rho_n}\right)$ as:
\begin{equation}
{R_n}\left( {\rho_n}  \right) = \frac{{{P_{tr}}{P_{s}}G}}{{\left( {1 - {P_{tr}}} \right){T_b} + {P_{tr}}{P_s}{T_s} + {P_{tr}}\left( {1 - {P_s}} \right){T_c}}}.\label{equ:WiFirate}
\end{equation}
Here, $G$ is the average payload length, $T_b$ is the backoff slot size, $T_s$ is the successful transmission slot size, $T_c$ is the collision slot size, $P_{tr}=1 - {\left( {1 - \varphi } \right)^{\rho_n} }$, $P_s=\frac{{\rho_n}\varphi{\left( {1 - \varphi } \right)^{{\rho_n}-1} }}{1 - {\left( {1 - \varphi } \right)^{\rho_n} }}$, and $\varphi$ is the transmission probability.
We choose the power consumption function from \cite{jung2012adaptive}, and define $P_n^W\left( {\rho_n}  \right)$ as
\begin{equation}
P_n^W\left( {\rho_n}  \right) = \frac{{\left( {1 - {P_{tr}}} \right){E_b} + {P_{tr}}{P_s}{E_s} + \sum\limits_{j = 2}^{\rho_n}  {{P_{c,j}}{E_{c,j}}\left(\rho_n\right)} }}{{\left( {1 - {P_{tr}}} \right){T_b} + {P_{tr}}{P_s}{T_s} + {P_{tr}}\left( {1 - {P_s}} \right){T_c}}},\label{equ:WiFipower}
\end{equation}
where ${P_{c,j}} =  {\binom{{\rho_n}}{j}} {\varphi ^j}{\left( {1 - \varphi } \right)^{{\rho_n}  - j}}$ and $E_b$, $E_s$, and $E_{c,j}\left(\rho_n\right)$ are the energy consumptions of the backoff slot, successful transmission, and $j$ collided transmissions, respectively. Table \ref{table:simulation} summarizes other system parameters.{\footnote{Parameter $\varphi$ is not given in Table \ref{table:simulation}, as it is obtained by solving a non-linear system \cite{bianchi2000performance}.}}
{{
\subsubsection{Users}
We assume that user $l$'s initial location ${S_l}\left(0\right)$ is uniformly chosen from set $\cal S$. For all later frames, \emph{i.e.}, $t\in{\cal T}_k, k>0$, user $l$ moves according to a Markovian process.
Similarly, for user $l$'s traffic arrival $A_l\left(t\right)$, we generate it based on an ergodic Markov chain. Unless specified otherwise, the mean traffic arrival rate per user is set to be $2$ Mbps.}}
\begin{figure*}[t]
  \centering
  \subfigure[Power vs Parameter $V$.]{
    \label{JSAC:1:a}
    \includegraphics[scale=0.36]{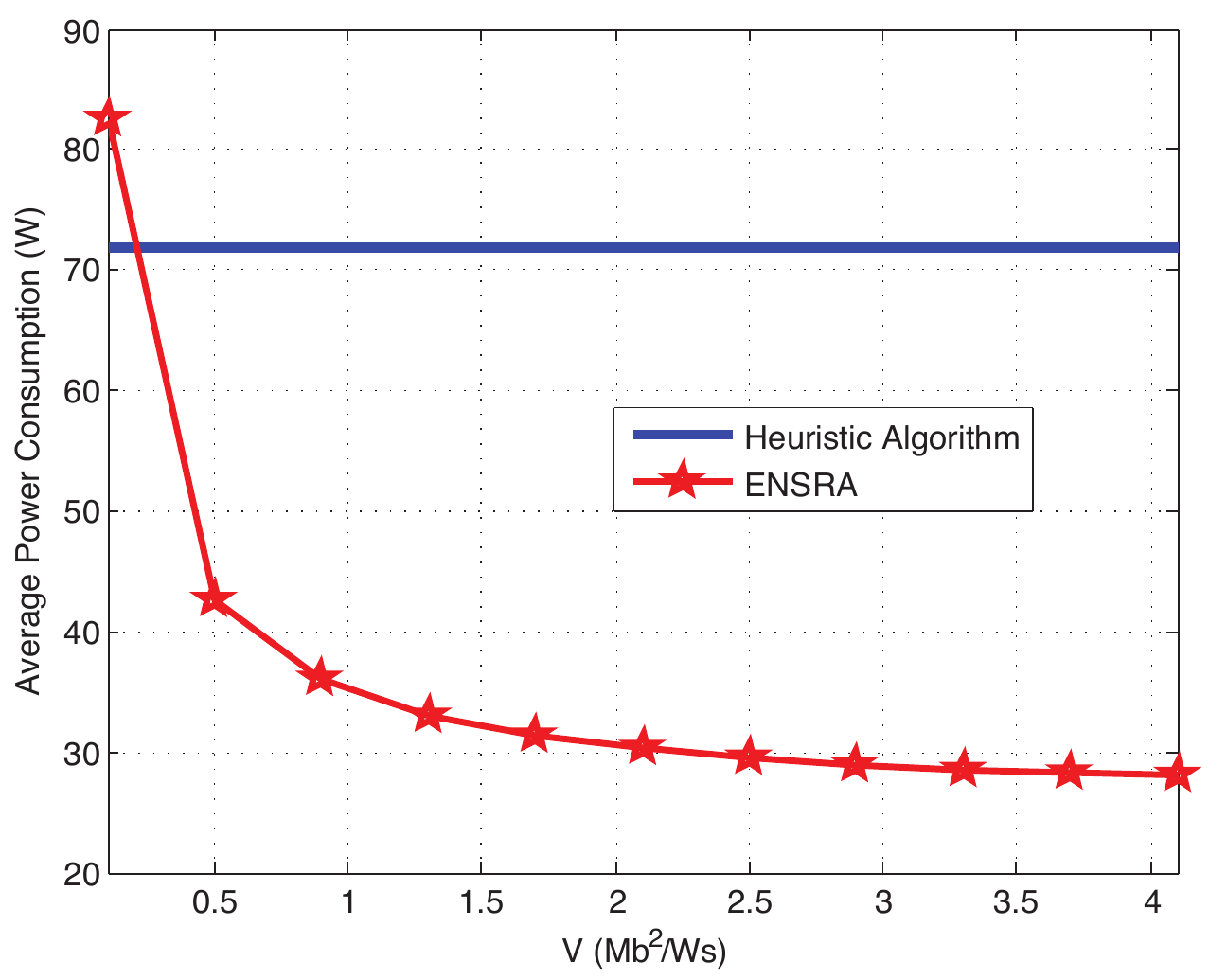}}
  \subfigure[Traffic Delay vs Parameter $V$.]{
    \label{JSAC:1:b}
    \includegraphics[scale=0.36]{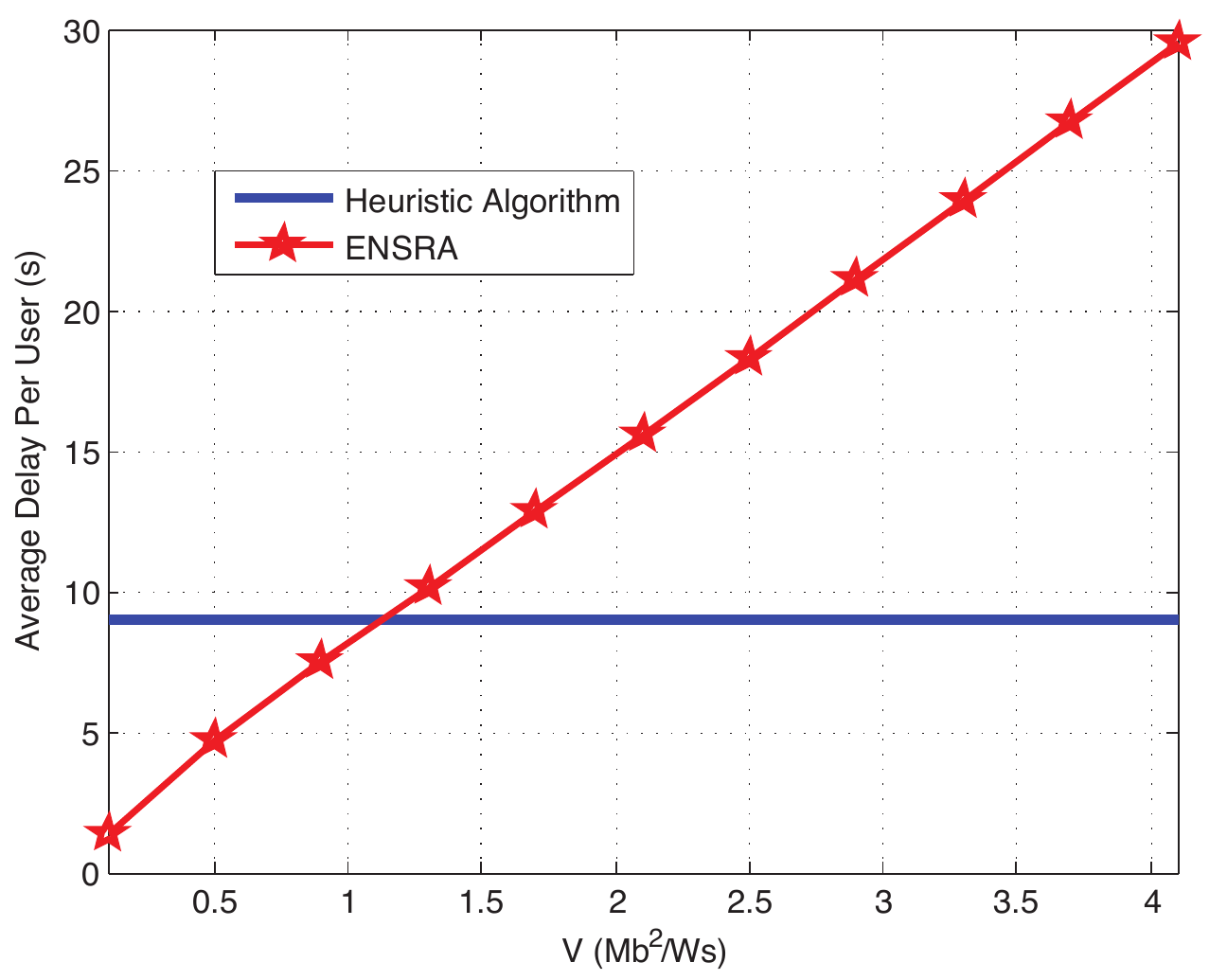}}
  \subfigure[Data Offloading vs Parameter $V$.]{
    \label{JSAC:1:c}
    \includegraphics[scale=0.36]{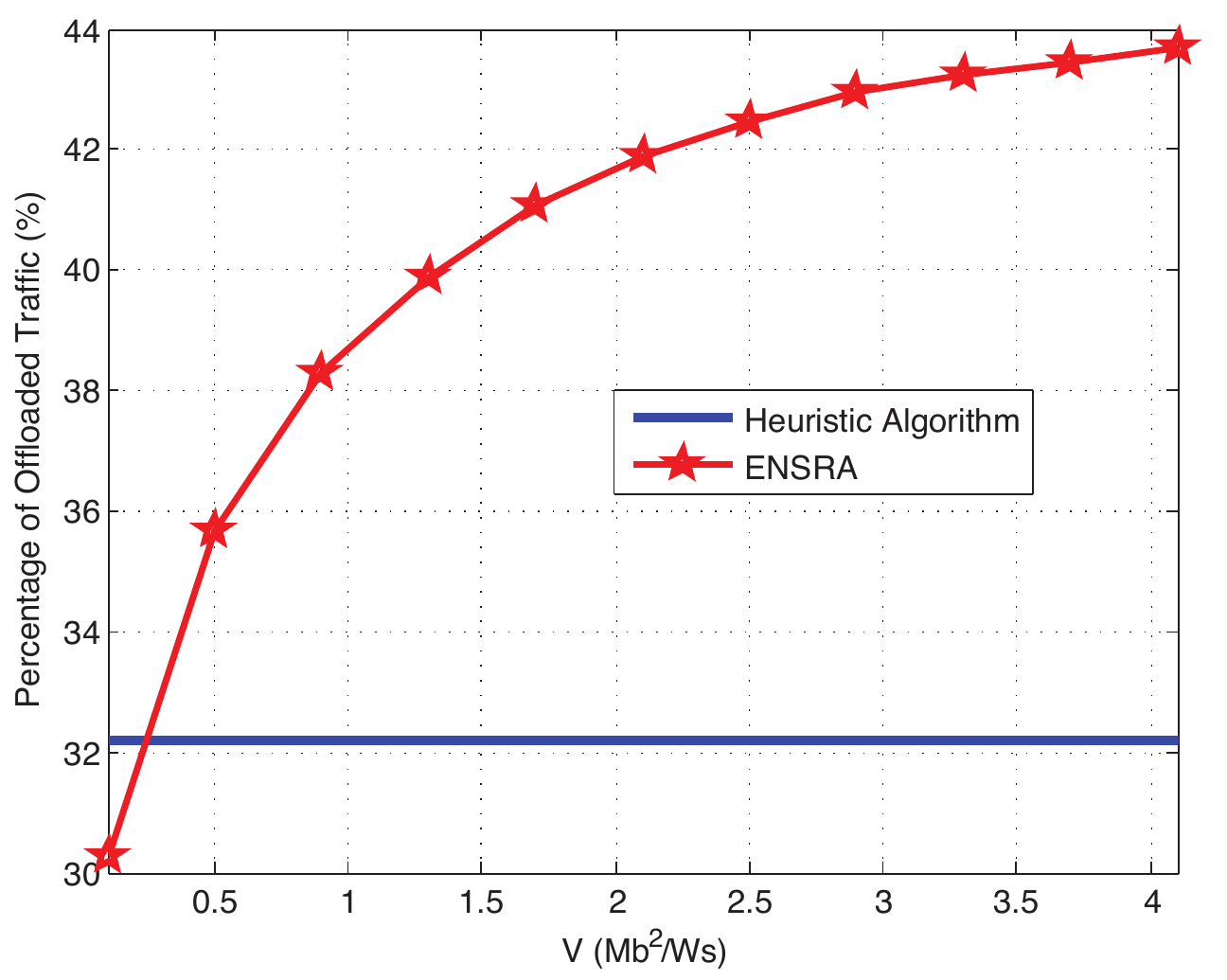}}
  \caption{Comparison of \textsf{ENSRA} and Heuristic Algorithm.}
  \label{fig:JSAC:1}
  \vspace{-0.5cm}
\end{figure*}
\begin{figure*}[t]
  \centering
  \subfigure[Power Consumption vs Parameter $V$.]{
    \label{JSAC:2:a}
    \includegraphics[scale=0.38]{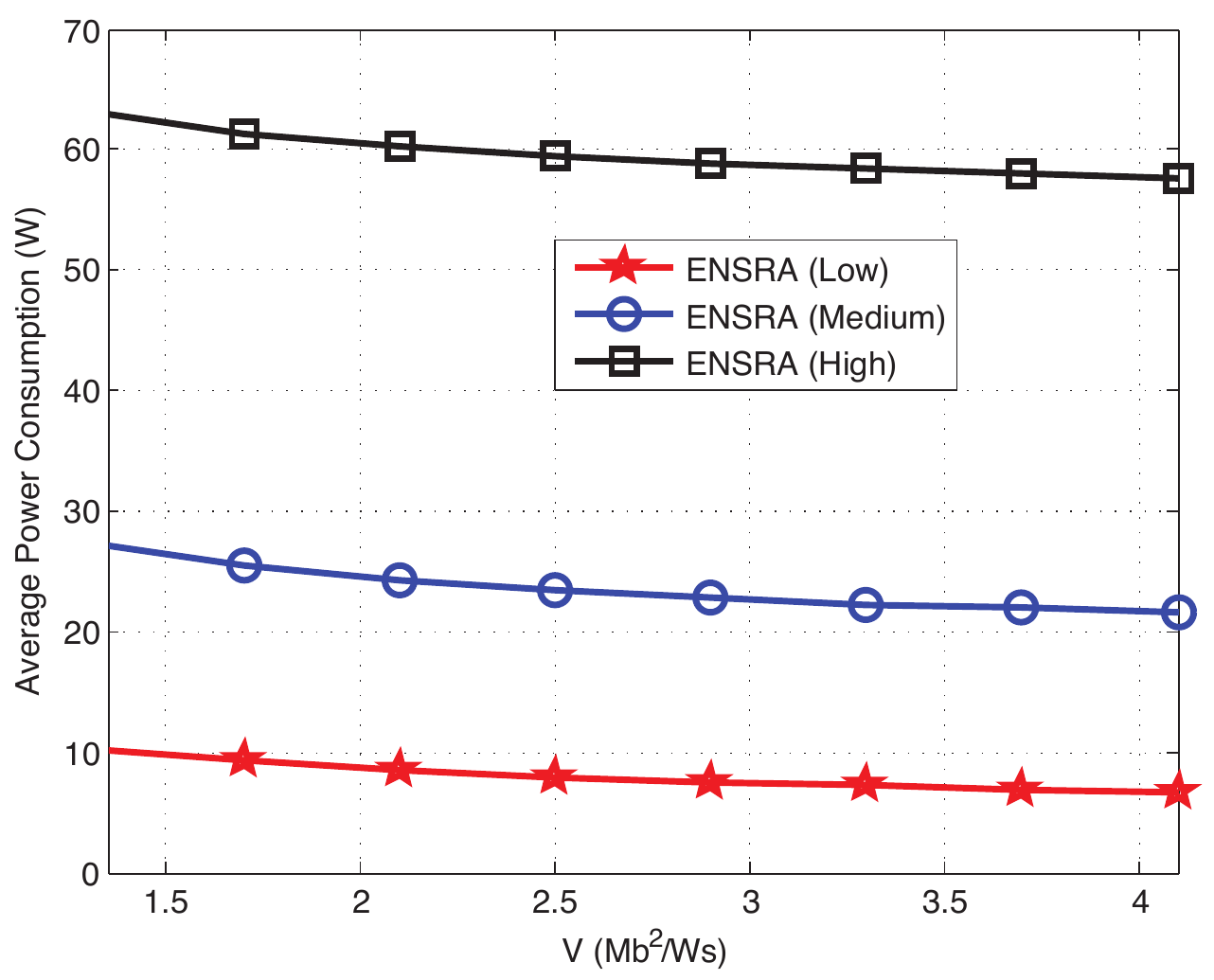}}
  \subfigure[Traffic Delay vs Parameter $V$.]{
    \label{JSAC:2:b}
    \includegraphics[scale=0.38]{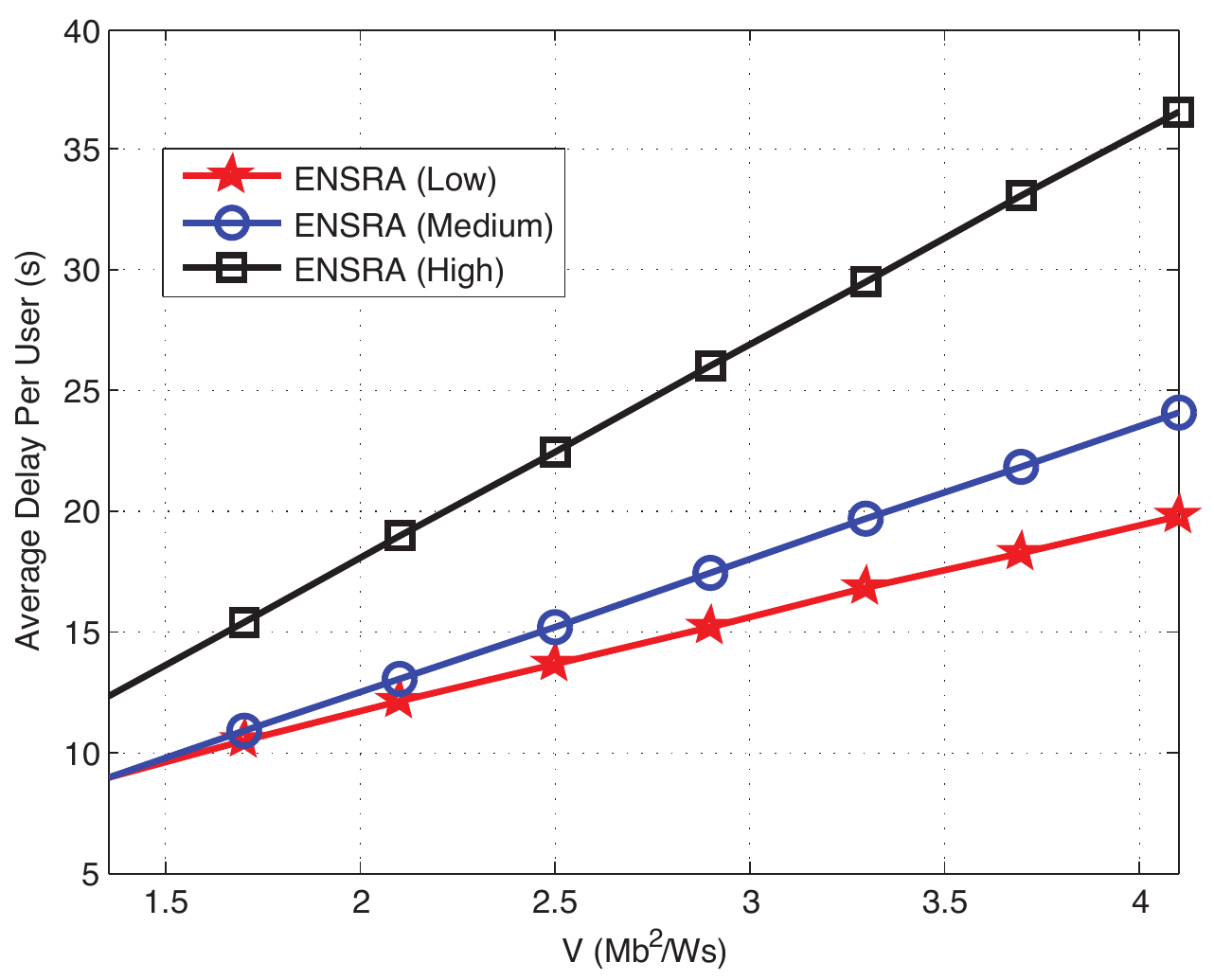}}
  \subfigure[Data Offloading vs Parameter $V$.]{
    \label{JSAC:2:c}
    \includegraphics[scale=0.38]{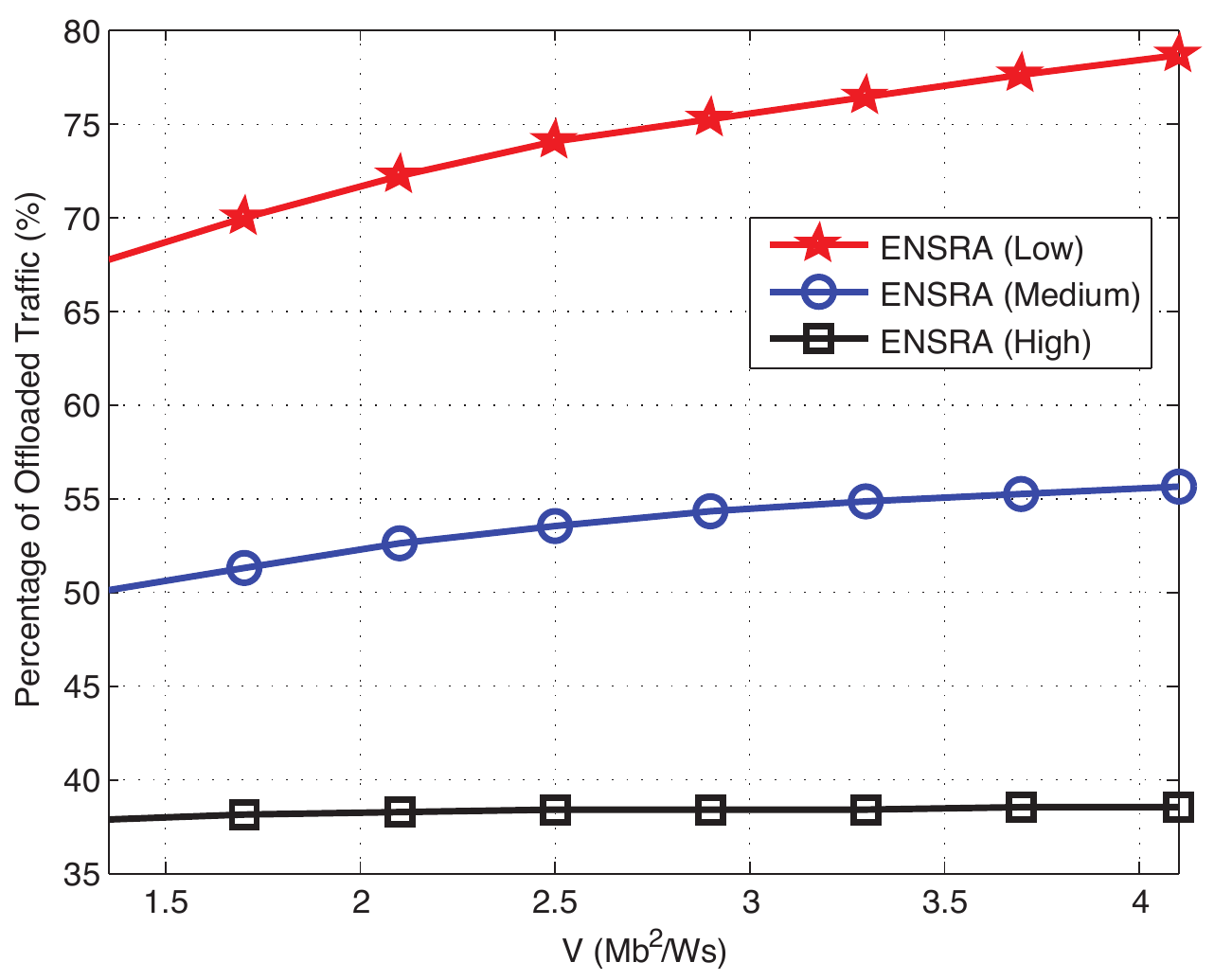}}
  \caption{\textsf{ENSRA}'s Performance under Low, Medium, and High Workloads.}
  \label{fig:JSAC:2}
  \vspace{-0.5cm}
\end{figure*}
\subsection{Simulation results}\label{subsec:simulationresults}
\subsubsection{Comparison between \textsf{ENSRA} and heuristic algorithm}
We compare \textsf{ENSRA} with the following \emph{heuristic algorithm}.

\emph{Heuristic algorithm:} {{At the beginning of each frame, the operator first assigns the users who are only covered by the macrocell network or have $d_l\left(t\right)$ smaller than $100 {\rm~m}$. Then the operator sequentially checks the available Wi-Fi networks for each of the remaining users, and assigns each user to the Wi-Fi network with the lowest number of connected users;}} at every time slot, the operator determines the resource allocation based on a heuristic method \cite{huang2009downlink}.{\footnote{Specifically, the operator first allocates the subchannels by assuming the total power is evenly allocated to all subchannels, then allocates the power based on the determined subchannel allocation.}}

In Figure \ref{fig:JSAC:1}, we compare \textsf{ENSRA} under different parameter $V$ with the heuristic algorithm. In Figure \ref{JSAC:1:a}, we plot the total power consumption of \textsf{ENSRA} against $V$. We observe that, as $V$ increases, \textsf{ENSRA}'s total power consumption decreases. According to (\ref{equ:pENSRA}), the upper bound of $P_{av}^{\textsf{ENSRA}}$ decreases with the increasing of $V$, which is consistent with our observation here. Figure \ref{JSAC:1:a} also shows the total power consumption of the heuristic algorithm, which is independent of $V$. We notice that \textsf{ENSRA} consumes less power than the heuristic algorithm for any $V>0.2{}$ ${{{\rm Mb}^2}}/{\rm W \cdot s}$.

{{In Figure \ref{JSAC:1:b}, we plot the average traffic delay per user under \textsf{ENSRA} against $V$.{\footnote{{In the simulation, we first obtain the average queue length per user. Based on Little's law, we compute the average traffic delay per user as the ratio between the average queue length and the mean traffic arrival rate, \emph{i.e.}, $2$ Mbps.}}} As $V$ increases, the average delay of \textsf{ENSRA} increases, which is consistent with the result in (\ref{equ:QENSRA}).}} Compared with the heuristic algorithm, \textsf{ENSRA} generates less delay for any $V<1.1{}$ ${{{\rm Mb}^2}}/{\rm W \cdot s}$. Figure \ref{JSAC:1:a} and Figure \ref{JSAC:1:b} imply that, if the operator chooses $0.2{{{~\rm Mb}^2}}/{\rm W \cdot s}\le V \le 1.1~{{{\rm Mb}^2}}/{\rm W \cdot s}{}$, \textsf{ENSRA} outperforms the heuristic algorithm in both the power and delay. For example, \textsf{ENSRA} with $V=0.5$ ${{{\rm Mb}^2}}/{\rm W \cdot s}{}$ saves $40.8{}$\% power and $47.8{}$\% delay over the heuristic algorithm.

In Figure \ref{JSAC:1:c}, we plot the percentage of the traffic served in Wi-Fi against $V$. According to (\ref{equ:ENSRA}), a larger $V$ implies that the operator focuses more on the power consumption than the traffic delay, and \textsf{ENSRA} will delay users' traffic to Wi-Fi networks to reduce the power cost. Hence, in Figure \ref{JSAC:1:c}, the percentage of the traffic served in Wi-Fi increases with $V$.
\subsubsection{\textsf{ENSRA}'s performance under different workloads}
In Figure \ref{fig:JSAC:2}, We compare \textsf{ENSRA}'s performance under low, medium, and high workloads (the mean traffic arrival rate per user equals $1$ Mbps, $2$ Mbps, and $3$ Mbps, respectively). In Figure \ref{JSAC:2:a}, we observe that \textsf{ENSRA} consumes more power under a higher workload. This is because the minimum power consumption required to stabilize the system increases with the traffic arrival rates. In Figure \ref{JSAC:2:b}, we find that \textsf{ENSRA} generates a larger delay under a higher workload. This is consistent with the reality that users experience severe traffic delay during the peak hours. Figure \ref{JSAC:2:c} shows that \textsf{ENSRA} offloads a larger percent of traffic to Wi-Fi under a lower workload. The reason is as follows: under the low workload, the operator can offload users' traffic to the lower cost Wi-Fi networks without causing much delay; while under the high workload, the operator has to fully utilize the cellular and Wi-Fi networks to serve users' high traffic demand.
\subsubsection{Comparison between \textsf{ENSRA} and \textsf{GP-ENSRA}}
\begin{figure*}[t]
  \centering
  \subfigure[Power-Delay Tradeoff.]{
    \label{JSAC:3:a}
    \includegraphics[scale=0.38]{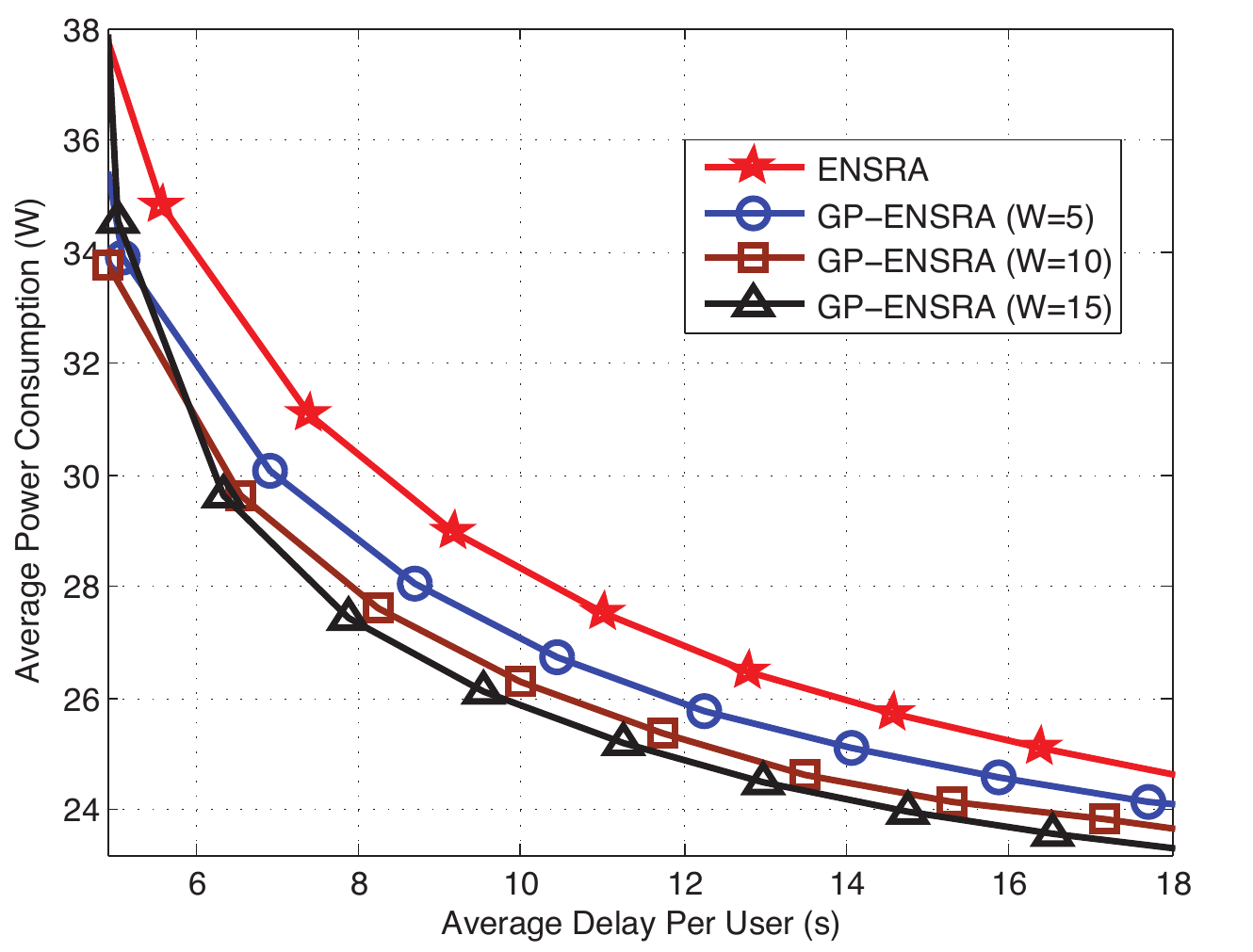}}
  \subfigure[Data Offloading vs Average Delay.]{
    \label{JSAC:3:b}
    \includegraphics[scale=0.38]{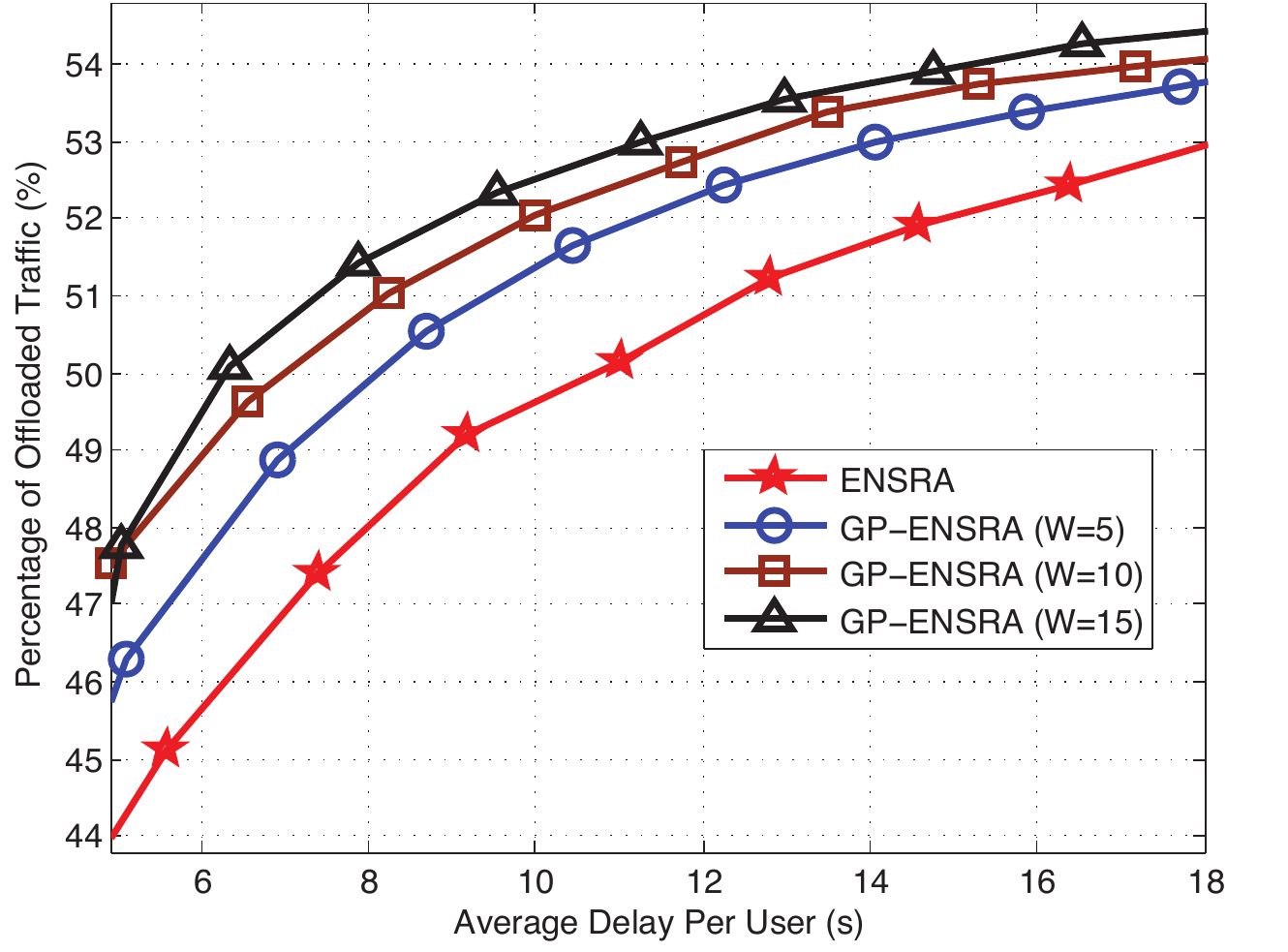}}
  \subfigure[Power-Delay Tradeoff under Prediction Errors.]{
    \label{JSAC:3:c}
    \includegraphics[scale=0.38]{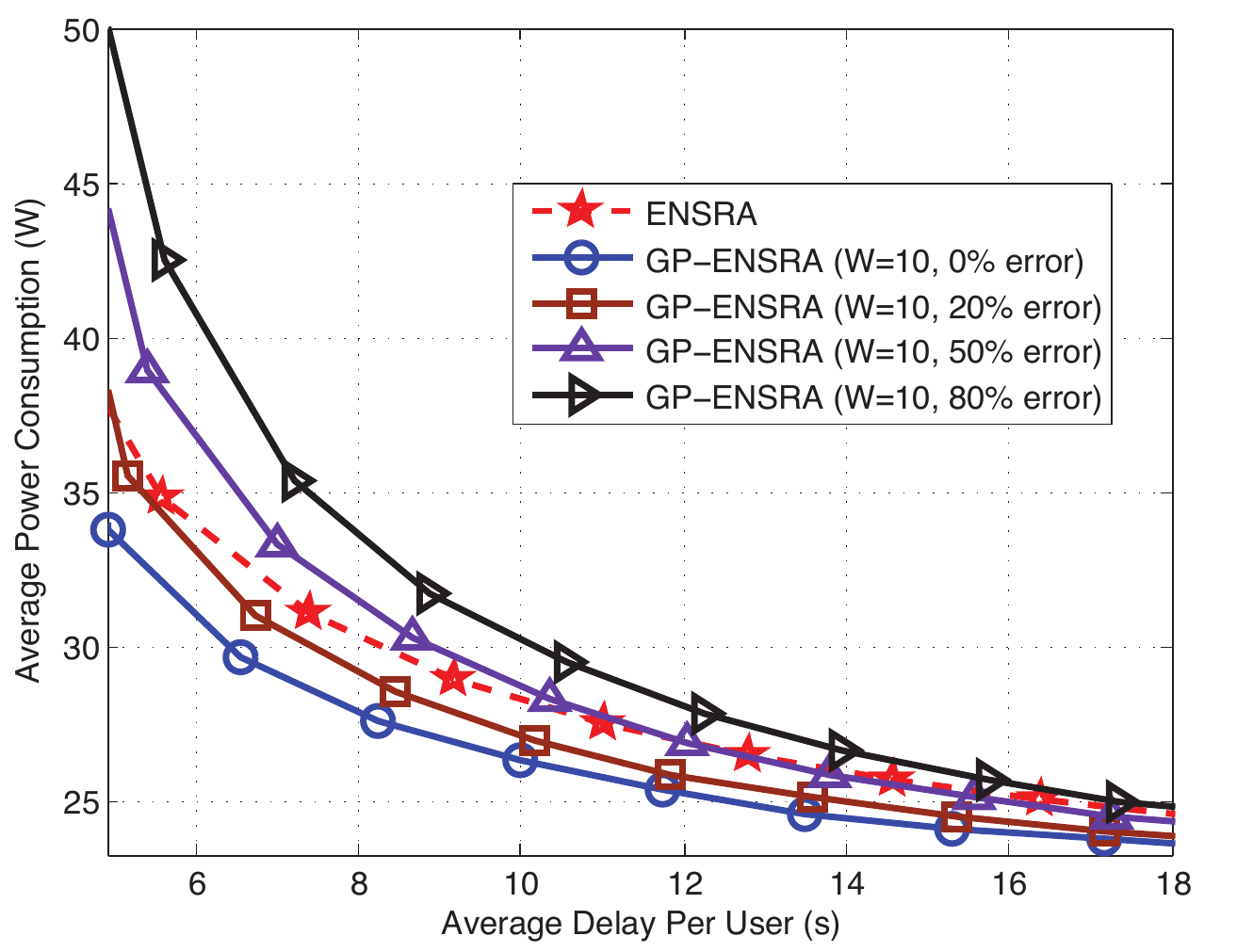}}
  \caption{Comparison of \textsf{ENSRA} and \textsf{GP-ENSRA}.}
  \label{fig:JSAC:3}
  \vspace{-0.5cm}
\end{figure*}
\begin{figure*}[t]
  \centering
  \subfigure[Power vs Parameter $V$.]{
    \label{JSAC:4:a}
    \includegraphics[scale=0.38]{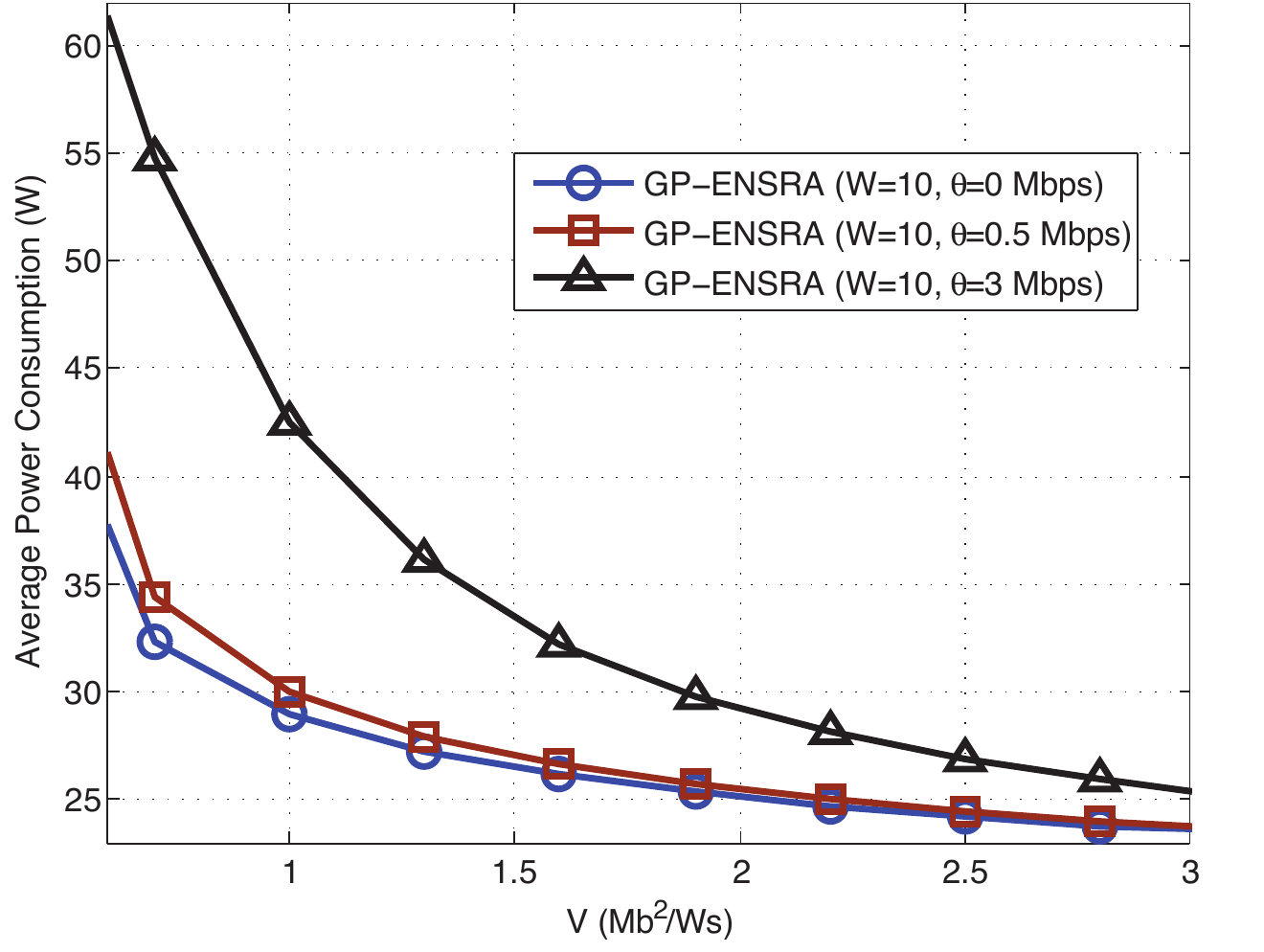}}
  \subfigure[Traffic Delay vs Parameter $V$.]{
    \label{JSAC:4:b}
    \includegraphics[scale=0.38]{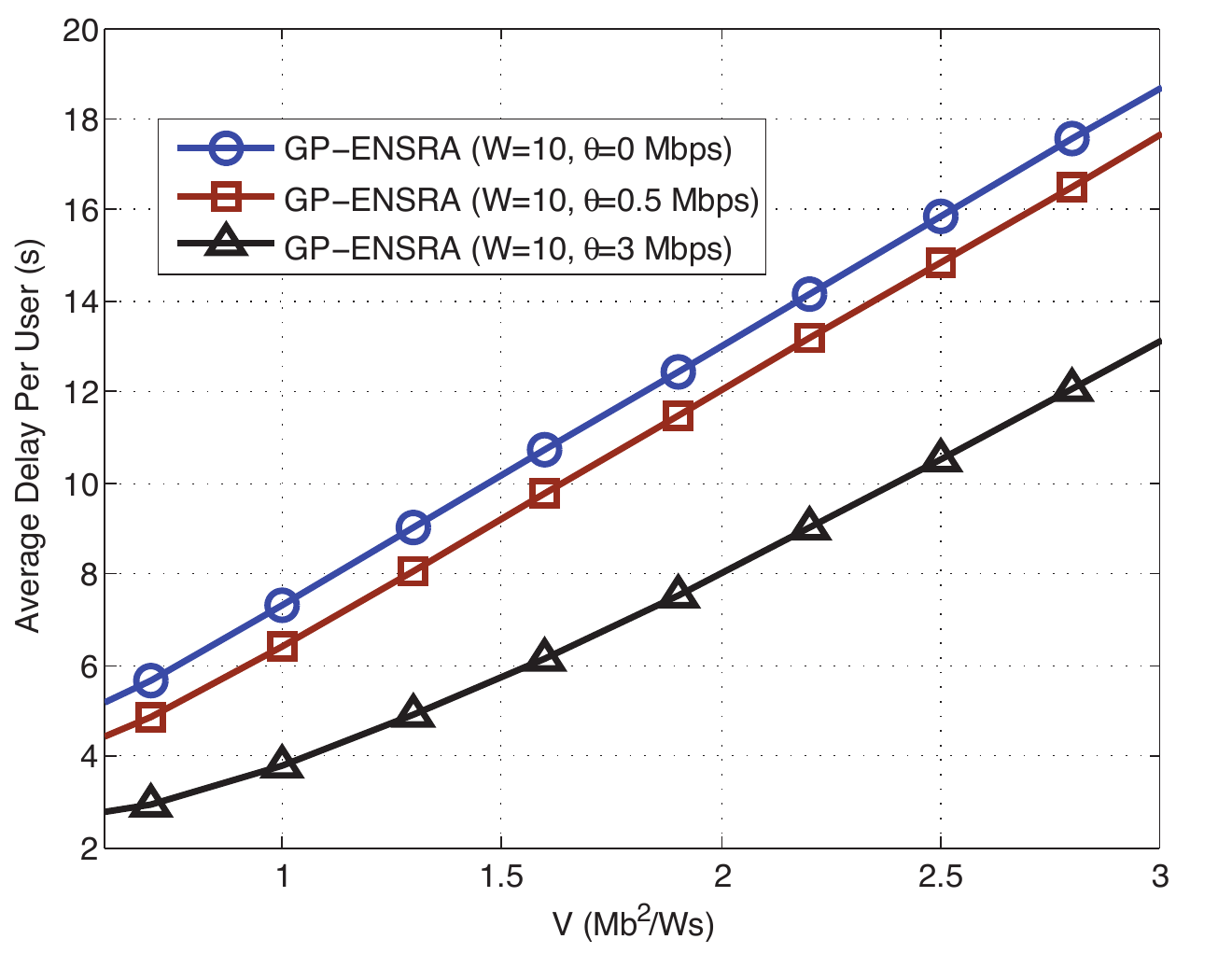}}
  \subfigure[Power-Delay Tradeoff.]{
    \label{JSAC:4:c}
    \includegraphics[scale=0.38]{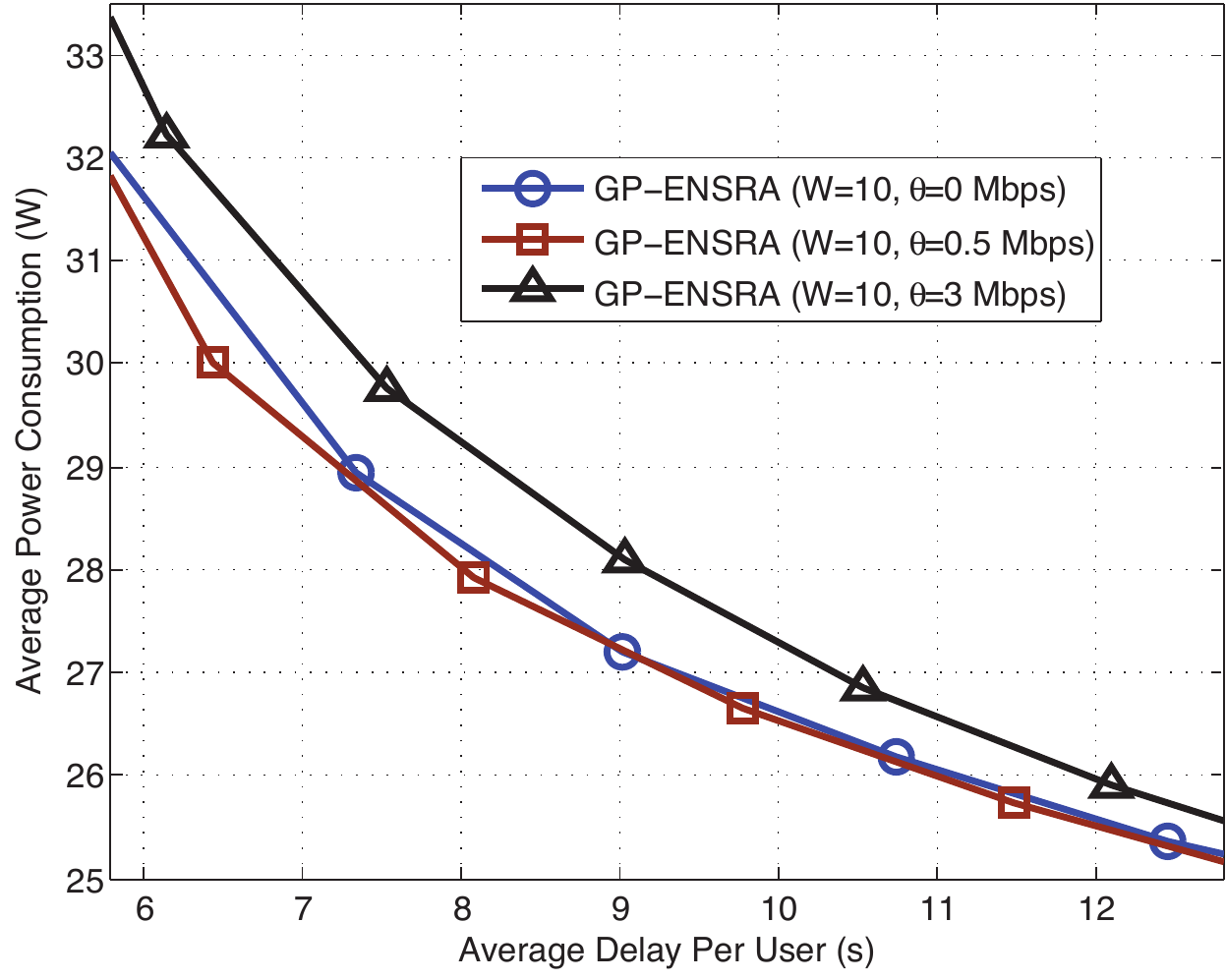}}
  \caption{Comparison of \textsf{GP-ENSRA} with Different $\theta$.}
  \label{fig:JSAC:4}
  \vspace{-0.5cm}
\end{figure*}
{{In Figure \ref{JSAC:3:a}, we plot the average total power consumption against the average traffic delay per user for \textsf{ENSRA} and \textsf{GP-ENSRA}.}} We obtain these power-delay tradeoff curves by varying $V$.
{{Comparing \textsf{ENSRA} with \textsf{GP-ENSRA}, we observe that when the traffic delay is above $6$ s, \textsf{GP-ENSRA} always generates a smaller power consumption than \textsf{ENSRA} under the same traffic delay.{\footnote{{When the generated traffic delay is restricted to a small value (\emph{e.g.}, smaller than $6$ s), the performance improvement of \textsf{GP-ENSRA} over \textsf{ENSRA} is not obvious. The reason is that in order to generate a small delay, the operator has to serve the traffic immediately even if the users' channel conditions and Wi-Fi availabilities in the future frames are better.}}}}}
{{For example, when the generated traffic delay is $8{}$ s, the power consumptions of \textsf{ENSRA} and \textsf{GP-ENSRA} with window size $W=15$ are $30.4{}$ W and $27.4$ W, respectively. Hence, the power saving of \textsf{GP-ENSRA} with $W=15$ over \textsf{ENSRA} is $9.9{}$\%.}}
The performance improvement of \textsf{GP-ENSRA} is more obvious in terms of the delay saving. For example, when the operator pursues a power consumption of $26{}$ W, the average traffic delays under \textsf{ENSRA} and \textsf{GP-ENSRA} with window size $W=15$ are $13.9{}$ s and $9.7{}$ s, respectively. This shows that \textsf{GP-ENSRA} with window size $W=15$ saves $30.2{}$\% delay over \textsf{ENSRA}.
{{In Figure \ref{JSAC:3:a}, we also observe that the performance improvement increases with the size of the prediction window.}}{\footnote{The improvement of \textsf{GP-ENSRA} over \textsf{ENSRA} is influenced by the variance of system randomness. For example, if users' locations change frequently every several frames, knowing users' new locations and Wi-Fi availabilities in the next few frames are crucial. In this case, \textsf{GP-ENSRA} outperforms \textsf{ENSRA} significantly. We have simulated a wide range of system parameters. {{Under the same traffic delay, \textsf{GP-ENSRA} usually reduces the power consumption over \textsf{ENSRA} by $5\sim10{}$\%.}} Under the same power consumption, \textsf{GP-ENSRA} usually reduces the traffic delay over \textsf{ENSRA} by $20\sim30{}$\%.}}

In Figure \ref{JSAC:3:b}, we compare the percentages of the traffic offloaded to Wi-Fi under \textsf{ENSRA} and \textsf{GP-ENSRA}. We plot the percentage of the traffic served in Wi-Fi against the average traffic delay. When generating the same traffic delay, \textsf{GP-ENSRA} offloads a larger percentage of traffic than \textsf{ENSRA}. {{The reason is that the predictive information helps the operator design a network selection and resource allocation strategy that utilizes Wi-Fi networks more efficiently to reduce the total power consumption.}}

{{
In Figure \ref{JSAC:3:c}, we investigate the power-delay performance of \textsf{GP-ENSRA} under the prediction errors. For example, \textsf{GP-ENSRA} with $20$\% prediction error means that for each information (\emph{i.e.}, users' locations, channel conditions, and traffic arrivals) of the future frames, with $0.8$ probability the operator accurately predicts its value, while with $0.2$ probability the operator obtains an incorrect value of the information.{\footnote{{The incorrect value is randomly picked from all possible values of the random event.}}} In Figure \ref{JSAC:3:c}, we plot the average power consumption against the average traffic delay per user for \textsf{ENSRA} and \textsf{GP-ENSRA} with window size $W=10$ under different percentages of the prediction errors. We observe that the power-delay performance of \textsf{GP-ENSRA} declines as the percentage of the prediction errors increases. However, \textsf{GP-ENSRA} with $20\%$ prediction error still achieves a better power-delay tradeoff than the non-predictive algorithm \textsf{ENSRA}, which shows the robustness of \textsf{GP-ENSRA} against the prediction errors.{\footnote{{Notice that the prediction errors only exist in the future frames, and the operator can still obtain the accurate information of the current frame.}}}
}}
{{
\subsubsection{Influence of Parameter $\theta$ in \textsf{GP-ENSRA}}
In Figure \ref{fig:JSAC:4}, we compare \textsf{GP-ENSRA} with window size $W=10$ under different parameter $\theta$. In Figure \ref{JSAC:4:a}, we plot the power consumption against $V$ for \textsf{GP-ENSRA} with different $\theta$, and observe that the power consumption of \textsf{GP-ENSRA} increases with $\theta$. This is because \textsf{GP-ENSRA} is the approximation of \textsf{P-ENSRA}, and the upper bound of the power consumption of \textsf{P-ENSRA} in (\ref{equ:pENSRA:pre}) increases with $\theta$.{\footnote{{Recall that $P\left(\theta\right)$ stands for the minimum power required to stabilize the traffic arrival vector ${\mathbb{E}}\left\{{{\bm A}\left(t\right)}\right\}+\theta \cdot {\bm 1}$, hence it is easy to verify that $P\left(\theta\right)$ increases with $\theta$.}}} In Figure \ref{JSAC:4:b}, we plot the average delay against $V$ for \textsf{GP-ENSRA} with different $\theta$. We observe that \textsf{GP-ENSRA} with a large $\theta$ generates a smaller traffic delay. As we explained in Section V-B, with a large $\theta$, the operator assigns large weights to the transmission rates of the earlier frames within the prediction window. The large assigned weights push the operator to serve the traffic in the earlier frames rather than later frames, which eventually decreases the average traffic delay.
From Figures \ref{JSAC:4:a} and \ref{JSAC:4:b}, we conclude that the increase of $\theta$ has two impacts: (i) it increases the power consumption; (ii) it decreases the traffic delay.
In Figure \ref{JSAC:4:c}, we plot the power-delay tradeoff for \textsf{GP-ENSRA} with different $\theta$. 
We find that \textsf{GP-ENSRA}'s power-delay performance first improves with $\theta$ (from $\theta=0{~\rm Mbps}$ to $\theta=0.5{~\rm Mbps}$) and then declines with $\theta$ (from $\theta=0.5{~\rm Mbps}$ to $\theta=3{~\rm Mbps}$). This is because, when $0~{\rm Mbps} \le\theta \le 0.5{~}{\rm Mbps}$, the aforementioned impact (ii) plays the dominant role; while when $\theta>0.5{~}{\rm Mbps}$, the aforementioned impact (i) plays the dominant role.
}}
\section{Conclusion} \label{sec:conclusion}
In this paper, we studied the online network selection and resource allocation problem in the stochastic integrated cellular and Wi-Fi networks. {{We first proposed the \textsf{ENSRA} algorithm, which can generate a close-to-optimal power consumption at the expense of an increase in the average traffic delay.}} We then proposed the \textsf{P-ENSRA} algorithm and the \textsf{GP-ENSRA} algorithm by incorporating the prediction of the system randomness into the network selection and resource allocation. {{Simulation results showed that the future information helps the operator achieve a much better power-delay performance in the large delay regime.}}

In our future work, we plan to address more challenges related to the power and channel allocation. For example, instead of the continuous power allocation, practical systems usually adopt discrete power control with a limited number of power levels and modulation coding schemes \cite{liarm}. The discrete power control problem is in general NP-hard. Furthermore, in a practical OFDM system, imperfect carrier synchronization and channel estimation may result in ``self-noise'' \cite{huang2009downlink}. We intend to incorporate the consideration of the discrete power control and ``self-noise'' into our algorithm design. Moreover, we want to consider the modified Shannon capacity bounds in \cite{mogensen2007lte} to better model the maximum achievable data rate of the macrocell network.

\bibliographystyle{IEEEtran}
\bibliography{IEEEabrv,bare_conf}

\begin{thebibliography}{10}
\providecommand{\url}[1]{#1}
\csname url@samestyle\endcsname
\providecommand{\newblock}{\relax}
\providecommand{\bibinfo}[2]{#2}
\providecommand{\BIBentrySTDinterwordspacing}{\spaceskip=0pt\relax}
\providecommand{\BIBentryALTinterwordstretchfactor}{4}
\providecommand{\BIBentryALTinterwordspacing}{\spaceskip=\fontdimen2\font plus
\BIBentryALTinterwordstretchfactor\fontdimen3\font minus
  \fontdimen4\font\relax}
\providecommand{\BIBforeignlanguage}[2]{{%
\expandafter\ifx\csname l@#1\endcsname\relax
\typeout{** WARNING: IEEEtran.bst: No hyphenation pattern has been}%
\typeout{** loaded for the language `#1'. Using the pattern for}%
\typeout{** the default language instead.}%
\else
\language=\csname l@#1\endcsname
\fi
#2}}
\providecommand{\BIBdecl}{\relax}
\BIBdecl

\bibitem{yu2014predictive}
H.~Yu, M.~H. Cheung, L.~Huang, and J.~Huang, ``Predictive delay-aware network
  selection in data offloading,'' in \emph{Proc. of IEEE GLOBECOM}, Austin, TX,
  December 2014, pp. 1376--1381.

\bibitem{fehske2011global}
A.~Fehske, G.~Fettweis, J.~Malmodin, and G.~Bicz{\'o}k, ``The global footprint
  of mobile communications: The ecological and economic perspective,''
  \emph{IEEE Communications Magazine}, vol.~49, no.~8, pp. 55--62, August 2011.

\bibitem{oh2011toward}
E.~Oh, B.~Krishnamachari, X.~Liu, and Z.~Niu, ``Toward dynamic energy-efficient
  operation of cellular network infrastructure,'' \emph{IEEE Communications
  Magazine}, vol.~49, no.~6, pp. 56--61, June 2011.

\bibitem{tutorialA}
M.~Ismail, W.~Zhuang, E.~Serpedin, and K.~Qaraqe, ``A survey on green mobile
  networking: from the perspectives of network operators and mobile users,''
  \emph{IEEE Communications Surveys \& Tutorials}, vol.~17, no.~3, pp.
  1535--1556, August 2015.

\bibitem{tutorialB}
J.~B. Rao and A.~O. Fapojuwo, ``A survey of energy efficient resource
  management techniques for multicell cellular networks,'' \emph{IEEE
  Communications Surveys \& Tutorials}, vol.~16, no.~1, pp. 154--180, First
  Quarter 2014.

\bibitem{ismail2011network}
M.~Ismail and W.~Zhuang, ``Network cooperation for energy saving in green radio
  communications,'' \emph{IEEE Wireless Communications}, vol.~18, no.~5, pp.
  76--81, October 2011.

\bibitem{auer2011much}
G.~Auer, V.~Giannini, C.~Desset, I.~Godor, P.~Skillermark, M.~Olsson, M.~A.
  Imran, D.~Sabella, M.~J. Gonzalez, O.~Blume, and A.~Fehske, ``How much energy
  is needed to run a wireless network?'' \emph{IEEE Wireless Communications},
  vol.~18, no.~5, pp. 40--49, October 2011.

\bibitem{neely2006energy}
M.~J. Neely, ``Energy optimal control for time-varying wireless networks,''
  \emph{IEEE Transactions on Information Theory}, vol.~52, no.~7, pp.
  2915--2934, July 2006.

\bibitem{yao2012data}
Y.~Yao, L.~Huang, A.~Sharma, L.~Golubchik, and M.~J. Neely, ``Data centers
  power reduction: {A} two time scale approach for delay tolerant workloads,''
  in \emph{Proc. of IEEE INFOCOM}, Orlando, FL, March 2012, pp. 1431--1439.

\bibitem{nicholson2008breadcrumbs}
A.~J. Nicholson and B.~D. Noble, ``Breadcrumbs: forecasting mobile
  connectivity,'' in \emph{Proc. of ACM MobiCom}, San Francisco, CA, September
  2008, pp. 46--57.

\bibitem{paul2011understanding}
U.~Paul, A.~P. Subramanian, M.~M. Buddhikot, and S.~R. Das, ``Understanding
  traffic dynamics in cellular data networks,'' in \emph{Proc. of IEEE
  INFOCOM}, Shanghai, China, April 2011, pp. 882--890.

\bibitem{arslan2007channel}
M.~K. Ozdemir and H.~Arslan, ``Channel estimation for wireless {OFDM}
  systems,'' \emph{IEEE Communications Surveys \& Tutorials}, vol.~9, no.~2,
  pp. 18--48, 2007.

\bibitem{neely2010stochastic}
M.~J. Neely, \emph{Stochastic Network Optimization With Application to
  Communication and Queueing Systems}.\hskip 1em plus 0.5em minus 0.4em\relax
  Morgan \& Claypool Publishers, 2010.

\bibitem{venturino2014energy}
L.~Venturino, A.~Zappone, C.~Risi, and S.~Buzzi, ``Energy-efficient scheduling
  and power allocation in downlink {OFDMA} networks with base station
  coordination,'' \emph{IEEE Transactions on Wireless Communications}, vol.~14,
  no.~1, pp. 1--14, January 2015.

\bibitem{xiong2012energy}
C.~Xiong, G.~Y. Li, S.~Zhang, Y.~Chen, and S.~Xu, ``Energy-efficient resource
  allocation in {OFDMA} networks,'' \emph{IEEE Transactions on Communications},
  vol.~60, no.~12, pp. 3767--3778, December 2012.

\bibitem{meshkati2009energy}
F.~Meshkati, H.~V. Poor, S.~C. Schwartz, and R.~V. Balan, ``Energy-efficient
  resource allocation in wireless networks with quality-of-service
  constraints,'' \emph{IEEE Transactions on Communications}, vol.~57, no.~11,
  pp. 3406--3414, November 2009.

\bibitem{neely2007optimal}
M.~J. Neely, ``Optimal energy and delay tradeoffs for multiuser wireless
  downlinks,'' \emph{IEEE Transactions on Information Theory}, vol.~53, no.~9,
  pp. 3095--3113, September 2007.

\bibitem{li2014energy}
Y.~Li, M.~Sheng, Y.~Shi, X.~Ma, and W.~Jiao, ``Energy efficiency and delay
  tradeoff for time-varying and interference-free wireless networks,''
  \emph{IEEE Transactions on Wireless Communications}, vol.~13, no.~11, pp.
  5921--5931, November 2014.

\bibitem{li2015throughput}
Y.~Li, M.~Sheng, C.-X. Wang, X.~Wang, Y.~Shi, and J.~Li, ``Throughput--delay
  tradeoff in interference-free wireless networks with guaranteed energy
  efficiency,'' \emph{IEEE Transactions on Wireless Communications}, vol.~14,
  no.~3, pp. 1608--1621, March 2015.

\bibitem{lakshminarayanatransmit}
S.~Lakshminarayana, M.~Assaad, and M.~Debbah, ``Transmit power minimization in
  small cell networks under time average {QoS} constraints,'' \emph{IEEE
  Journal on Selected Areas in Communications}, 2015.

\bibitem{ATTWiFi}
\url{http://www.fiercewireless.com/press-releases/12-billion-customer-connections-made-nearly-30000-att-wi-fi-hot-spots-2011}.

\bibitem{docomo5G}
DOCOMO, ``Docomo {5G} white paper,'' Tech. Rep., July 2014.

\bibitem{huang2009downlink}
J.~Huang, V.~G. Subramanian, R.~Agrawal, and R.~A. Berry, ``Downlink scheduling
  and resource allocation for {OFDM} systems,'' \emph{IEEE Transactions on
  Wireless Communications}, vol.~8, no.~1, pp. 288--296, January 2009.

\bibitem{shen2005adaptive}
Z.~Shen, J.~G. Andrews, and B.~L. Evans, ``Adaptive resource allocation in
  multiuser {OFDM} systems with proportional rate constraints,'' \emph{IEEE
  Transactions on Wireless Communications}, vol.~4, no.~6, pp. 2726--2737,
  November 2005.

\bibitem{oh2013dynamic}
E.~Oh, K.~Son, and B.~Krishnamachari, ``Dynamic base station switching-on/off
  strategies for green cellular networks,'' \emph{IEEE Transactions on Wireless
  Communications}, vol.~12, no.~5, pp. 2126--2136, May 2013.

\bibitem{bianchi2000performance}
G.~Bianchi, ``Performance analysis of the {IEEE} 802.11 distributed
  coordination function,'' \emph{IEEE Journal on Selected Areas in
  Communications}, vol.~18, no.~3, pp. 535--547, March 2000.

\bibitem{jung2012adaptive}
B.~H. Jung, H.~Jin, and D.~K. Sung, ``Adaptive transmission power control and
  rate selection scheme for maximizing energy efficiency of {IEEE} 802.11
  stations,'' in \emph{Proc. of IEEE International Symposium on Personal Indoor
  and Mobile Radio Communications (PIMRC)}, September 2012, pp. 266--271.

\bibitem{Michael2014}
M.~H. Cheung, R.~Southwell, and J.~Huang, ``Congestion-aware network selection
  and data offloading,'' in \emph{Proc. of IEEE CISS}, Princeton, NJ, March
  2014, pp. 1--6.

\bibitem{haoran2015JSAC}
H.~Yu, M.~H. Cheung, L.~Huang, and J.~Huang, ``Power-delay tradeoff with
  predictive scheduling in integrated cellular and {Wi-Fi} networks,''
  \emph{arXiv Tech Report arXiv:1512.06428}, 2015.

\bibitem{huang2010max}
L.~Huang and M.~J. Neely, ``Max-weight achieves the exact $[{O} (1/{V}), {O}
  ({V})]$ utility-delay tradeoff under markov dynamics,'' \emph{arXiv preprint
  arXiv:1008.0200}, 2010.

\bibitem{huang2013backpressure}
L.~Huang, S.~Zhang, M.~Chen, and X.~Liu, ``When backpressure meets predictive
  scheduling,'' in \emph{Proc. of ACM MobiHoc}, Philadelphia, PA, August 2014.

\bibitem{rappaport1996wireless}
T.~S. Rappaport, \emph{Wireless communications: Principles and practice}.\hskip
  1em plus 0.5em minus 0.4em\relax New Jersey: Prentice Hall, 1996.

\bibitem{gajic2014competition}
V.~Gaji{\'c}, J.~Huang, and B.~Rimoldi, ``Competition of wireless providers for
  atomic users,'' \emph{IEEE/ACM Transactions on Networking (TON)}, vol.~22,
  no.~2, pp. 512--525, April 2014.

\bibitem{liarm}
S.~Li, Z.~Shao, and J.~Huang, ``Arm: Anonymous rating mechanism for discrete
  power control,'' in \emph{Proc. of IEEE WiOpt}, Mumbai, India, May 2015.

\bibitem{mogensen2007lte}
P.~Mogensen, W.~Na, I.~Z. Kov{\'a}cs, F.~Frederiksen, A.~Pokhariyal,
  K.~Pedersen, T.~Kolding, K.~Hugl, and M.~Kuusela, ``{LTE} capacity compared
  to the {Shannon} bound,'' in \emph{Proc. of IEEE Vehicular Technology
  Conference}, Dublin, Ireland, April 2007, pp. 1234--1238.

\end{thebibliography}
\begin{IEEEbiography}
[{\includegraphics[width=1in,height=1.25in,clip,keepaspectratio]{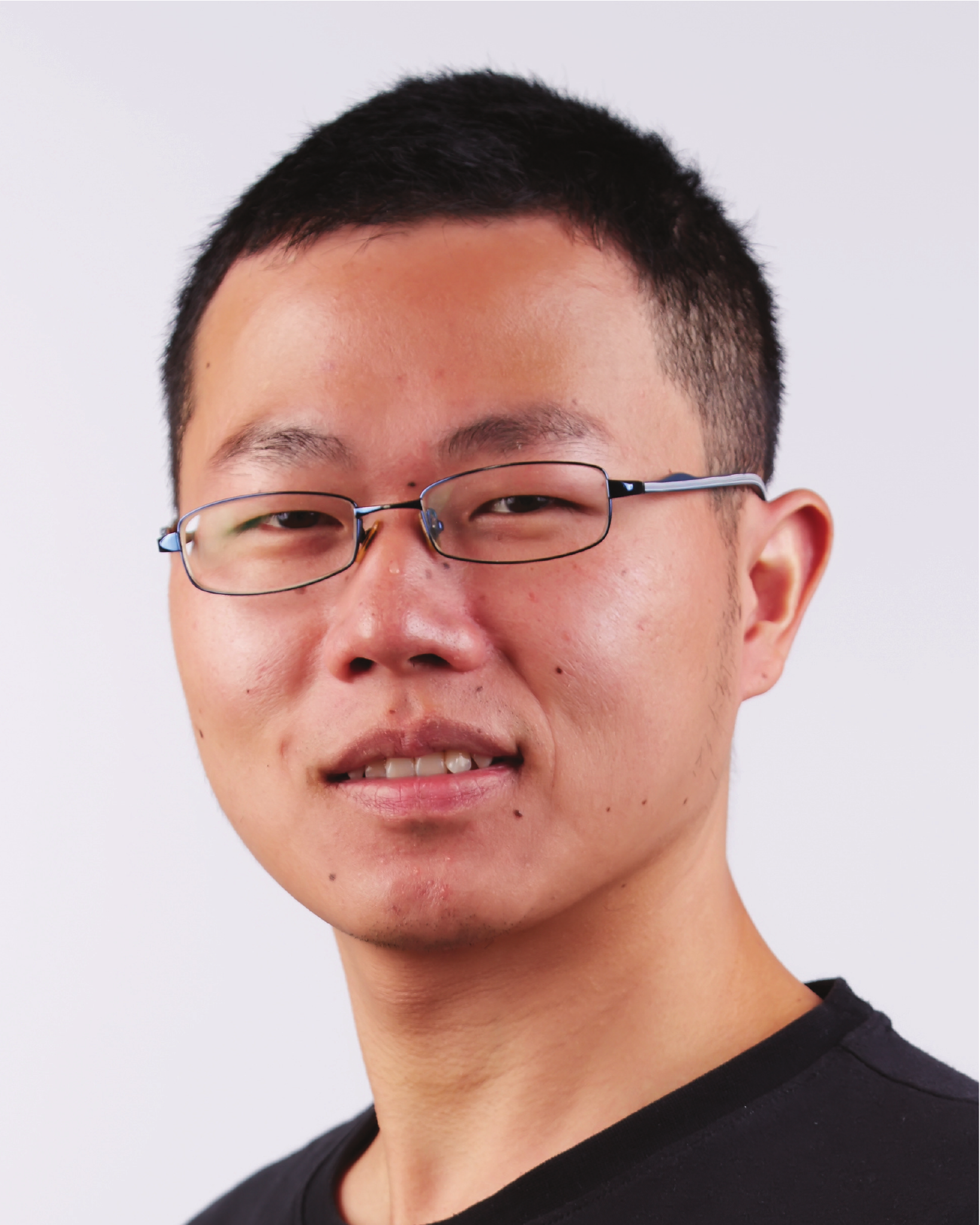}}]
{Haoran Yu} is a Ph.D. student in the Department of Information Engineering at the Chinese University of Hong Kong (CUHK). He is also a visiting student in the Yale Institute for Network Science (YINS) and the Department of Electrical Engineering at Yale University. His research interests lie in the field of wireless communications and network economics, with current emphasis on mobile data offloading, cellular/Wi-Fi integration, LTE in unlicensed spectrum, and economics of public Wi-Fi networks. 
\end{IEEEbiography}

\begin{IEEEbiography}
[{\includegraphics[width=1in,height=1.25in,clip,keepaspectratio]{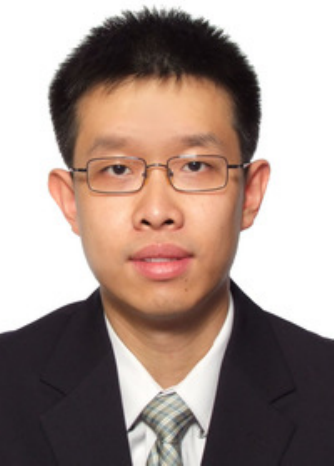}}]
{Man Hon Cheung} received the B.Eng. and M.Phil. degrees in Information Engineering from the Chinese University of Hong Kong (CUHK) in 2005 and 2007, respectively, and the Ph.D. degree in Electrical and Computer Engineering from the University of British Columbia (UBC) in 2012.
 Currently, he is a postdoctoral fellow in the Department of Information Engineering in CUHK.
 He received the IEEE Student Travel Grant for attending {\it IEEE ICC 2009}. He was awarded the Graduate Student International Research Mobility Award by UBC, and the Global Scholarship Programme for Research Excellence by CUHK.
 He serves as a Technical Program Committee member in {\it IEEE ICC}, {\it Globecom}, and {\it WCNC}.
 His research interests include the design and analysis of wireless network protocols using optimization theory, game theory, and dynamic programming, with current focus on mobile data offloading, mobile crowd sensing, and network economics.
\end{IEEEbiography}

\begin{IEEEbiography}
[{\includegraphics[width=1in,height=1.25in,clip,keepaspectratio]{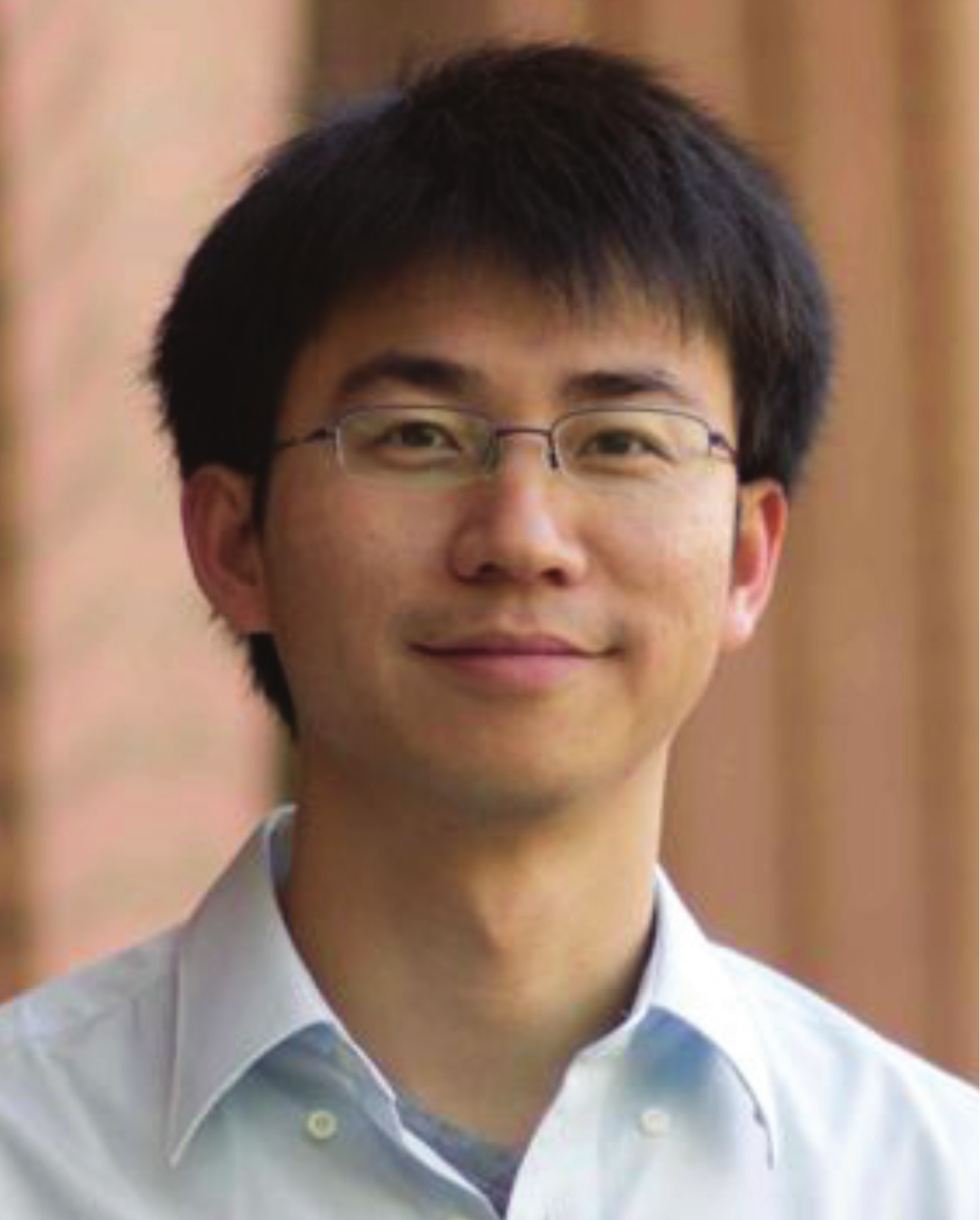}}]
{Longbo Huang} received his Ph.D. in Electrical Engineering from the University of Southern California (USC) in 2011. He then worked as a postdoctoral researcher in the Electrical Engineering and Computer Sciences department (EECS) at University of California at Berkeley (UC Berkeley) from  2011 to 2012. Since 2012, Dr. Huang has been an assistant professor at the Institute for Interdisciplinary Information Sciences (IIIS) at Tsinghua University, Beijing, China. Dr. Huang was a visiting scholar at the LIDS lab at MIT and at the EECS department at UC Berkeley. He was also a visiting professor at the Chinese University of Hong Kong (CUHK) and at Bell-labs France. Dr. Huang was selected into China's Youth 1000-talent program in 2013, and received the Google Research Award and the Microsoft Research Asia (MSRA) Collaborative Research Award in 2014. Dr. Huang was selected into the MSRA StarTrack Program in 2015. His paper in ACM MobiHoc 2014 was selected as a Best Paper Finalist. Dr. Huang has served/serves as the TPC Vice Chair for Submissions for WiOpt 2016, and as TPC members for top-tier IEEE and ACM conferences including ACM Sigmetrics/Performance, MobiHoc, INFOCOM, WiOpt, and E-Energy. Dr. Huang's current research interests are in the areas of learning and optimization for networked systems, mobile networks, data center networking, and smart grid.
\end{IEEEbiography}

\begin{IEEEbiography}[{\includegraphics[width=1in,height=1.25in,clip,keepaspectratio]{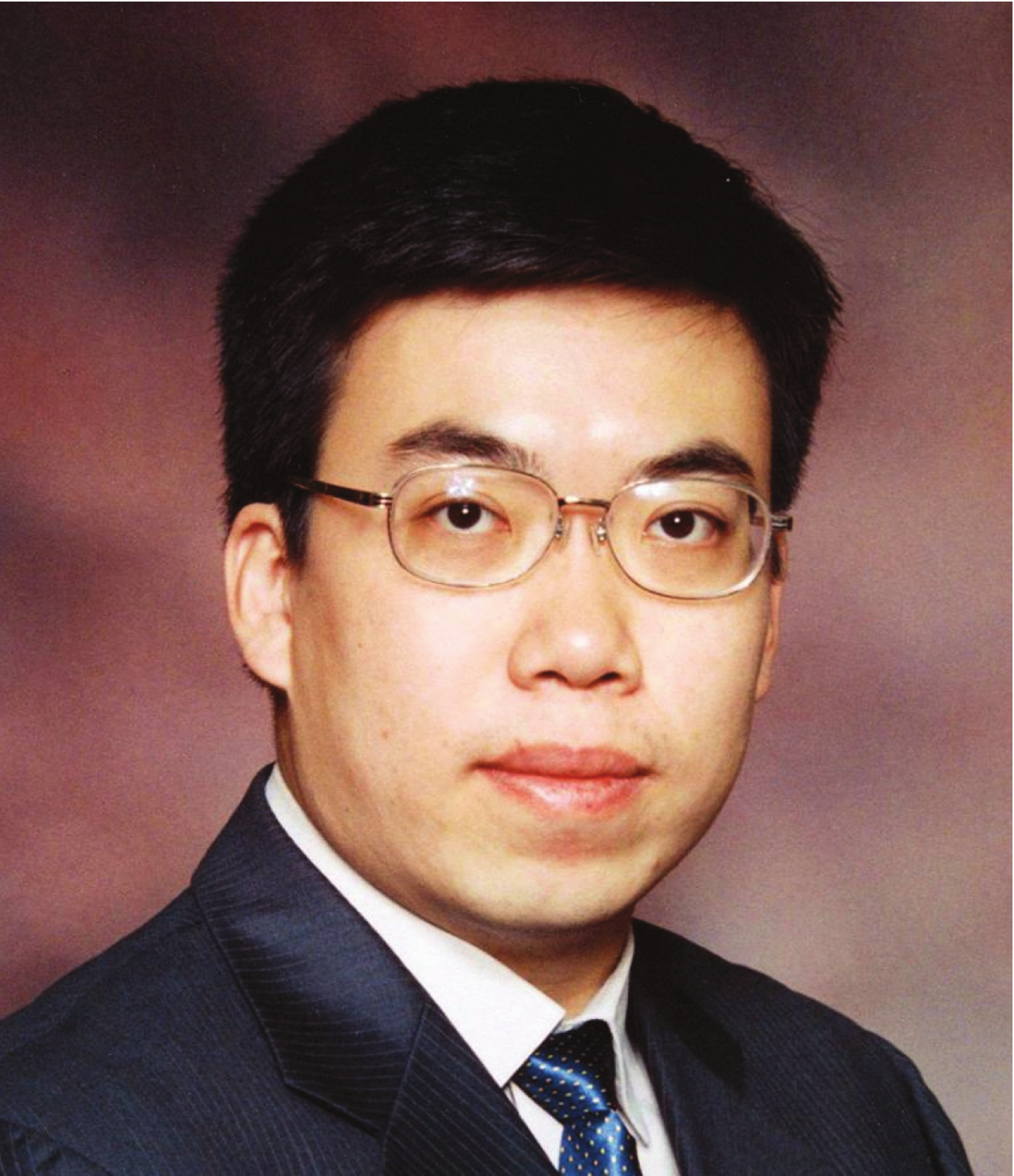}}]
{Jianwei Huang} (S'01-M'06-SM'11-F'16) is an Associate Professor and Director of the Network Communications and Economics Lab (ncel.ie.cuhk.edu.hk), in the Department of Information Engineering at the Chinese University of Hong Kong. He received the Ph.D. degree from Northwestern University in 2005, and worked as a Postdoc Research Associate in Princeton during 2005-2007. He is the co-recipient of 8 international Best Paper Awards, including IEEE Marconi Prize Paper Award in Wireless Communications in 2011. He has co-authored four books: "Wireless Network Pricing," "Monotonic Optimization in Communication and Networking Systems,"  "Cognitive Mobile Virtual Network Operator Games,'' and "Social Cognitive Radio Networks". He has served as an Associate Editor of IEEE Transactions on Cognitive Communications and Networking, IEEE Transactions on Wireless Communications, and IEEE Journal on Selected Areas in Communications - Cognitive Radio Series. He is the Vice Chair of IEEE ComSoc Cognitive Network Technical Committee and the Past Chair of IEEE ComSoc Multimedia Communications Technical Committee. He is a Fellow of IEEE (Class of 2016) and a Distinguished Lecturer of IEEE Communications Society.
\end{IEEEbiography}

\newpage

\appendix
\subsection{Solution to Problem (\ref{equ:resourceallocation})}
We first consider the following relaxed problem:
\begin{align}
\begin{split}
&\max~- V {\kappa \sum\limits_{m=1}^M {\sum\limits_{l\in{\cal L}_0} {{p_{lm}}\left( \tau \right)} }}\\
&+ \frac{B}{M}\sum\limits_{m=1}^{M} {\sum\limits_{l\in{\cal L}_0} {{Q_l}\left( kT \right){x_{lm}}\left( \tau \right)\log \left( {1 + \frac{{{p_{lm}}\left( \tau \right)H_{lm}^2\left( \tau \right)}}{{x_{lm}\left(\tau\right)}{{N_0}\frac{B}{M}}}} \right)} }\\
&{\rm{s.t.}}~~~~~\sum\limits_{m=1}^M {\sum\limits_{l\in{\cal L}_0} {{p_{lm}}\left( \tau \right)} }  \le {P_{\max }^C},\sum\limits_{l\in{\cal L}_0} {{x_{lm}}\left( \tau \right)}  \le 1,\forall m,\\
&{\rm{var.}}~~~~~{x_{lm}}\left( \tau \right) \in \left[ {0,1} \right],{p_{lm}}\left( \tau \right) \ge 0,\forall l\in {\cal L}_0,m\in{\cal M}.
\end{split}\label{equ:relaxedresourceallocation}
\end{align}
The differences between problems (\ref{equ:relaxedresourceallocation}) and (\ref{equ:resourceallocation}) are: (1) the noise term ${{{N_0}{B}/{M}}}$ is replaced by ${{x_{lm}\left(\tau\right)}{{N_0}{B}/{M}}}$; (2) variable ${x_{lm}}\left( \tau \right) \in \left\{ {0,1} \right\}$ is replaced by ${x_{lm}}\left( \tau \right) \in \left[ {0,1} \right]$. The physical meaning of the relaxation is that we allow the subchannels to be time-shared among users. It is easy to prove that, if ${\bm x}^*\left({\tau}\right)$ and ${\bm p}^*\left({\tau}\right)$ are optimal solutions to problem (\ref{equ:relaxedresourceallocation}) and ${x}^*_{lm}\left({\tau}\right)\in \left\{ {0,1} \right\}$ for all $l\in{\cal L}_0,m\in{\cal M}$, ${\bm x}^*\left({\tau}\right)$ and ${\bm p}^*\left({\tau}\right)$ are also optimal solutions to problem (\ref{equ:resourceallocation}).

Based on \cite{huang2009downlink}, problem (\ref{equ:relaxedresourceallocation}) is convex and satisfies Slater's condition. We use $\lambda$ to denote the Lagrange multiplier for constraint $\sum\limits_{m=1}^M {\sum\limits_{l\in{\cal L}_0} {{p_{lm}}\left( \tau \right)} }  \le {P_{\max }^C}$, and $\mu_m$ to denote the Lagrange multiplier for constraint $\sum\limits_{l\in{\cal L}_0} {{x_{lm}}\left( \tau \right)}\le1$. The optimal Lagrange multiplier $\lambda^*$ is solved by the following problem (see \cite{huang2009downlink} for details):
\begin{align}
& {\mathop {\max }\limits_{\lambda  \ge 0}  \lambda {P_{\max }^C} + \sum\limits_{m=1}^M {\mu _m^*\left( \lambda  \right)}},\label{equ:appendix:dualsolution}
\end{align}
where function ${\mu _m^*\left( \lambda  \right)}$ is given as
\begin{align}
\mu _m^*\left( \lambda  \right) = \mathop {\max }\limits_{l\in{\cal L}_0} {\mu _{lm}}\left( \lambda  \right),
\end{align}
and
\begin{align}
\nonumber
& {\mu _{lm}}\left( \lambda  \right) \triangleq -\left( {V \kappa + \lambda } \right)\frac{B}{M}{\left( {\frac{{{Q_l}\left( kT \right)}}{{V \kappa + \lambda }} - \frac{{{N_0}}}{{H_{lm}^2\left( \tau \right)}}} \right)^ + }\\
& +\frac{B}{M}{Q_l}\left( kT \right)\log \left( {1 + \frac{{H_{lm}^2\left( \tau \right)}}{{{N_0}}}{{\left( {\frac{{{Q_l}\left( kT \right)}}{{V \kappa  + \lambda }} - \frac{{{N_0}}}{{H_{lm}^2\left( \tau \right)}}} \right)}^ + }} \right).
\end{align}
Problem (\ref{equ:appendix:dualsolution}) can be optimally solved by using an iterated one dimensional search, \emph{e.g.}, the Golden Section method.

We define ${{\cal A}_m} \triangleq \left\{ {l\in{\cal L}_0:{\mu _{lm}}\left( {{\lambda ^*}} \right) = \mathop {\max }\limits_{l\in{\cal L}_0} {\mu _{lm}}\left( {{\lambda ^*}} \right)} \right\}$ for all $m$ under $\lambda ^*$. We then refer to a subchannel allocation as an extreme point if for all $m\in\cal M$, it satisfies:
\begin{itemize}
\item $x_{lm}\left( \tau \right) = 0$ if $l\notin {\cal A}_m$ and $l\in{\cal L}_0$;
\item $x_{lm}\left( \tau \right) \in \left\{0,1\right\}$ if $l\in {\cal A}_m$;
\item $\sum\limits_{l\in{\cal L}_0} {x_{lm}\left( \tau \right)}=1$.
\end{itemize}
Notice that the extreme point implies that the subchannels are not shared among the users and there is exactly one user per subchannel. We then represent such an extreme point by a function $f:{\cal M}\rightarrow \cal L$, where $f\left(m\right)\in{\cal A}_m$ indicates the user who is allocated to subchannel $m$. If an extreme point optimizes problem (\ref{equ:relaxedresourceallocation}), it will also optimize problem (\ref{equ:resourceallocation}). Next we discuss the following two cases.

\textbf{Case a:} If there is an extreme point $f$ satisfies:
\begin{align}
\sum\limits_{m\in\cal M} {\frac{B}{M}{{\left( {\frac{{{Q_l}\left( kT \right)}}{{V\kappa + {\lambda ^*}}} - \frac{{{N_0}}}{{H_{f\left( m \right)m}^2\left( \tau \right)}}} \right)}^ + }}  = {P_{\max }^C},\label{equ:appendix:condition}
\end{align}
then such an extreme point $f$ is optimal to problem (\ref{equ:relaxedresourceallocation}) \cite{huang2009downlink}. Therefore, it is also optimal to problem (\ref{equ:resourceallocation}). Based on $f$, it is easy to obtain ${\bm x}^*\left( \tau \right)$. Furthermore, we can compute the optimal power allocation ${\bm p}^*\left( \tau \right)$ by
\begin{align}
\nonumber
& {p_{lm}^*}\left( \tau \right) = \\
&{x_{lm}^*}\left( \tau \right)\frac{B}{M}{\left( {\frac{{{Q_l}\left( kT \right)}}{{V\kappa + \lambda^* }} - \frac{{{N_0}}}{{H_{lm}^2\left( \tau \right)}}} \right)^ + }, \forall l\in{\cal L}_0,m\in{\cal M}.
\end{align}
Vectors ${\bm x}^*\left( \tau \right)$ and ${\bm p}^*\left( \tau \right)$ are the optimal solutions to problem (\ref{equ:resourceallocation}).

\textbf{Case b:} If there is no extreme point $f$ satisfies (\ref{equ:appendix:condition}), we need to approximately solve problem (\ref{equ:resourceallocation}) by picking the extreme point in a heuristic manner. According to \cite{huang2009downlink}, we pick the extreme point $f$, for which $\sum\limits_{m\in{\cal M}_0} {\frac{B}{M}{{\left( {\frac{{{Q_l}\left( kT \right)}}{{V\kappa + {\lambda ^*}}} - \frac{{{N_0}}}{{H_{f\left( m \right)m}^2\left( \tau \right)}}} \right)}^ + }}$ is closest to $P_{max}^C$ without exceeding it.

Based on the chosen extreme point $f$, we can quickly determine vector ${\bm x}^*\left( \tau \right)$. Next we re-optimize the power allocation ${\bm p} ^*\left( \tau \right)$ for the given extreme point $f$. Apparently, we have ${p_{lm}^*}\left( \tau \right) = 0$ for $l\ne f\left( m \right)$. For the value of ${p_{f\left(m\right)m}^*}\left( \tau \right)$, we consider the following problem:
\begin{align}
\begin{split}
&\max~\!\frac{B}{M}\!\sum\limits_{m\in\cal M} \!{{Q_{f\left( m \right)}}\!\left( kT \right)\!\log \!\left( {1\! +\! \frac{{{p_{f\left( m \right)m}}\left( \tau \right)H_{f\left( m \right)m}^2\left( \tau \right)}}{{{N_0}\frac{B}{M}}}} \right)}  \\
&{~~~~~~~~}- V\kappa\sum\limits_{m\in\cal M} {{p_{f\left( m \right)m}}\left( \tau \right)} \\
&{\rm{s.t.}}~~\sum\limits_{m\in\cal M} { {{p_{f\left( m \right)m}}\left( \tau \right)} }  \le {P_{\max }^C},\\
&{\rm{var.}}~~~~~~{p_{f\left( m \right)m}}\left( \tau \right) \ge 0, \forall m\in\cal M.\label{equ:appendix:reoptimize}
\end{split}
\end{align}
Problem (\ref{equ:appendix:reoptimize}) is a convex optimization problem, and its optimal solution is described as follows:
\begin{itemize}
\item If $\!\sum\limits_{m\in \cal M}\!\! {\frac{B}{M}\!\left(\! {\frac{{{Q_{f\left( m \right)}}\left( kT \right)}}{V\kappa}\! -\! \frac{{{N_0}}}{{H_{f\left( m \right)m}^2\left( \tau \right)}}} \right)}  \le {P_{\max }^C}$, ${p_{f\left(m\right)m}^*}\left( \tau \right)$, $m\in\cal M$, is:
\begin{align}
{p_{f\left( m \right)m}^*}\left( \tau \right) = \frac{B}{M}{\left( {\frac{{{Q_{f\left( m \right)}}\left( kT \right)}}{{V\kappa}} - \frac{{{N_0}}}{{H_{f\left( m \right)m}^2\left( \tau \right)}}} \right)^ + };
\end{align}
\item Otherwise, ${p_{f\left(m\right)m}^*}\left( \tau \right)$, $m\in\cal M$, is:
\begin{align}
{p_{f\left( m \right)m}^*}\left( \tau \right) = \frac{B}{M}{\left( {\frac{{{Q_{f\left( m \right)}}\left( kT \right)}}{{V\kappa}+\vartheta} - \frac{{{N_0}}}{{H_{f\left( m \right)m}^2\left( \tau \right)}}} \right)^ + },
\end{align}
where $\vartheta$ equals {${\sum\limits_{m\in\cal M} {\frac{B}{M}\left( {\frac{{{Q_{f\left( m \right)}}\left( kT \right)}}{{V \kappa + \vartheta }} - \frac{{{N_0}}}{{H_{f\left( m \right)m}^2\left( \tau \right)}}} \right)} ^ + } = {P_{\max }^C}$}.
\end{itemize}
Vectors ${\bm x}^*\left( \tau \right)$ and ${\bm p}^*\left( \tau \right)$ are the solutions to problem (\ref{equ:resourceallocation}).

Summarizing \textbf{Case a} and \textbf{Case b}, we solve problem (\ref{equ:resourceallocation}).${\rm{~}}\Box$
\subsection{Proof of Lemma \ref{lemma:nonpreDrift}}
Recall the queueing dynamics (\ref{equ:queueing}):
\begin{align}
\nonumber
{Q_l}\left( {t + 1} \right)=& \left[{Q_l}\left( t \right)-{{r_l}\bigl( {{\bm \alpha} \left( {{t_T}} \right),{{\bm x}^l}\left( t \right),{{\bm p}^l}\left( t \right)} \bigr)} \right]^{+} \\
&+ {A_l}\left( t \right),\forall l \in\mathcal{L},t\geq 0.
\vspace{-0.2cm}
\end{align}
We then have:
\begin{align}
\nonumber
& {Q_l}{\left( {t + 1} \right)^2} \le {Q_l}{\left( t \right)^2} + {{r_l}\bigl( {{\bm \alpha} \left( {{t_T}} \right),{{\bm x}^l}\left( t \right),{{\bm p}^l}\left( t \right)} \bigr)^2} + {A_l}{\left( t \right)^2}\\
\nonumber
& - 2{Q_l}\left( t \right){{r_l}\bigl( {{\bm \alpha} \left( {{t_T}} \right),{{\bm x}^l}\left( t \right),{{\bm p}^l}\left( t \right)} \bigr)} + 2{Q_l}\left( t \right){A_l}\left( t \right)\\
\nonumber
& \le {Q_l}{\left( t \right)^2}- 2{Q_l}\left( t \right){{r_l}\bigl( {{\bm \alpha} \left( {{t_T}} \right),{{\bm x}^l}\left( t \right),{{\bm p}^l}\left( t \right)} \bigr)} \\
& + 2{Q_l}\left( t \right){A_l}\left( t \right)+A_{\max}^2+r_{\max}^2.
\end{align}
Therefore, we compute the upper bound of ${\Delta_T}\left( kT \right)$ as (\ref{equ:appendix:driftupperbound}).
\begin{figure*}
\begin{align}
\nonumber
& {\Delta_T}\left( kT \right)= {\mathbb {E}}\left\{ {\frac{1}{2}\sum\limits_{l = 1}^L {{Q_l}{{\left( kT+T \right)}^2}} - \frac{1}{2}\sum\limits_{l = 1}^L {{Q_l}{{\left( kT \right)}^2}} \left| {{\bm Q}\left( kT \right)} \right.} \right\}\\
\nonumber
& = {\mathbb {E}}\left\{\sum\limits_{\tau=kT}^{kT+T-1} {\left({\frac{1}{2}\sum\limits_{l = 1}^L {{Q_l}{{\left( \tau+1 \right)}^2}} - \frac{1}{2}\sum\limits_{l = 1}^L {{Q_l}{{\left( \tau \right)}^2}}}\right)} \left| {{\bm Q}\left( kT \right)} \right. \right\}\\
& \le {\mathbb {E}} \left\{ \sum\limits_{\tau=kT}^{kT +T -1} {\left(B_1+{\sum\limits_{l=1}^L {\left( - {Q_l} \left( \tau \right){{r_l}\bigl( {{\bm \alpha}  \left( {{kT}} \right),{{\bm x}^l} \left( \tau \right),{{\bm p}^l} \left( \tau \right)} \bigr)} +  {Q_l} \left( \tau \right) {A_l}\left( \tau \right) \right)}} \right)}  \left| {{\bm Q}\left( kT \right)} \right. \right\}.\label{equ:appendix:driftupperbound}
\end{align}
\hrulefill
\end{figure*}
Adding $ V {\mathbb E}\left\{\sum\limits_{\tau=kT}^{kT+T-1} {  {P\left({\bm \alpha}\left(kT\right),{\bm p}\left(\tau\right)\right) \left| {{\bm Q}\left( kT \right)} \right.}  }\right\}$ to both sides of (\ref{equ:appendix:driftupperbound}), we prove Lemma \ref{lemma:nonpreDrift}.${\rm{~~~~~~~~~~~~~~~~~~~~~~~~~~~~~~}}\Box$
\subsection{Proof of Lemma \ref{lemma:nonpreDrift:b}}
In Lemma \ref{lemma:nonpreDrift}, we have shown (\ref{equ:appendix:lemma2}).
\begin{figure*}
\begin{align}
\nonumber
& D_T\left(kT\right) \le {B_1}T + V {\mathbb E}\left\{\sum\limits_{\tau=kT}^{kT+T-1} {  {P\left({\bm \alpha}\left(kT\right),{\bm p}\left(\tau\right)\right) \left| {{\bm Q}\left( kT \right)} \right.}  }\right\}\\
&+ {\mathbb E} \left\{ {{\sum\limits_{l = 1}^L{\sum\limits_{\tau=kT}^{kT+T-1} { {{Q_l}\left( \tau \right)\left( {{A_l}\left( \tau \right) - {{r_l\left( {{\bm \alpha} \left( {kT} \right),{{\bm x}^l}\left( \tau \right),{\bm p}^l\left( \tau \right)} \right)}}} \right)}} \left| {{\bm Q}\left( kT \right)} \right.}}}\right\}.\label{equ:appendix:lemma2}
\end{align}
\hrulefill
\end{figure*}
According to the queueing dynamics (\ref{equ:queueing}), we have the following relation for all $l\in{\cal L}, \tau\in{\cal T}_k$:
\begin{align}
{Q_l}\left( {kT} \right)\! -\! \left( {\tau\! -\! kT} \right){r_{\max }} \le {Q_l}\left( \tau  \right) \le {Q_l}\left( {kT} \right)\! +\! \left( {\tau\!-\!kT} \right){A_{\max }}.
\end{align}
Therefore, we obtain
\begin{align}
\nonumber
&{Q_l}\left( \tau \right) {{A_l}\left( \tau \right)}\le {Q_l}\left( {kT} \right) {{A_l}\left( \tau \right)}+\left( {\tau-kT} \right){A_{\max }} {{A_l}\left( \tau \right)}\\
&\le {Q_l}\left( {kT} \right) {{A_l}\left( \tau \right)}+\left( {\tau-kT} \right){A_{\max }^2}.\label{equ:appendix:lemma2:a}
\end{align}
and
\begin{align}
\nonumber
&-{Q_l}\left( \tau \right) {{r_l\left( {{\bm \alpha} \left( {kT} \right),{{\bm x}^l}\left( \tau \right),{\bm p}^l\left( \tau \right)} \right)}}\\
\nonumber
& \le -{Q_l}\left( {kT} \right) {{{r_l\left( {{\bm \alpha} \left( {kT} \right),{{\bm x}^l}\left( \tau \right),{\bm p}^l\left( \tau \right)} \right)}}}\\
\nonumber
&+\left( {\tau-kT} \right){r_{\max }} {{{r_l\left( {{\bm \alpha} \left( {kT} \right),{{\bm x}^l}\left( \tau \right),{\bm p}^l\left( \tau \right)} \right)}}}\\
& \le -{Q_l}\left( {kT} \right) {{{r_l\left( {{\bm \alpha} \left( {kT} \right),{{\bm x}^l}\left( \tau \right),{\bm p}^l\left( \tau \right)} \right)}}}+\left( {\tau-kT} \right){r_{\max }^2}.\label{equ:appendix:lemma2:b}
\end{align}
Based on (\ref{equ:appendix:lemma2}), (\ref{equ:appendix:lemma2:a}), and (\ref{equ:appendix:lemma2:b}), we obtain (\ref{equ:awake:A}).
\begin{figure*}
\begin{align}
\nonumber
& D_T\left(kT\right) \le T\left( {T - 1} \right){B_1} + T{B_1}+ V {\mathbb E}\left\{\sum\limits_{\tau=kT}^{kT+T-1} {  {P\left({\bm \alpha}\left(kT\right),{\bm p}\left(\tau\right)\right) \left| {{\bm Q}\left( kT \right)} \right.}  }\right\}\\
& + {\mathbb E} \left\{ {{\sum\limits_{l = 1}^L{\sum\limits_{\tau=kT}^{kT+T-1} { {{Q_l}\left( kT \right)\left( {{A_l}\left( \tau \right) - {{r_l\left( {{\bm \alpha} \left( {kT} \right),{{\bm x}^l}\left( \tau \right),{\bm p}^l\left( \tau \right)} \right)}}} \right)}} \left| {{\bm Q}\left( kT \right)} \right.}}}\right\}.\label{equ:awake:A}
\end{align}
\hrulefill
\end{figure*}
After arrangement, we prove Lemma \ref{lemma:nonpreDrift:b}.${\rm{~~~~~~~~~~~~~~~~~~~~~~~~~~~~~~~~~~~~~~~}}\Box$
\subsection{Proof of Theorem \ref{theorem:ENSRA}}
Recall the assumption we made in (\ref{equ:eta}), \emph{i.e.}, there exists an $\eta>0$ such that
\begin{equation}
{\mathbb{E}}\left\{{{\bm A}\left(t\right)}\right\}+\eta \cdot {\bm 1} \in\Lambda.
\end{equation}
According to \cite{neely2010stochastic}, for any $\varepsilon  \in \left[ {0,\eta } \right]$, there exists a stationary randomized algorithm that determines network selection and resource allocation independent of queue backlog and yields for frame ${\cal T}_k$,
\begin{align}
\nonumber
&\!\sum\limits_{\tau=kT}^{kT+T-1}\!\!\!\left({\mathbb E}\left\{{A_l\left(\tau\right)}\right\}\!-{\mathbb E}\left\{ {{r_l\!\left( {{\bm \alpha} \left( {kT} \right),{{\bm x}^l}\!\left( \tau \right)\!,{\bm p}^l\!\left( \tau \right)} \right)}\!\left| {{\bm Q}\left( kT \right)} \!\right.} \right\}\right)\!\\
& \le-\varepsilon T,\forall l \in \cal{L},\label{equ:appendix:theorem:b}\\
& {\mathbb E}\left\{\sum\limits_{\tau=kT}^{kT+T-1} {  {P\left({\bm \alpha}\left(kT\right),{\bm p}\left(\tau\right)\right) \left| {{\bm Q}\left( kT \right)} \right.}  }\right\}=P\left(\varepsilon\right)T.\label{equ:appendix:theorem:c}
\end{align}
where $P\left(\varepsilon\right)$ is defined in Section \ref{subsec:P-ENSRA}.

Next we study $D_T\left(kT\right)$ term under \textsf{ENSRA}. Based on Lemma \ref{lemma:nonpreDrift:b}, we have (\ref{equ:appendix:theorem:a}).
\begin{figure*}
\begin{align}
\nonumber
& D_T\left(kT\right) \le {B_2}T + V {\mathbb E}\left\{\sum\limits_{\tau=kT}^{kT+T-1} {  {P\left({\bm \alpha}\left(kT\right),{\bm p}\left(\tau\right)\right) \left| {{\bm Q}\left( kT \right)} \right.}  }\right\}\\
&+ {\mathbb E} \left\{ {{\sum\limits_{l = 1}^L{{Q_l}\left( kT \right) \sum\limits_{\tau=kT}^{kT+T-1} { {\left( {{A_l}\left( \tau \right) - {{r_l\left( {{\bm \alpha} \left( {kT} \right),{{\bm x}^l}\left( \tau \right),{\bm p}^l\left( \tau \right)} \right)}}} \right)}} \left| {{\bm Q}\left( kT \right)} \right.}}}\right\}.\label{equ:appendix:theorem:a}
\end{align}
\hrulefill
\end{figure*}
Since \textsf{ENSRA} minimizes the right hand side of (\ref{equ:appendix:theorem:a}), we compare its value under \textsf{ENSRA} with that under a stationary randomized algorithm. By utilizing (\ref{equ:appendix:theorem:b}) and (\ref{equ:appendix:theorem:c}), we obtain:
\begin{align}
D_T\left(kT\right) \le {B_2}T + V T P\left(\varepsilon\right)- T \varepsilon{\mathbb E} \left\{ {{\sum\limits_{l = 1}^L{{Q_l}\left( kT \right) \left| {{\bm Q}\left( kT \right)} \right.}}}\right\}.\label{equ:appendix:theorem:d}
\end{align}

Taking the expectation of (\ref{equ:appendix:theorem:d}) with respect to the distribution of ${\bm Q}\left( kT \right)$ and using the law of iterated expectation yields (\ref{equ:awake:B}).
\begin{figure*}
\begin{align}
\nonumber
& \mathbb{E}\left\{\frac{1}{2}\sum\limits_{l = 1}^L {\left( {{Q_l}{{\left( {kT + T} \right)}^2} - {Q_l}{{\left( kT \right)}^2}} \right)}+V\sum\limits_{\tau=kT}^{kT+T-1}{P\left({\bm \alpha}\left(kT\right),{\bm p}\left(\tau\right)\right)}\right\}\\
& \le B_2 T-\varepsilon T{\sum\limits_{l = 1}^L{\mathbb E}\left\{ {{Q_l}\left( kT \right)}\right\}}+V T P\left(\varepsilon\right).\label{equ:awake:B}
\end{align}
\hrulefill
\end{figure*}
Summing over time slots $k\in\left\{0,1,\ldots,K-1\right\}$ and dividing by $KT$ yields (\ref{equ:appendix:theorem:e}).
\begin{figure*}
\begin{align}
\nonumber
& \frac{1}{{2KT}}\sum\limits_{l = 1}^L {\mathbb E}{\left\{ { {{Q_l}{{\left( KT \right)}^2} - {Q_l}{{\left( 0 \right)}^2}} } \right\}} \!\! +\!\! \frac{V}{KT}\sum\limits_{k = 0}^{K - 1}\sum\limits_{\tau = kT}^{kT+T - 1} {\mathbb E}{\left\{ {P\left({\bm \alpha}\left(kT\right),{\bm p}\left(\tau\right)\right)} \right\}} \\
& \le B_2 - \frac{\varepsilon}{K} \sum\limits_{k=0}^{K-1}{\sum\limits_{l = 1}^L{\mathbb E}\left\{ {{Q_l}\left( kT \right)}\right\}}+V P\left(\varepsilon\right).\label{equ:appendix:theorem:e}
\end{align}
\hrulefill
\end{figure*}

{\bf{Part A: Proof of (\ref{equ:pENSRA})}}

Because ${\bm Q}\left(0\right)={\bm 0}$ and $Q_l\left(kT\right)\ge0$ for all $l$ and $k$, we have:
\begin{align}
\frac{V}{KT}\sum\limits_{k = 0}^{K - 1}\sum\limits_{\tau = kT}^{kT+T - 1} {\mathbb E}{\left\{ {P\left({\bm \alpha}\left(kT\right),{\bm p}\left(\tau\right)\right)} \right\}}\le B_2 +V P\left(\varepsilon\right).
\end{align}
After arrangement, we have
\begin{align}
\frac{1}{KT}\sum\limits_{k = 0}^{K - 1}\sum\limits_{\tau = kT}^{kT+T - 1} {\mathbb E}{\left\{ {P\left({\bm \alpha}\left(kT\right),{\bm p}\left(\tau\right)\right)} \right\}}\le \frac{B_2}{V} +P\left(\varepsilon\right).
\end{align}
Taking limits of the above inequality as $K\to \infty$ yields
\begin{align}
P_{av}^{\textsf{ENSRA}}\le \frac{B_2}{V} +P\left(\varepsilon\right).
\end{align}
The above inequality holds for all $\varepsilon\in\left[0,\eta\right]$, taking $\varepsilon=0$ yields (\ref{equ:pENSRA}).

{\bf{Part B: Proof of (\ref{equ:QENSRA})}}

Because ${\bm Q}\left(0\right)={\bm 0}$ and $Q_l\left(KT\right)\ge0$ for all $l$, we have the following relation from (\ref{equ:appendix:theorem:e}):
\begin{align}
\frac{\varepsilon}{K} \sum\limits_{k=0}^{K-1}{\sum\limits_{l = 1}^L{\mathbb E}\left\{ {{Q_l}\left( kT \right)}\right\}}\le B_2+V P\left(\varepsilon\right).
\end{align}
Dividing both sides by $\varepsilon$ and taking limits of the above inequality as $K\to \infty$ yields
\begin{align}
Q_{av,T}^{\textsf{ENSRA}}\le \frac{{B_2}+V P\left(\varepsilon\right)}{{\varepsilon}}\le\frac{{B_2}+V P_{\max}}{{\varepsilon}}.
\end{align}
The above inequality holds for all $\varepsilon\in\left[0,\eta\right]$, taking $\varepsilon=\eta$ yields (\ref{equ:QENSRA}).

Summarizing \textbf{Part A} and \textbf{Part B}, we complete the proof.${\rm{~}}\Box$
\subsection{ENSRA under Incomplete Channel Condition Information}
\begin{figure*}
\begin{align}
\begin{split}
&\min~V {\mathbb E}\left\{{P\left({\bm \alpha}\left(kT\right),{\bm p}^*\left({{\bm \alpha}\left(kT\right),\tau}\right)\right)}\right\}-\\
&{~~~~~~~}\sum\limits_{l = 1}^L {{Q_l}\left( {kT} \right) {\mathbb E}\left\{{r_l\left( {{\bm \alpha} \left( {kT} \right),{{\bm x}^{l,*}}\left( {\bm \alpha}\left(kT\right),\tau\right),{\bm p}^{l,*}\left( {\bm \alpha}\left(kT\right),\tau\right)} \right)}\right\} }\\
&{\rm{s.t.}}~~~~~~~~~~{\rm constraint~~ (\ref{equ:feasibility})},\\
&{\rm{variable}}~~~~{\bm \alpha}\left( kT\right).\label{equ:appendix:revise:selection}
\end{split}
\end{align}
\hrulefill
\begin{align}
\begin{split}
&\min~V {P^C\left({\bm p}\left(t\right)\right)}  - \sum\limits_{l\in{\cal L}_0} {{Q_l}\left( {kT} \right) {r_l^C\left( {{{\bm x}^l}\left( t \right),{\bm p}^l\left( t \right)} \right)} } \\
&{\rm{s.t.}}~~~~{\rm constraints~~ (\ref{equ:subchannelconstraint}),(\ref{equ:powerbudeget})},\\
&{\rm{var.}}~~~~{\bm x}^l\left( t\right), {\bm p}^l\left( t\right),\forall l\in{\cal L}_0.\label{equ:appendix:revise:allocation}
\end{split}
\end{align}
\hrulefill
\end{figure*}

\begin{algorithm}[h]\small
\caption{Revised Energy-Aware Network Selection and Resource Allocation (R-ENSRA)}
\begin{algorithmic}[1]\label{algo:appendix:R-ENSRA}
\STATE Set $t=0$ and ${\bm Q}\left(0\right)={\bm 0}$;
\STATE {\bf while} {$t<t_{end}$} {\bf do}
\STATE $ \ \ \ $ {\bf if} $\mod\left(t,T\right)=0$ $//$ \emph{Compute the network selection for the frame if $t$ is the beginning of the frame.}
\STATE $ \ \ \ \ \ \ $ Set $k=\frac{t}{T}$ and solve problem (\ref{equ:appendix:revise:selection}) to determine ${\bm \alpha}\left( kT\right)$;
\STATE $ \ \ \ $ {\bf end if}
\STATE $ \ \ \ $ Solve problem (\ref{equ:appendix:revise:allocation}) to determine ${\bm x}\left(t\right), {\bm p}\left(t\right)$; $//$ \emph{Compute the resource allocation every slot.}
\STATE $ \ \ \ $ Update ${\bm Q}\left(t+1\right)$, according to (\ref{equ:queueing});
\STATE $ \ \ \ $ $t\leftarrow t+1$.
\STATE {\bf end while}
\end{algorithmic}
\end{algorithm}

We propose a revised network selection and resource allocation algorithm in Algorithm \ref{algo:appendix:R-ENSRA} to tackle the incomplete channel condition. At the beginning time slot of each frame, the operator optimizes an expected function in (\ref{equ:appendix:revise:selection}) to determine the network selection. At every time slot, the operator observes the channel condition and solves (\ref{equ:appendix:revise:allocation}) to determine the resource allocation.

Notice that, we have assumed that the channel conditions ${\bm H}\left(t\right)$ are independent and identically distributed over $t$. Therefore, in (\ref{equ:appendix:revise:selection}), we only need to optimize the expectation with respect to the channel condition over one time slot instead of the whole frame. The probability distribution of the channel conditions can be approximated by the historical information \cite{yao2012data}.
\subsection{Proof of Lemma \ref{lemma:pre2}}
Applying the result of Lemma \ref{lemma:nonpreDrift:b} (the upper bound of the ``drift-plus-penalty'' term for a frame), we can easily prove Lemma \ref{lemma:pre2} (the upper bound of the ``drift-plus-penalty'' term for a window).
\subsection{Proof of Lemma \ref{lemma:pre3}}
According to the objective function in \textsf{P-ENSRA}, \textsf{P-ENSRA} minimizes (\ref{equ:appendix:lemma4:a}) for each window.
\begin{figure*}
\begin{align}
\nonumber
& {\mathbb E}\!\left\{\!{{\sum\limits_{l = 1}^L \!\sum\limits_{w=0}^{W-1}{\!{Q_l}\!\left( {h}WT\!+\!wT \right) \!\!\!\!\!\!\sum\limits_{\tau=\left({h}W+w\right)T}^{\left({h}W+w+1\right)T-1}\!\!\!\! { {\Bigl( {\!{A_l}\left( \tau \right)\!+\theta-\!{{r_l\bigl( {{{\bm \alpha}}\!\left( {{h}WT+wT} \right),{{\bm x}^{l}}\!\left( \tau \right),{\bm p}^{l}\!\left( \tau \right)} \bigr)}}} \Bigr)\!}} \left| {{\bm Q}\left( {h}WT \right)} \right.}}}\!\!\right\}\\
& + V  {\mathbb E}\left\{ { {{\sum\limits_{w = 0}^{W-1}{\sum\limits_{\tau = \left({h}W+w\right)T}^{\left({h}W+w+1\right)T - 1} {P\bigl({\boldsymbol \alpha}\left({h}WT+wT\right),{\boldsymbol p}\left(\tau\right)\bigr)}}} \left| {{\bm Q}\left( {h}WT \right)} \right.}  }\right\}.\label{equ:appendix:lemma4:a}
\end{align}
\hrulefill
\end{figure*}
Hence, the value of (\ref{equ:appendix:lemma4:a}) under \textsf{P-ENSRA} is not greater than that under any randomized algorithm. Recall (\ref{equ:appendix:theorem:b}) and (\ref{equ:appendix:theorem:b}), we consider the randomized algorithm that satisfies (\ref{longequ:sleep:A}) and (\ref{longequ:sleep:B}).
\begin{figure*}
\begin{align}
&\sum\limits_{\tau=\left(hW+w\right)T}^{\left(hW+w+1\right)T-1}\left({\mathbb E}\left\{{A_l\left(\tau\right)}\right\}-{\mathbb E}\left\{ {{r_l\left( {{\bm \alpha} \left( {hWT+wT} \right),{{\bm x}^l}\left( \tau \right),{\bm p}^l\left( \tau \right)} \right)}\left| {{\bm Q}\left( hWT \right)} \right.} \right\}\right)\le-\theta T,\forall l \in \cal{L},\label{longequ:sleep:A} \\
& {\mathbb E}\left\{\sum\limits_{\tau=\left(hW+w\right)T}^{\left(hW+w+1\right)T-1} {  {P\left({\bm \alpha}\left(hWT+wT\right),{\bm p}\left(\tau\right)\right) \left| {{\bm Q}\left( hWT \right)} \right.}  }\right\}=P\left(\theta\right)T.\label{longequ:sleep:B}
\end{align}
\hrulefill
\end{figure*}
According to these two inequalities, the value of (\ref{equ:appendix:lemma4:a}) under such a randomized algorithm is not greater than $VTWP\left(\theta\right)$. Therefore, the value of (\ref{equ:appendix:lemma4:a}) under \textsf{P-ENSRA} is also not greater than $VTWP\left(\theta\right)$. In other words, we have (\ref{longequ:sleep:C}) for \textsf{P-ENSRA}.
\begin{figure*}
\begin{align}
\nonumber
& {\mathbb E}\!\left\{\!{{\sum\limits_{l = 1}^L \!\sum\limits_{w=0}^{W-1}{\!{Q_l}\!\left( {h}WT\!+\!wT \right) \!\!\!\!\!\!\sum\limits_{\tau=\left({h}W+w\right)T}^{\left({h}W+w+1\right)T-1}\!\!\!\! { {\Bigl( {\!{A_l}\left( \tau \right)\!+\theta-\!{{r_l\bigl( {{{\bm \alpha}^*}\!\left( {{h}WT+wT} \right),{{\bm x}^{l,*}}\!\left( \tau \right),{\bm p}^{l,*}\!\left( \tau \right)} \bigr)}}} \Bigr)\!}} \left| {{\bm Q}\left( {h}WT \right)} \right.}}}\!\!\right\}\\
& + V  {\mathbb E}\left\{ { {{\sum\limits_{w = 0}^{W-1}{\sum\limits_{\tau = \left({h}W+w\right)T}^{\left({h}W+w+1\right)T - 1} {P\bigl({\boldsymbol \alpha}^*\left({h}WT+wT\right),{\boldsymbol p}^{*}\left(\tau\right)\bigr)}}} \left| {{\bm Q}\left( {h}WT \right)} \right.}  }\right\}\le VTWP\left(\theta\right).\label{longequ:sleep:C}
\end{align}
\hrulefill
\end{figure*}
Subtracting $\theta T {\mathbb{E}}\left\{ {\sum\limits_{l = 1}^L {\sum\limits_{w = 0}^{W-1} { {{Q_l}\left( {h}WT+wT  \right)} }} \left| {{\bm Q}\left( {{h}WT} \right)} \right.} \right\}$ from both sides, we prove Lemma \ref{lemma:pre3}.${\rm{~~~~~~~~~~~~~~~~~~~~~~~~~~~~~~~~~}}\Box$
\subsection{Proof of Theorem \ref{theorem:PENSRA}}
According to Lemma \ref{lemma:pre2} and Lemma \ref{lemma:pre3}, \textsf{P-ENSRA} guarantees that
\begin{align}
\nonumber
& D_{WT}\left( {h}WT \right) \le {B_2}WT \\
& -\theta T {\mathbb{E}}\left\{ {\sum\limits_{l = 1}^L {\sum\limits_{w = 0}^{W-1} { {{Q_l}\left( {h}WT+wT  \right)} }} \left| {{\bm Q}\left( {{h}WT} \right)} \right.} \right\} + VWTP\left( \theta  \right).\label{equ:appendix:theorem2:a}
\end{align}
Taking the expectation of (\ref{equ:appendix:theorem2:a}) with respect to the distribution of ${\bm Q}\left( hWT \right)$ and using the law of iterated expectation yields
\begin{align}
\nonumber
& {\mathbb {E}}\left\{ {\frac{1}{2}\sum\limits_{l = 1}^L {{Q_l}{{\left( hWT+WT \right)}^2}} - \frac{1}{2}\sum\limits_{l = 1}^L {{Q_l}{{\left( hWT \right)}^2}}} \right\}+ \\
\nonumber
&V  {\mathbb E}\left\{ { {{\sum\limits_{w = 0}^{W-1}{\sum\limits_{\tau = \left({h}W+w\right)T}^{\left({h}W+w+1\right)T - 1} {P\bigl({\boldsymbol \alpha}\left({h}WT+wT\right),{\boldsymbol p}\left(\tau\right)\bigr)}}}}  }\right\}\\
& \le {B_2}WT + VWTP\left( \theta  \right) -\theta T {\mathbb{E}}\left\{ {\sum\limits_{l = 1}^L {\sum\limits_{w = 0}^{W-1} { {{Q_l}\left( {h}WT+wT  \right)} }}} \right\}.
\end{align}
Summing over $h\in\left\{0,1,\ldots,H-1\right\}$ and dividing by $HWT$ yields
\begin{align}
\nonumber
& \frac{1}{HWT}{\mathbb {E}}\left\{ {\frac{1}{2}\sum\limits_{l = 1}^L {{Q_l}{{\left( HWT \right)}^2}} - \frac{1}{2}\sum\limits_{l = 1}^L {{Q_l}{{\left( 0 \right)}^2}}} \right\}+ \\
\nonumber
&\frac{1}{HWT} V  {\mathbb E}\left\{ { {{\sum
\limits_{h=0}^{H-1}\!\sum\limits_{w = 0}^{W-1}\!{\sum\limits_{\tau = \left({h}W+w\right)T}^{\left({h}W+w+1\right)T - 1} {\!\!\!\!P\bigl({\boldsymbol \alpha}\left({h}WT+wT\right),{\boldsymbol p}\left(\tau\right)\bigr)}}}}  }\!\right\}\\
& \le {B_2} + VP\left( \theta  \right) -\frac{\theta}{HW} {\mathbb{E}}\left\{ {\sum\limits_{h=0}^{H-1}\sum\limits_{w = 0}^{W-1}{\sum\limits_{l = 1}^{L} { {{Q_l}\left( {h}WT+wT  \right)} }}} \right\}.\label{equ:appendix:theorem2:b}
\end{align}

{\bf{Part A: Proof of (\ref{equ:pENSRA:pre})}}

Because ${\bm Q}\left(0\right)={\bm 0}$ and $Q_l\left(hWT+wT\right)\ge0$ for all $l$, $h$, and $w$, we have (\ref{equ:awake:C}).
\begin{figure*}
\begin{align}
\frac{1}{HWT} V  {\mathbb E}\left\{ { {{\sum
\limits_{h=0}^{H-1}\sum\limits_{w = 0}^{W-1}{\sum\limits_{\tau = \left({h}W+w\right)T}^{\left({h}W+w+1\right)T - 1} {P\bigl({\boldsymbol \alpha}\left({h}WT+wT\right),{\boldsymbol p}\left(\tau\right)\bigr)}}}}  }\right\} \le {B_2} + VP\left( \theta  \right).\label{equ:awake:C}
\end{align}
\hrulefill
\end{figure*}
After arrangement, we have (\ref{equ:awake:D}).
\begin{figure*}
\begin{align}
\frac{1}{HWT} {\mathbb E}\left\{ { {{\sum
\limits_{h=0}^{H-1}\sum\limits_{w = 0}^{W-1}{\sum\limits_{\tau = \left({h}W+w\right)T}^{\left({h}W+w+1\right)T - 1} {P\bigl({\boldsymbol \alpha}\left({h}WT+wT\right),{\boldsymbol p}\left(\tau\right)\bigr)}}}}  }\right\} \le \frac{B_2}{V} + P\left( \theta  \right).\label{equ:awake:D}
\end{align}
\hrulefill
\end{figure*}
Taking limits of the above inequality as $H\to \infty$ yields (\ref{equ:pENSRA:pre}):
\begin{align}
P_{av}^{\textsf{P-ENSRA}}\le \frac{B_2}{V} +P\left(\theta\right).
\end{align}

{\bf{Part B: Proof of (\ref{equ:QENSRA:pre})}}

Because ${\bm Q}\left(0\right)={\bm 0}$ and $Q_l\left(HWT\right)\ge0$ for all $l$, we have the following relation from (\ref{equ:appendix:theorem2:b}):
\begin{align}
\frac{\theta}{HW} {\mathbb{E}}\left\{ {\sum\limits_{h=0}^{H-1}\sum\limits_{w = 0}^{W-1}{\sum\limits_{l = 1}^{L} { {{Q_l}\left( {h}WT+wT  \right)} }}} \right\} \le {B_2} + VP\left( \theta  \right).
\end{align}
Dividing both sides by $\theta$ and taking limits of the above inequality as $H\to \infty$ yields (\ref{equ:QENSRA:pre}):
\begin{align}
Q_{av,T}^{\textsf{P-ENSRA}}\le \frac{{B_2}+V P\left(\theta\right)}{{\theta}}.
\end{align}

Summarizing \textbf{Part A} and \textbf{Part B}, we complete the proof.${\rm{}}\Box$
\subsection{Iteration Ending Condition in Algorithm \ref{algo:GP-ENSRA}}
Here we prove line \ref{line:while} in Algorithm \ref{algo:GP-ENSRA} is guaranteed to be achievable. We first show the value of the objective function in problem (\ref{equ:P-ENSRA}) is bounded. Recall problem (\ref{equ:P-ENSRA}).
According to (\ref{equ:arrivalbound}), (\ref{equ:ratebound}), and (\ref{equ:boundedtotalpower}), ${P\bigl({\boldsymbol \alpha}\left({h}WT+wT\right),{\boldsymbol p}\left(\tau\right)\bigr)}$, ${r_l\bigl({\boldsymbol \alpha} \left( {{h}WT+wT} \right), {{{\boldsymbol x}^l}\left( \tau \right),{\boldsymbol p}^l\left( \tau \right)} \bigr)}$, and $A_l\left(\tau \right)$ are both upper and lower bounded. Furthermore, we have
\begin{align}
0\le{{Q_l}\left( {{h}WT+wT} \right)}\le \left(hWT+wT\right)A_{\max}.
\end{align}
Therefore, it is easy to show the value of the objective function in problem (\ref{equ:P-ENSRA}) is also both upper and lower bounded. We denote the bounds as $F_{\max}$ and $F_{\min}$. Since in Algorithm \ref{algo:GP-ENSRA}, we use $F^i$ to denote the objective function's value in the $i$-th iteration, we have $F_{\min}\le F^i \le F_{\max}$ for all $i$.

Based on the updating rule (line \ref{line:A} and line \ref{line:C}), we find $F^{i}$ is always non-increasing in $i$. Hence, if there does not exist a finite $i$ such that ${F^{i-1}} - {F^{i}} \le \epsilon$, we will have $F^{i-1}-F^{i}>\epsilon$ for $i=3,4,\ldots,3+M$, where $M=\lfloor \frac{F_{\max}-F_{\min}}{\epsilon} \rfloor$. However, this generates $F^{2}-F^{3+M}>\left(M+1\right) \epsilon>F_{\max}-F_{\min}$, which contradicts with the fact that $F_{\min}\le F^i \le F_{\max}$ for all $i$. Therefore, there exists a finite $i$ such that ${F^{i-1}} - {F^{i}} \le \epsilon$. In other words, line \ref{line:while} in Algorithm \ref{algo:GP-ENSRA} is guaranteed to be achievable, and we complete the proof.${\rm{~~~~~~~~~~~~~~~~~~~~~~~~}}\Box$
\subsection{GP-ENSRA under Heavy Traffic Region}
We study the problem in line \ref{line:A} of Algorithm \ref{algo:GP-ENSRA} under the following condition:
\begin{align}
{{Q_l}\left( {{h}WT} \right)}\ge WT r_{\max},\forall l\in{\cal L}.
\end{align}

Recall that in Algorithm \ref{algo:GP-ENSRA}, we use ${{\bm \beta}\left({h}WT\!+\!wT\right)}\!=\!\bigl({{\bm \alpha}\left({h}WT\!+\!wT\right)},{{\bm x}\left(\tau\right)},{{\bm p }\left(\tau\right)},\tau\!\in{\cal T}_{{h}W\!+\!w}\bigr)$ to represent the operator's operations over frame ${\cal T}_{{h}W+w}$, $w=0,1,\ldots,W-1$. Based on (\ref{equ:P-ENSRA}), the problem in line \ref{line:A} of Algorithm \ref{algo:GP-ENSRA} is formulated as (\ref{equ:appendix:heavy:a}).
\begin{figure*}
\begin{align}
\begin{split}
&\min~V\sum\limits_{w' = 0}^{W-1}{\sum\limits_{\tau = \left({h}W+w'\right)T}^{\left({h}W+w'+1\right)T - 1} {P\bigl({\boldsymbol \alpha}\left({h}WT+w'T\right),{\boldsymbol p}\left(\tau\right)\bigr)}} \\
&{~~~~~~} - \sum\limits_{l = 1}^L {\sum\limits_{w' = 0}^{W-1} {{{Q_l}\left( {{h}WT+w'T} \right)}\sum\limits_{\tau = \left({h}W+w'\right)T}^{\left({h}W+w'+1\right)T- 1}{r_l\bigl({\boldsymbol \alpha} \left( {{h}WT+w'T} \right), {{{\boldsymbol x}^l}\left( \tau \right),{\boldsymbol p}^l\left( \tau \right)} \bigr)} }} \\
&{~~~~~~}+ \sum\limits_{l = 1}^L {\sum\limits_{w' = 0}^{W-1} {{{Q_l}\left( {{h}WT+w'T} \right)}\sum\limits_{\tau = \left({h}W+w'\right)T}^{\left({h}W+w'+1\right)T- 1} {\left(A_l\left(\tau \right)+\theta\right)} }}\\
&{\rm{s.t.}}~~~~~{\rm constraints~~ (\ref{equ:feasibility}),(\ref{equ:subchannelconstraint}),(\ref{equ:constraintforresource:a}),(\ref{equ:powerbudeget}),(\ref{equ:constraintforresource:b})},\\
&{\rm{var.}}~~~~{\boldsymbol \alpha}\left( {h}WT+wT\right),{\boldsymbol x}\left( \tau\right), {\boldsymbol p}\left( \tau\right), \tau\in{\cal T}_{hW+w}.
\end{split}\label{equ:appendix:heavy:a}
\end{align}
\hrulefill
\end{figure*}
The difference between problems (\ref{equ:appendix:heavy:a}) and (\ref{equ:P-ENSRA}) is that, problem (\ref{equ:appendix:heavy:a}) only considers the optimization over the operations of one frame, instead of the whole window. In other words, in problem (\ref{equ:appendix:heavy:a}), operations ${\boldsymbol \alpha}\left( {h}WT+w'T\right)$ and ${\boldsymbol x}\left( \tau\right), {\boldsymbol p}\left( \tau\right), \tau\in{\cal T}_{hW+w'}$ with $w'\ne w,w'=0,1,\ldots,W-1$ are treated as given constants.

Removing the terms that are independent of ${\boldsymbol \alpha}\left( {h}WT+wT\right),{\boldsymbol x}\left( \tau\right), {\boldsymbol p}\left( \tau\right), \tau\in{\cal T}_{hW+w}$, we can rewrite problem (\ref{equ:appendix:heavy:a}) as (\ref{equ:appendix:heavy:c}).
\begin{figure*}
\begin{align}
\begin{split}
&\min~V{\sum\limits_{\tau = \left({h}W+w\right)T}^{\left({h}W+w+1\right)T - 1} {P\bigl({\boldsymbol \alpha}\left({h}WT+wT\right),{\boldsymbol p}\left(\tau\right)\bigr)}} \\
&{~~~~~~} - \sum\limits_{l = 1}^L { {{{Q_l}\left( {{h}WT+wT} \right)}\sum\limits_{\tau = \left({h}W+w\right)T}^{\left({h}W+w+1\right)T- 1}{r_l\bigl({\boldsymbol \alpha} \left( {{h}WT+wT} \right), {{{\boldsymbol x}^l}\left( \tau \right),{\boldsymbol p}^l\left( \tau \right)} \bigr)} }} \\
&{~~~~~~} - \sum\limits_{l = 1}^L {\sum\limits_{w' = w+1}^{W-1} {{{Q_l}\left( {{h}WT+w'T} \right)}\sum\limits_{\tau = \left({h}W+w'\right)T}^{\left({h}W+w'+1\right)T- 1}{r_l\bigl({\boldsymbol \alpha} \left( {{h}WT+w'T} \right), {{{\boldsymbol x}^l}\left( \tau \right),{\boldsymbol p}^l\left( \tau \right)} \bigr)} }} \\
&{~~~~~~}+ \sum\limits_{l = 1}^L {\sum\limits_{w' = w+1}^{W-1} {{{Q_l}\left( {{h}WT+w'T} \right)}\sum\limits_{\tau = \left({h}W+w'\right)T}^{\left({h}W+w'+1\right)T- 1} {\left(A_l\left(\tau \right)+\theta\right)} }}\\
&{\rm{s.t.}}~~~~~{\rm constraints~~ (\ref{equ:feasibility}),(\ref{equ:subchannelconstraint}),(\ref{equ:constraintforresource:a}),(\ref{equ:powerbudeget}),(\ref{equ:constraintforresource:b})},\\
&{\rm{var.}}~~~~{\boldsymbol \alpha}\left( {h}WT+wT\right),{\boldsymbol x}\left( \tau\right), {\boldsymbol p}\left( \tau\right), \tau\in{\cal T}_{hW+w}.
\end{split}\label{equ:appendix:heavy:c}
\end{align}
\hrulefill
\end{figure*}
According to the queueing dynamics (\ref{equ:queueing}) and the fact that ${{Q_l}\left( {{h}WT} \right)}\ge WT r_{\max},\forall l\in{\cal L}$, we can express ${{Q_l}\left( {{h}WT+w'T} \right)}$ as (\ref{equ:appendix:heavy:b}).
\begin{figure*}
\begin{align}
\nonumber
& {{Q_l}\left( {{h}WT+w'T} \right)}={{Q_l}\left( {{h}WT+wT} \right)}+\sum\limits_{w''=w+1}^{w'-1}\sum\limits_{\tau = \left({h}W+w''\right)T}^{\left({h}W+w''+1\right)T- 1}{\left(A_l\left(\tau \right)+\theta\right)}\\
\nonumber
& -\sum\limits_{w''=w+1}^{w'-1}\sum\limits_{\tau = \left({h}W+w''\right)T}^{\left({h}W+w''+1\right)T- 1}{{r_l\bigl({\boldsymbol \alpha} \left( {{h}WT+w''T} \right), {{{\boldsymbol x}^l}\left( \tau \right),{\boldsymbol p}^l\left( \tau \right)} \bigr)}}\\
& +\sum\limits_{\tau = \left({h}W+w\right)T}^{\left({h}W+w+1\right)T- 1}{\left(A_l\left(\tau \right)+\theta\right)}-\sum\limits_{\tau = \left({h}W+w\right)T}^{\left({h}W+w+1\right)T- 1}{{r_l\bigl({\boldsymbol \alpha} \left( {{h}WT+wT} \right), {{{\boldsymbol x}^l}\left( \tau \right),{\boldsymbol p}^l\left( \tau \right)} \bigr)}}.\label{equ:appendix:heavy:b}
\end{align}
\hrulefill
\end{figure*}
Only the last term in (\ref{equ:appendix:heavy:b}) depends on the operation during frame ${\cal T}_{hW+w}$. Now we define the following constants:
\begin{align}
\nonumber
& C_{R,l}\triangleq \\
& {\sum\limits_{w' = w+1}^{W-1} \!\!\!{\sum\limits_{\tau = \left({h}W+w'\right)T}^{\left({h}W+w'+1\right)T- 1}{\!\!\!\!r_l\bigl({\boldsymbol \alpha} \left( {{h}WT+w'T} \right), {{{\boldsymbol x}^l}\left( \tau \right),{\boldsymbol p}^l\left( \tau \right)} \bigr)} }},\\
& C_{A,l}\triangleq {\sum\limits_{w' = w+1}^{W-1} {\sum\limits_{\tau = \left({h}W+w'\right)T}^{\left({h}W+w'+1\right)T- 1} {\left(A_l\left(\tau \right)+\theta\right)} }}.
\end{align}
Then we can further rewrite problem (\ref{equ:appendix:heavy:c}) as (\ref{equ:appendix:heavy:d}).
\begin{figure*}
\begin{align}
\begin{split}
&\min~V{\sum\limits_{\tau = \left({h}W+w\right)T}^{\left({h}W+w+1\right)T - 1} {P\bigl({\boldsymbol \alpha}\left({h}WT+wT\right),{\boldsymbol p}\left(\tau\right)\bigr)}} \\
&{~~~~~~} - \sum\limits_{l = 1}^L { {\left({{Q_l}\left( {{h}WT+wT} \right)}-C_{R,l}+C_{A,l}\right)\sum\limits_{\tau = \left({h}W+w\right)T}^{\left({h}W+w+1\right)T- 1}{r_l\bigl({\boldsymbol \alpha} \left( {{h}WT+wT} \right), {{{\boldsymbol x}^l}\left( \tau \right),{\boldsymbol p}^l\left( \tau \right)} \bigr)} }} \\
&{\rm{s.t.}}~~~~~{\rm constraints~~ (\ref{equ:feasibility}),(\ref{equ:subchannelconstraint}),(\ref{equ:constraintforresource:a}),(\ref{equ:powerbudeget}),(\ref{equ:constraintforresource:b})},\\
&{\rm{var.}}~~~~{\boldsymbol \alpha}\left( {h}WT+wT\right),{\boldsymbol x}\left( \tau\right), {\boldsymbol p}\left( \tau\right), \tau\in{\cal T}_{hW+w}.\label{equ:appendix:heavy:d}
\end{split}
\end{align}
\hrulefill
\end{figure*}
Comparing problem (\ref{equ:appendix:heavy:d}) with problem (\ref{equ:ENSRA}), we find they are essentially the same optimization problems. In problem (\ref{equ:appendix:heavy:d}), we modify the queue backlog term by constants $C_{R,l}$ and $C_{A,l}$, which are related to the future transmission rates and traffic arrivals. Now, we have shown that, if we have ${{Q_l}\left( {{h}WT} \right)}\ge WT r_{\max},\forall l\in{\cal L}$, the problem in line \ref{line:A} can be solved as problem (\ref{equ:ENSRA}).${\rm{~~~~~~~~~~~~~~~~~~~~~~~~~~~~~~~~~~~~~~~~~~~~~~~~~~~~~~}}\Box$

\end{document}